\documentclass[11pt,a4paper]{article}
\pdfoutput=1
\usepackage[hmargin=3.0cm,vmargin=3.0cm]{geometry}

\usepackage{caption,subcaption}

\usepackage[table]{xcolor}

\usepackage{bm,amsmath,amsfonts,amssymb,systeme}
\usepackage{graphicx}
\usepackage{subcaption}
\graphicspath{{figures/}} % Location of the graphics files

\usepackage{algorithm}
\usepackage[noend]{algpseudocode}
\usepackage[sort,square,numbers]{natbib}

\usepackage{hyperref}
\usepackage{authblk}
\hypersetup{
colorlinks=true,
citecolor=red
}

\DeclareMathOperator{\E}{\mathbb{E}}
\DeclareMathOperator{\K}{\mathcal{K}}
\DeclareMathOperator{\FF}{\mathcal{F}}
\DeclareMathOperator{\LL}{\mathcal{L}}

\DeclareMathOperator{\U}{\mathcal{U}}

\begin{document}\sloppy

\title{Optimal Bayesian experimental design for subsurface flow problems\footnote{The current manuscipt is a preprint of the paper published in Computer Methods in Applied Mechanics and Engineering. DOI: \href{https://www.sciencedirect.com/science/article/pii/S0045782520303935}{10.1016/j.cma.2020.113208}}}

\author[a]{Tarakanov, Alexander \thanks{alexander.tarakanov@manchester.ac.uk}}
\author[b]{Elsheikh, Ahmed H. \thanks{a.elsheikh@hw.ac.uk}}
\affil[a]{Department of Mathematics,
          University of Manchester, Manchester, UK}
\affil[b]{School of Energy, Geoscience, Infrastructure and Society,
          Heriot-Watt University, Edinburgh, UK}
% \author{Tarakanov, Alexander \and Elsheikh, Ahmed H.}

% \author[HWU]{Alexander\corref{cor1}}
% \ead{Tarakanov, Alexander <a.tarakanov@hw.ac.uk>}

% \author[HWU]{Ahmed H. Elsheikh}
% \ead{a.elsheikh@hw.ac.uk}
% \address[HWU]{
%   School of Energy, Geoscience, Infrastructure and Society,\\
%   Heriot-Watt University, Edinburgh, UK \\
% }
% \cortext[cor1]{Corresponding author}

\maketitle

\begin{abstract}

% In recent years, there has been an increasing interest in application of optimal Bayesian experimental design to complex physical and engineering systems. 
Optimal Bayesian design techniques provide an estimate for the best parameters of an experiment in order to maximize the value of measurements prior to the actual collection of data. In other words, these techniques explore the space of possible observations and determine an experimental setup that produces maximum information about the system parameters on average. Generally, optimal Bayesian design formulations result in multiple high-dimensional integrals that are difficult to evaluate without incurring significant computational costs as each integration point corresponds to solving a coupled system of partial differential equations.

In the present work, we propose a novel approach for development of polynomial chaos expansion (PCE) surrogate model for the design utility function. In particular, we demonstrate how the orthogonality of PCE basis polynomials can be utilized in order to replace the expensive integration over the space of possible observations by direct construction of PCE approximation for the expected information gain. This novel technique enables the derivation of a reasonable quality response surface for the targeted objective function with a computational budget comparable to several single-point evaluations. Therefore, the proposed technique reduces dramatically the overall cost of optimal Bayesian experimental design. We evaluate this alternative formulation utilizing PCE on few numerical test cases with various levels of complexity to illustrate the computational advantages of the proposed approach.
\end{abstract}

%=========================================
\section{Introduction}
The extraordinary progress in available computational power and the recent advances in computational methods made a significant impact on engineering applications with the wide adoption of simulation based design approaches. Nowadays, it is common to represent any given physical, economical, biological system as a mathematical model. Such a system could be generally formulated by:
\begin{equation}
    \label{eq:mathematical_model}
    \mathbf{m} = f(\mathbf{\theta}, \mathbf{d})
\end{equation}
where $\mathbf{m}$ is a vector of observable quantities, $\mathbf{\theta}$ is a vector of model parameters, $\mathbf{d}$ are the controlled parameters (aka. design) and $f$ is a linear or nonlinear operator describing the system dynamics. For instance, numerical simulations are widely used for partial replacement of real experiments in geophysics~\citep{2003GeoJI.155..411V}, subsurface flow problems~\citep{doi:10.1002/2016WR018598, doi:10.1029/2009WR008312} and pharmacokinetics~\citep{RYAN201526} studies. Representation of real-world systems as done in Eq.~\eqref{eq:mathematical_model} is quite generic and could be applied to a large number of practically important problems with mathematical models of different complexity, starting from simplistic algebraic models~\citep{doi:10.1111/j.0006-341X.2004.00148.x} and ending by a set of ordinary differential equations (ODE)~\citep{KRAMER20101645} or partial differential equations (PDE)~\citep{doi:10.1002/2014WR015740}.

Despite possible differences in nature and complexity of the system of interest, model calibration is most likely an unavoidable task. At this step, the poorly known model parameters $\mathbf{\theta}$ are inferred from measured data $\mathbf{m}$. However, given a partially observed system with strong nonlinearities, the model parameters could only be statistically estimated where the prior distribution of the model parameters could be updated to the posterior distribution given $\mathbf{m}$ for any given values of design parameters $\mathbf{d}$ using a Bayesian framework.
In this setting, it is reasonable to seek specific set of design parameters $\mathbf{d}$ that could minimize the uncertainties in the estimated model parameters $\mathbf{\theta}$ or provide a better set of parameters maximizing the accuracy of the model predictions~\citep{10.2307/2246015}. The case of partially observed system governed by strong nonlinearities is very pronounced in subsurface flow modeling and finding the optimal design parameters $\mathbf{d}$ that maximize a certain criterion for an observed dataset is generally a challenging task~\citep{10.2307/2246015}.

Historically, a variety of schemes for model parameters estimation have been developed~\citep{Elsheikh2012, ELSHEIKH2014515, ELSHEIKH201310, ELSHEIKH201314, ELSHEIKH201514}. Similarly various techniques for experimental design could be adopted~\citep{CEULEMANS200241, box_behnken, well_test} depending on the objectives of measurements and constraints on the experimental setup. For instance, sampling based methods for experimental design allow one to investigate all possible combinations of model parameters and investigate their interactions. Such methods are known as full factorial design~\citep{montgomery2012design}. However, full factorial design is prone to redundant calculations as the model response might not be sensitive to some model parameters or might be insensitive to some of the interaction terms of the model parameters. Efficient sampling methods for experimental design have been developed to account for the model sensitivity including fractional factorial design~\citep{montgomery2012design}, Placket-Burman design~\citep{10.1093/biomet/33.4.305}, Box-Brehnken design~\citep{doi:10.1080/00401706.1960.10489912}, Latin hypercube sampling~\citep{doi:10.1080/00401706.1979.10489755} and central composite design~\citep{ASADI2017335}. Sampling techniques demonstrate excellent performance and robustness in various practical applications. However, the main focus of all of these methods is to minimize the number of experimental samples needed to infer the model parameters. Therefore, such methods do not guarantee the optimality of the experimental scheme in terms of accuracy or reliable model predictions. This issue was resolved by applying Bayesian methods to data acquisition problems.

Bayesian-like approaches provide a solid mathematical foundation with strong theoretical guarantees for solving data acquisition problems. However, Bayesian techniques relies on high-dimensional integrations that could be performed analytically for a very limited class of problems. Therefore, numerical methods are frequently implemented in order to utilize Bayesian solutions for experimental design~\citep{TEREJANU2012178}. Unfortunately, the computational costs to solve an optimal Bayesian experimental design can be too high even for modern computing systems~\citep{doi:10.1198/1061860032012}. Therefore, in most of the cases direct Bayesian approach to optimal design is applied to systems that can be described by either algebraic model or by ODE~\citep{Walker2019BayesianDO, a694a2872b9d4b709fc6705616ff718f}. Examples of systems that are modeled through the numerical solution of PDEs are less common in the literature on optimal Bayesian experimental design ~\citep{HUAN2013288}, because of the computational costs associated with multidimensional integration. Due to the high practical importance of systems governed by PDEs, several approaches have been proposed in order to perform optimal Bayesian experimental design for such systems.

One approach is to utilize surrogate modeling techniques to replace the expensive numerical simulator with a relatively cheap proxy model. For instance in~\citep{10.2307/2291522}, optimal Bayesian experimental design has been conducted with model-specific response surfaces. Along the same line, Gaussian process surrogates~\citep{doi:10.1002/2016WR018598, HU201792} were utilized to reduce the computational cost of solving the optimal Bayesian experimental design problem. Similarly, Polynomial chaos expansion (PCE) surrogates~\citep{10.2307/2371268} have been utilized to accelerate the solutions of Bayesian experimental design problems~\citep{doi:10.1137/15M1043303, doi:10.1061/AJRUA6.0000966}.

A second common approach in the published literature relies on introducing algorithmic improvement to reduce the cost of evaluation of multi-dimensional integrals that appear in the intermediate calculations of Bayesian experimental design work-flows. For instance, advanced MCMC methods can significantly reduce the cost of design utility function evaluation~\citep{doi:10.1080/07362994.2019.1705168}. Even more significant cost reduction can be achieved if the problem specific knowledge is exploited. In many practical cases certain simplifying assumptions can be made that allow one to partially replace the numerical integration steps with analytical methods. For instance, locally linearized objective functions were utilized by~\citet{MOSBACH20121303} and tabular approximation of the posterior distribution for certain cases were utilized by~\citet{doi:10.1080/00401706.2012.680399}. If high number of measurements with low level of noise are available, then Laplace approximation techniques can be successfully adopted for estimating the Bayesian evidence factors~\citep{long2013fast, beck2018fast}.

In the current manuscript, we combine two techniques for reducing the computational costs of solving the Bayesian experimental design problems. PCE-based response surface is utilized to replace the numerically expensive PDE solver. In addition to that, we introduce a novel approach to build a PCE surrogate for design utility function relying on the orthogonality properties of PCE basis functions. We demonstrate that integration over the space of possible observations is equivalent to projection in the space of polynomial functions. Given that relation between integration and projection, the response surface for the design utility function is built using few evaluations of the expected information gain. The overall computational cost of solving for the optimal design parameters is significantly reduced in comparison to direct implementations of existing optimal Bayesian experimental design methods. The proposed approach is applied to subsurface-flow problems where the model parameters are highly uncertain and the system dynamics are governed by coupled system of nonlinear equations that is computationally expensive to solve. We also note, that the introduced formulation is quite generic and can be applied to solve Bayesian experimental design problems appearing in different engineering fields without significant modifications.

The rest of the paper is organized as follows: in section~\ref{sec:methodology}, the mathematical formulation of Bayesian experimental design is introduced followed by an introduction to the proposed method. Section~\ref{sec:num_ex} evaluate the proposed technique on a number of test cases and in the last section~\ref{sec:conclusion} the advantage and limitations of introduced PCE-based Bayesian experimental design are discussed.

\section{Methodology}
\label{sec:methodology}
In this section, we present the mathematical details of the Bayesian experimental design generally along with the details of PCE proxy modeling. In subsection \ref{subsec:bed_intro}, the mathematical formulation of optimal Bayesian experimental design is introduced. Subsection \ref{subsec:overview_of_polynomial_chaos} presents an overview of PCE. Finally, in subsection \ref{subsec:bed_numerics}, we present the numerics behind the proposed efficient method to find optimal experimental designs.

\subsection{Bayesian Experimental Design}
\label{subsec:bed_intro}
Optimal experimental design is concerned with finding of the best experimental setup for a given physical system or mathematical model in order to achieve extremum value of a certain well-defined quantitative criterion. Common examples of the criterion concerned are based on the variance of model parameters~\citep{RYAN201526, doi:10.1111/biom.12081, f4a8ebf309f7474787235f2bdbb14498}. It is important to note that the optimal solution reflects features not only of physical system itself but also it can be highly sensitive to the selection of the optimality criterion~\citep{doi:10.1111/j.0006-341X.2004.00148.x}. Therefore, depending on ultimate goals of the experiment, whether it is the precise parameter estimation or the accurate model predictions, the conditions for optimal experiment can be different.

In the present study, real-world physical system are abstracted using a mathematical model of the form:
\begin{equation}
    \label{eq:system_bed}
    \mathbf{m} = f(\mathbf{\theta}, \mathbf{d}) + \mathbf{\eta}
\end{equation}
where all the notations are similar to Eq.~\eqref{eq:mathematical_model} and $\mathbf{\eta}$ is the measurement noise or the model errors. In the present work we adopt Bayesian approach to optimal experimental design for model parameter estimations following a D-optimality criterion~\citep{doi:10.1002/sim.3395}. In Bayesian framework, model parameters and measured values are considered as random variables with certain probability distribution. Therefore, it is natural to quantify the significance of a given experiment through information theory techniques. In such setting it is assumed that there is a prior information about model parameters that is expressed through prior probability distribution:
%============
\begin{equation}
    \label{eq:prior_distibution}
    p(\mathbf{\theta}) = p(\mathbf{\theta} | \mathbf{d})
\end{equation}
%============
where $p(\mathbf{\theta})$ and $p(\mathbf{\theta} | \mathbf{d})$ denotes the prior distribution for a given design $\mathbf{d}$. Equation~\eqref{eq:prior_distibution} expresses a common assumption that the prior distribution of the model parameters $\mathbf{\theta}$ does not depend on the design $\mathbf{d}$. The prior distribution $p(\mathbf{\theta}|\mathbf{d})$ is typically selected from relatively simple distribution like uniform or normal~\citep{doi:10.1029/2018WR023615, PAPADIMITRIOU2017972, LUNA2018943, LAINEZAGUIRRE2015312}. For a given model parameters $\mathbf{\theta}$, design $\mathbf{d}$ the probability distribution of observations is determined by the likelihood function, which is simply a conditional probability $p(\mathbf{m}|\mathbf{\theta}, \mathbf{d})$ of observing a given model output $\mathbf{m}$ for given the model parameters $\mathbf{\theta}$ and the design $\mathbf{d}$. In the present work, the likelihood function is assumed to be a Gaussian of the from:
%============
\begin{equation}
    \label{eq:likelihood_fun}
    p(\mathbf{m}|\mathbf{\theta}, \mathbf{d}) = \frac{1}{{(2 \pi \sigma^2)}^{\text{dim}(\mathbf{m})/2}} \exp{\bigg (-\frac{(\mathbf{m} - f(\mathbf{\theta}, \mathbf{d}))^2}{2\sigma^2} \bigg )}
\end{equation}
%============
where $p(\mathbf{m}|\mathbf{\theta}, \mathbf{d})$ is a conditional probability distribution of observing $\mathbf{m}$ for given a $\mathbf{\theta}$ and $\mathbf{d}$, $\sigma$ is the standard deviation of the distribution and $\text{dim}(\mathbf{m})$ is the dimension of $\mathbf{m}$. The joint distribution of the model parameters and observations can be derived from the definition of conditional distributions:
%============
\begin{equation}
    \label{eq:joint_distribution}
    p(\mathbf{m}, \mathbf{\theta} | \mathbf{d}) = p(\mathbf{m} | \mathbf{\theta}, \mathbf{d}) p(\mathbf{\theta} | \mathbf{d})
\end{equation}
%============
where $p(\mathbf{m}, \mathbf{\theta} | \mathbf{d})$ is a joint probability distribution of observations and model parameters. The well-known Bayes' Theorem~\citep{Held:2013:ASI:2601513} admits an alternative expression for the joint probability distribution in Eq.~\eqref{eq:joint_distribution}:
%============
\begin{equation}
    \label{eq:joint_distribution_inverse}
    p(\mathbf{m}, \mathbf{\theta} | \mathbf{d}) = p(\mathbf{\theta} | \mathbf{m}, \mathbf{d}) p(\mathbf{m} | \mathbf{d})
\end{equation}
%============
where $p(\mathbf{m}|\mathbf{d})$ is the Bayesian evidence factor and $p(\mathbf{\theta}|\mathbf{m}, \mathbf{d})$ is the posterior distribution of the model parameters. Equation~\eqref{eq:joint_distribution} and Eq.~\eqref{eq:joint_distribution_inverse} can be utilized to derive a well-known expression for the Bayesian evidence:
%============
\begin{equation}
    \label{eq:bayesian_evidence}
    p(\mathbf{m} | \mathbf{d}) = \int p(\mathbf{m} | \mathbf{\theta}, \mathbf{d}) p(\mathbf{\theta} | \mathbf{d}) \,d\mathbf{\theta}
\end{equation}
%============
The posterior probability distribution function can be computed via Eq.~\eqref{eq:joint_distribution} - Eq.~\eqref{eq:bayesian_evidence} as follows:
%============
\begin{equation}
    \label{eq:posterior_distribution}
    p(\mathbf{\theta} | \mathbf{m}, \mathbf{d}) = \frac{p(\mathbf{m}| \mathbf{\theta}, \mathbf{d}) p(\mathbf{\theta} | \mathbf{d}) }{p(\mathbf{m} | \mathbf{d})}
\end{equation}
%============
The Posterior distribution defined in Eq.~\eqref{eq:posterior_distribution} provides information about the probability distribution of the model parameters $\mathbf{\theta}$ that are in agreement with observed data. Therefore, the difference between the posterior and the prior distribution indicates the value of measurements. In the ideal cases, the posterior distribution should be narrow with high peak in the region of true model parameters. In information theory setting, the distance between distribution could be quantified via the Kullback-Leibler (KL) divergence~\citep{kullback1951}:
%============
\begin{equation}
    \label{eq:kl_divergence}
    D_{\text{KL}}(\mathbf{m}, \mathbf{d}) = \int p(\mathbf{\theta}| \mathbf{m}, \mathbf{d}) \log \bigg( \frac{p(\mathbf{\theta}| \mathbf{m}, \mathbf{d})}{p(\mathbf{\theta}| \mathbf{d})} \bigg) \,d\mathbf{\theta}
\end{equation}
%============
where $D_{\text{KL}}(\mathbf{m}, \mathbf{d})$ is a value of KL-divergence or information gain for posterior distribution corresponding to given measurements $\mathbf{m}$ and design $\mathbf{d}$. Therefore, KL-divergence can be computed only after an experiment has been conducted. Obviously, this is not possible at the stage when the design of experiment is only planned and values of observations have not been collected yet. This issue is resolved in the theory of Bayesian experimental design by considering the mean value of the KL-divergence defined in Eq.~\eqref{eq:kl_divergence} with respect to all possible observations. The mean KL-divergence is defined as:
%============
\begin{equation}
    \label{eq:expected_kl_divergence}
    U(\mathbf{d}) = \int D_{\text{KL}}(\mathbf{m}, \mathbf{d}) p(\mathbf{m}|\mathbf{d}) \,d\mathbf{m}
\end{equation}
%============
where $U(\mathbf{d})$ is mean-value of KL-divergence (aka. expected information gain). In the present work, we consider $U(\mathbf{d})$ as the utility or an objective function that represents the quality of a given experimental design $\mathbf{d}$ and in order to determine the optimal parameters of an experiment, the following problem should be solved:
%============
\begin{equation}
    \label{eq:d_optimal_design}
    \mathbf{d}_* = \underset{\mathbf{d}}{\arg\max} \big(U(\mathbf{d})\big)
\end{equation}
%============
where parameters $\mathbf{d}_*$ maximize the value of expected information gain defined in Eq.\eqref{eq:expected_kl_divergence}. Therefore $\mathbf{d}_*$ corresponds to the best possible experimental design in the framework of D-optimality criterion.

\subsection{Polynomial Chaos Expansion}
\label{subsec:overview_of_polynomial_chaos}
Polynomial chaos expansion (PCE) is a response surface technique that utilizes decomposition of multivariate function as a series of orthogonal polynomials. In PCE, a square-integrable function $f(\mathbf{x}) = f(x_1, ... , x_n)$ of a vector $\mathbf{x}$ with dimension $n$ is represented as follows~\citep{TARAKANOV2019108909}:
%============
\begin{equation}
    \label{eq:pce_expansion}
    f(\mathbf{x}) = \sum_A c_A p_A(\mathbf{x}) = \sum_{\alpha_1, ... , \alpha_n} c_{\alpha_1, ... , \alpha_n}p_{\alpha_1}^{(1)}(x_1) ... p_{\alpha_n}^{(n)}(x_n)
\end{equation}
%============
where $A = \alpha_1, ... , \alpha_n$ is a multi-index, $c_A$ is PCE coefficient, $p_{\alpha_k}^{(k)}(x_k)$ is a single-variate polynomial of degree $\alpha_k$ that depends only on k-th coordinate of vector $\mathbf{x}$. In certain problems, it is reasonable to utilize different types of orthogonal polynomials for different components of $\mathbf{x}$, which is reflected by the superscript in the notation $p_{\alpha_k}^{(k)}(x_k)$. Therefore, the basis polynomials can be expressed as:
%============
\begin{equation}
    \label{eq:basis_polynomials}
    p_A(\mathbf{x}) = p_{\alpha_1}^{(1)}(x_1) ... p_{\alpha_n}^{(n)}(x_n)
\end{equation}
%============
The essential part of PCE is the construction of orthogonal polynomials that form a basis in the space of functions of $\mathbf{x}$. For the purposes of uncertainty quantification (UQ) and sensitivity analysis (SA), the inner product in the space of polynomial basis functions should be related to
the statistics of the input data~\citep{PCE_Stat1}. In other words, the probability distribution with density $\K(\mathbf{x})$ induces an inner product on the space of square-integrable functions:
%============
\begin{equation}
    \label{eq:inner_product}
    \langle f_1, f_2 \rangle = \int \K (\mathbf{x}) f_1(\mathbf{x}) f_2(\mathbf{x})  \,d\mathbf{x} = \E[f_1, f_2]
\end{equation}
%============
where $f_1(\mathbf{x})$ and $f_2(\mathbf{x})$ are an arbitrary square-integrable functions with respect to probability distribution determined by the density function $\K(\mathbf{x})$. In the PCE framework, it is supposed that orthogonality of polynomial basis functions is in agreement with the inner-product introduced in Eq.~\eqref{eq:inner_product}:
%============
\begin{equation}
    \label{eq:orthogonality}
    \langle p_A(\mathbf{x}), p_B(\mathbf{x}) \rangle = \langle p_A(\mathbf{x}), p_A(\mathbf{x}) \rangle \delta_{AB} 
\end{equation}
%============
where $\delta_{AB}$ is a Kronecker symbol. It is possible to construct PCE basis functions for generic kernels and this could improve the quality of the constructed response surface~\citep{PCE_Preconditioning}. However, for many practical problems, the components of $\mathbf{x}$ are statistically independent or can be made such via appropriate transformation~\citep{rosenblatt1952}. Therefore, without loss of generality, kernel functions or densities of probability distribution of the following form can be considered:%============
\begin{equation}
    \label{eq:stat_independence}
    \K(\mathbf{x}) = \K_1 (x_1) ... \K_n (x_n)
\end{equation}
%============
where $\K_a(x_a)$ are single-variate distribution functions.

It can be shown that for statistically independent components of $\mathbf{x}$, the polynomials of type Eq.~\eqref{eq:basis_polynomials} satisfy Eq.~\eqref{eq:orthogonality} if single variate polynomials $p_{\alpha_a}^{(a)}(x_a)$ form family of orthogonal functions with respect to the inner product determined by the corresponding single-variate factor $\K_a(x_a)$ in the kernel function $\K(\mathbf{x})$. Therefore, for normally distributed variable with zero mean and unit variance Hermite polynomials should be used. Similarly, for the components of $\mathbf{x}$ that are uniformly distributed in the interval $[-1,1]$, Legendre polynomials should be utilized. 

It is simple to demonstrate that exact convergence of PCE to the approximated function can be achieved only for infinite number of terms in Eq.~\eqref{eq:pce_expansion}. In practical applications, various truncation techniques are adopted~\citep{classic_sparse_PCE, Hyperbolic_Truncation_Scheme}. The finite number of coefficients of truncated PCE could be determined by solving an error-minimization problem~\citep{DIAZ2018640, PCE_SVR, PCE_PCG, BAZARGAN2015, PCE_Residual_Based_Ranking} or by a collocation techniques~\citep{PC_Hermite, 71f9c0cab8754d25b8bf1a985052b1b2, Collocation_1, Babaei_Paper2}. In the current work regression-type approach is adopted and the mean-square error functional with Elastic-Net regularization is utilized~\citep{Regularized_Linear_Regression, TARAKANOV2019108909}. Therefore, PCE coefficients are estimated by solving:
%============
\begin{equation}
    \label{eq:minimization_problem}
    \mathbf{c_*} = \underset{\textbf{c}}{\arg\min} \LL(\mathbf{c}) = \underset{\textbf{c}}{\arg\min} \bigg( \FF(\textbf{c}) + \lambda_1 \sum_{A} |c_A| + \lambda_2 \sum_{A} c_A^2\bigg)
\end{equation}
%============
where $\mathbf{c}_*$ is the solution for the error minimization problem, $\lambda_1$ and $\lambda_2$ are the regularization hyper-parameters, $\FF(\mathbf{c})$ is a mean-square error functional and $\LL(\mathbf{c})$ is a regularized mean-square error functional. The values of the hyper-parameters $\lambda_1$ and $\lambda_2$ are determined through cross-validation~\citep{Scikit_Learn}. The mean-square error functional in Eq.~\eqref{eq:minimization_problem} is defined as:
%============
\begin{equation}
    \label{eq:mean_square_functional}
    \FF(\mathbf{c}) = \sum_i \frac{(y_i - \sum_A \textit{c}_Ap_A(\textbf{x}_i))^2} {N}
\end{equation}
%============
where the index $i$ takes all possible values from $0$ till $N$ the total number of training samples and $y_i$ is the quantity of interest (QoI) value at point $x_i$. In the current work, we truncate PCE by the total degree $d_{\text{poly}}$ of basis polynomial function. Therefore, the overall number of PCE coefficients $N_c$ that have to be calculated via Eq~\eqref{eq:minimization_problem} can be computed as:
%============
\begin{equation}
    \label{eq:dimension}
    N_c = \binom{n+d_{\text{poly}}}{n}
\end{equation}
%============
It is obvious that $N_c$ increases exponentially with the growth of the dimension $n$ and the polynomial degree $d_{\text{poly}}$. However, the PCE convergence rate is relatively high especially for smooth functions~\citep{refId0} and thus a tractable number of PCE terms can provide an accurate response surface.
In addition to that, the first regularization term in Eq.~\eqref{eq:minimization_problem} promotes sparsity of the PCE model and the second term improves the numerical stability of the regression problem. The latter simplifies the numerical solution for the regression problem defined in Eq.~\eqref{eq:minimization_problem} and enables faster function-evaluation, which is rather important for optimal Bayesian experimental design. In the present work, PCE surrogate models are developed with polychaos-learn library~\citep{polychaos-learn, polychaos-learn-git}. In particular, we utilize the variety of powerful regularization techniques and advanced methods for hyperparameters tuning (aka cross-validation) that are supported in polychaos-learn because of the integration with Scikit-learn~\citep{Scikit_Learn} library for machine-learning.

\subsection{Bayesian experimental design with polynomial chaos expansion}
\label{subsec:bed_numerics}
For a limited number of situations (e.g. linear forward models), the utility function defined in Eq.~\eqref{eq:expected_kl_divergence} could be estimated analytically~\citep{doi:10.1111/j.0006-341X.2004.00148.x}. Unfortunately, this is not the case for most of the practical systems. For instance, in subsurface flow models the function $f(\mathbf{\theta}, \mathbf{d})$ in Eq.~\eqref{eq:system_bed} relies on the solution of a coupled system of non-linear PDEs~\citep{LONG2015123}.
Therefore, analytical expressions for the utility function Eq.~\eqref{eq:expected_kl_divergence} do not generally exit. For these systems, numerical integration is the only feasible methods. One of the standard techniques for estimating the expected information gain is based on substituting the expressions for 
conditional probabilities (Eq.~\eqref{eq:joint_distribution} and Eq.~\eqref{eq:joint_distribution_inverse}) into Eq.~\eqref{eq:expected_kl_divergence} for the expected information gain. This results in the following~\citep{HUAN2013288}:
%============
\begin{equation}
    \label{eq:expected_kl_divergence_v1}
    U(\mathbf{d}) = \int p(\mathbf{m} | \mathbf{\theta}, \mathbf{d}) \log \big(p(\mathbf{m} | \mathbf{\theta}, \mathbf{d})\big) p(\mathbf{\theta}|\mathbf{d}) \,d\mathbf{\theta} \,d\mathbf{m} - \int p(\mathbf{m} | \mathbf{d}) \log \big(p(\mathbf{m}|\mathbf{d})\big) \,d\mathbf{m}
\end{equation}
%============
Each of the two terms in Eq.~\eqref{eq:expected_kl_divergence_v1} is then estimated separately. In generic cases, first term is not very computationally involved. Moreover, in the case of normally distributed noise, the first term in Eq.~\eqref{eq:expected_kl_divergence_v1} can be computed analytically:
\begin{equation}
    \label{eq:analytical_expression}
    \begin{gathered}
    \int p(\mathbf{m} | \mathbf{\theta}, \mathbf{d}) \log \big(p(\mathbf{m} | \mathbf{\theta}, \mathbf{d})\big) p(\mathbf{\theta}|\mathbf{d}) \,d\mathbf{\theta} \,d\mathbf{m} = \\
    = \int p\big(f(\mathbf{\theta}, \mathbf{d}) + \mathbf{\eta} | \mathbf{\theta}, \mathbf{d}\big) \log \Big(p\big(f(\mathbf{\theta}, \mathbf{d}) + \mathbf{\eta} | \mathbf{\theta}, \mathbf{d}\big)\Big) p(\mathbf{\theta}|\mathbf{d}) \,d\mathbf{\theta} \,d\mathbf{\eta}= \\
    = \int \mathcal{N}(\mathbf{\eta}, 0, \sigma) \log\big(\mathcal{N}(\mathbf{\eta}, 0, \sigma)\big) p(\mathbf{\theta} | \mathbf{d}) \,d\mathbf{\eta} \,d\mathbf{\theta} = \\
    = -\int \mathcal{N}(\mathbf{\eta}, 0, \sigma)  \Big( \frac{1}{2} \text{dim}(\mathbf{m}) \log(2\pi) + \text{dim}(\mathbf{m}) \log(\sigma) + \frac{|\mathbf{\eta}|^2}{2\sigma^2}  \Big) p(\mathbf{\theta} | \mathbf{d}) \,d\mathbf{\eta}\,d\mathbf{\theta} = \\
    = -\int \text{dim}(\mathbf{m}) \Big (  \frac{\log(2\pi) + 1}{2} +  \log(\sigma) \Big) p(\mathbf{\theta} | \mathbf{d}) \,d\mathbf{\theta} = \\
    = -\text{dim}(\mathbf{m}) \Big (  \frac{\log(2\pi) + 1}{2} +  \log(\sigma) \Big) \\
    \end{gathered}
\end{equation}
where $\text{dim}(\mathbf{m})$ is the dimension of observation vector and $\mathcal{N}(\mathbf{\eta}, 0, \sigma)$ is the density of $\text{dim}(\mathbf{m})$-dimensional normal distribution with zero mean and standard deviation $\sigma$ at the point $\mathbf{\eta}$. The most challenging part in the numerical calculation of $U(\mathbf{d})$ is due to the second term. One of the possible options is to utilize a Markov chain Monte Carlo (MCMC) over $\mathbf{m}$ that are in agreement with the probability distribution determined by Bayesian evidence factors $p(\mathbf{m}|\mathbf{d})$. In other words, if MCMC samples $m_i$ with $i = 1 \ ... \ N$ are generated from $p(\mathbf{m}|\mathbf{d})$, then the following equality holds~\citep{HUAN2013288}:
%============
\begin{equation}
    \label{eq:mcmc_int}
    \int p(\mathbf{m} | \mathbf{d}) \log (p(\mathbf{m}|\mathbf{d})) \,d\mathbf{m} \approx \frac{1}{N} \sum_{i=1}^{N} \log \big(p(\mathbf{m}_i|\mathbf{d})\big)
\end{equation}
%============
Here, the summation of $\log (p(\mathbf{m}_i|\mathbf{d}))$ is performed over MCMC samples generated from $p(\mathbf{m}|\mathbf{d})$ known as outer MCMC loop. The main challenge with this approach is that each calculation of $p(\mathbf{m} | \mathbf{d})$ requires either numerical integration with another MCMC chain or intensive Monte Carlo (MC) simulations. For instance, the Bayesian evidence could be estimated using MC samples generated from the prior distribution~\citep{HUAN2013288, doi:10.1002/2014WR016062}:
\begin{equation}
    \label{eq:b_ev_sampling_from_prior}
    p(\mathbf{m}|\mathbf{d}) \approx \frac{1}{N} \sum_{i=1}^{N} p(\mathbf{m}_i|\mathbf{\theta}_i, \mathbf{d})
\end{equation}
Alternatively, MCMC samples generated from posterior distribution could be utilized~\citep{doi:10.1002/2014WR016062}:
\begin{equation}
    \label{eq:b_ev_sampling_from_posterior}
    \frac{1}{p(\mathbf{m}|\mathbf{d})} \approx \frac{1}{N}\sum_{i=1}^{N} \frac{1}{p(\mathbf{m}|\mathbf{\theta}_i, \mathbf{d})}
\end{equation}
Moreover, alternative techniques for estimating the Bayesian evidence like thermodynamic integration~\citep{doi:10.1002/2014WR016062} or nested sampling approach~\citep{ELSHEIKH2013NS, ELSHEIKH201514} could be utilized. The MCMC loops in Eq.~\eqref{eq:b_ev_sampling_from_prior} and Eq.~\eqref{eq:b_ev_sampling_from_posterior} are referred to as the inner MCMC loops~\citep{HUAN2013288}. It is easy to see that Eq.~\eqref{eq:b_ev_sampling_from_prior}, Eq.~\eqref{eq:b_ev_sampling_from_posterior} and Eq.~\eqref{eq:mcmc_int} imply that each utility function evaluation requires an outer MCMC loop. Further, estimating the Bayesian evidence factors at each point of the outer MCMC loop requires an inner MCMC loop. Therefore, a single utility function evaluation corresponds to a very high computational cost. Moreover, multiple evaluations of utility function are required in order to determine the optimal parameters of the experimental design. Therefore, solving the optimization problem defined in Eq.~\eqref{eq:d_optimal_design} using the formulation defined in Eq.~\eqref{eq:expected_kl_divergence_v1} is generally a very computationally intensive task.

%High computational cost of optimization problem due to expensive function evaluation provides a motivation for application of surrogate modeling techniques.
In the current manuscript we propose a novel response surface method for the expected information gain calculation. The proposed technique is inspired by the following expression for the utility function:
\begin{equation}
    \label{eq:central_equation}
    \begin{split}
    & U(\mathbf{d}) = \int D_{\text{KL}}(\mathbf{m}, \mathbf{d}) p(\mathbf{m}|\mathbf{d}) \,d\mathbf{m} = \\
    & = \int D_{\text{KL}}(\mathbf{m}, \mathbf{d}) p(\mathbf{m}|\mathbf{\theta}, \mathbf{d}) p(\mathbf{\theta} | \mathbf{d}) \,d\mathbf{m} \,d\mathbf{\theta} = \\
    & = \int D_{\text{KL}}(f(\mathbf{\theta}, \mathbf{d}) + \mathbf{\eta}, \mathbf{d}) p(f(\mathbf{\theta}, \mathbf{d}) + \mathbf{\eta}|\mathbf{\theta}, \mathbf{d}) p(\mathbf{\theta} | \mathbf{d}) \,d\mathbf{\eta} \,d\mathbf{\theta} = \\
    & = \int D_{\text{KL}}(f(\mathbf{\theta}, \mathbf{d}) + \mathbf{\eta}, \mathbf{d}) \mathcal{N}(\mathbf{\eta},0,\sigma) p(\mathbf{\theta}|\mathbf{d}) \,d\mathbf{\eta} \,d\mathbf{\theta}
    \end{split}
\end{equation}
where $\mathcal{N}(\mathbf{\eta},0,\sigma)$ is the density of a normal distribution at the point $\mathbf{\eta}$ that has zero mean and standard deviation $\sigma$. Therefore, the utility function is computed by taking average of the following function: 
\begin{equation}
    \label{eq:dkl_notation}
    G(\mathbf{\theta}, \mathbf{\eta}, \mathbf{d}) = D_{\text{KL}}(f(\mathbf{\theta}, \mathbf{d}) + \mathbf{\eta}, \mathbf{d}).
\end{equation}
If we assume a uniform distribution for design parameter $\mathbf{d}$ in a certain domain, then it is possible to formulate a PCE for $G(\mathbf{\theta}, \mathbf{\eta}, \mathbf{d})$:
\begin{equation}
    \label{eq:dkl_pce}
    G(\mathbf{\theta}, \mathbf{\eta}, \mathbf{d}) = \sum_{A, B, \Gamma} c_{AB\Gamma} p_{A}(\mathbf{\theta}) q_{B}(\mathbf{\eta}) r_{\Gamma}(\mathbf{d})
\end{equation}
where $A, B, \Gamma$ are multi-indices for the orthogonal polynomials $p_A(\mathbf{\theta})$, $q_B(\mathbf{\eta})$ and $r_{\Gamma}(\mathbf{d})$ for the variables $\mathbf{\theta}$, $\mathbf{\eta}$ and $\mathbf{d}$, respectively and $c_{AB\Gamma}$ are the coefficients of PCE for $G(\mathbf{\theta}, \mathbf{\eta}, \mathbf{d})$. PCE expansion defined in Eq.~\eqref{eq:dkl_pce} dramatically simplifies the integration over $\mathbf{\theta}$ and $\mathbf{\eta}$ in Eq.~\eqref{eq:central_equation}. It is simple to see that due to orthogonality of basis polynomials the following equation is valid:
\begin{equation}
    \label{eq:pce_int}
    U(\mathbf{d}) = \int G(\mathbf{\theta}, \mathbf{\eta}, \mathbf{d}) \mathcal{N}(\mathbf{\eta},0,\sigma) p(\mathbf{\theta}|\mathbf{d}) d\,\mathbf{\eta} d\,\mathbf{\theta} = \sum_{\Gamma} c_{00\Gamma} p_0(\mathbf{\theta}) q_0(\mathbf{\eta}) r_{\Gamma}(\mathbf{d})
\end{equation}
In other words, PCE for $U(\mathbf{d})$ can be obtained from PCE for $G(\mathbf{\theta}, \mathbf{\eta}, \mathbf{d})$ simply by keeping only those terms that are constant with respect to $\mathbf{\theta}$ and $\mathbf{\eta}$. Moreover, it is simple to show that the polynomial of $\mathbf{d}$ in the right side of Eq.~\eqref{eq:pce_int} is a projection of $G(\mathbf{\theta}, \mathbf{\eta}, \mathbf{d})$ on the space of square-integrable functions that are constant with respect to $\mathbf{\theta}$ and $\mathbf{\eta}$. Therefore, the coefficients $c_{00\Gamma}$ can be computed via Eq.~\eqref{eq:minimization_problem} which is identical to the following:
\begin{equation}
    \label{eq:minimization_problem_b_ed}
    \mathbf{h}^* = \underset{\mathbf{h}}{\arg\min} \Bigg( \frac{1}{N} \sum_{i=1}^{N} \bigg(  G(\mathbf{\theta}_i, \mathbf{\eta}_i, \mathbf{d}_i) - \sum_{\Gamma} h_\Gamma r_\Gamma(\mathbf{d}_i) \bigg)^2 + \lambda_1 \sum_\Gamma |h_\Gamma| + \lambda_2 \sum_\Gamma h_\Gamma^2 \Bigg)
\end{equation}
where $\mathbf{h}^*$ is the minimization problem solution, $\mathbf{h}$ is the PCE coefficients vector of an arbitrary polynomial in $\mathbf{d}$ with components $h_\Gamma = c_{00\Gamma}$ for all possible $\Gamma$, $N$ is the number of generated samples, $i$ is an index for points in training set, $\mathbf{\theta}_i$, $\mathbf{\eta}_i$ and $\mathbf{d}_i$ are the arguments of $G(\mathbf{\theta}_i, \mathbf{\eta}_i, \mathbf{d}_i)$ that are sampled independently from prior, normal and uniform distributions respectively, $\lambda_1$ and $\lambda_2$ are regularization parameters determined via cross-validation and $G(\mathbf{\theta}_i, \mathbf{\eta}_i, \mathbf{d}_i)$ is the KL-divergence value defined in Eq.~\eqref{eq:dkl_notation}.

In the present work, PCE for the utility function is developed via Eq.~\eqref{eq:minimization_problem_b_ed}. In order to accomplish that task multiple evaluations of the KL-divergence between the prior and the posterior distributions are required, which can be computed numerically using an MCMC integration techniques applied to a transformed KL-divergence of the form:
\begin{equation}
    \label{eq:kl_div_num}
    \begin{split}
    & D_{\text{KL}}(\mathbf{m}, \mathbf{d}) = \int p(\mathbf{\theta}|\mathbf{m}, \mathbf{d}) \log \bigg (  \frac{p(\mathbf{\theta}|\mathbf{m}, \mathbf{d})}{p(\mathbf{\theta}| \mathbf{d})}  \bigg ) d\, \mathbf{\theta} = \\
    & = \int p(\mathbf{\theta}|\mathbf{m}, \mathbf{d}) \log (p(\mathbf{m}|\mathbf{\theta}, \mathbf{d})) d\, \theta - \log(p(\mathbf{m}|\mathbf{d}))
    \end{split}
\end{equation}
It is simple to see that the first term in the right part of Eq.~\eqref{eq:kl_div_num} can be computed numerically with MCMC sampling from the posterior distribution and the second term is the Bayesian evidence that can be computed via Eq.\eqref{eq:b_ev_sampling_from_prior} and Eq.\eqref{eq:b_ev_sampling_from_posterior}. Finally, the combination of Eq.~\eqref{eq:minimization_problem_b_ed} and Eq.~\eqref{eq:kl_div_num} are the major building blocks of the proposed proxy modeling technique for estimating the expected information gain.

%=================
\section{Numerical Examples}
\label{sec:num_ex}
In this section, we evaluate the proposed PCE-based approach to solve the optimal Bayesian experimental design on several numerical examples. The proposed technique is first validated against a simple analytical model that admits approximate analytical solution and then evaluated on subsurface flow problem.
\subsection{Test Case 1}
In the current test case, two models in the form of Eq.~\eqref{eq:system_bed} are considered:
\begin{equation}
    \label{eq:models_index}
    \mathbf{m}_a = f_a(\mathbf{\theta}, \mathbf{d}) + \eta_a
\end{equation}
where $a=1,~2$ is the model index, $\mathbf{m}_a$ are the observable data for model $a$,
$\mathbf{\theta}$ and $\mathbf{d}$ are one-dimensional design parameters,
$\mathbf{\eta}_a$ is a normally distributed noise with standard deviation $\sigma$.
The nonlinear functions $f_a(\mathbf{\theta}, \mathbf{d})$ are defined by:
\begin{equation}
    \label{eq:model_function_1}
    f_1(\mathbf{\theta}, \mathbf{d}) = \theta^3 d^2 + \theta \exp \big(|d - 0.2| \big)
\end{equation}
\begin{equation}
    \label{eq:model_function_2}
    f_2(\mathbf{\theta}, \mathbf{d}) = \theta^3 d^2 + \theta \exp \big(-20 (d - 0.2)^2 \big)
\end{equation}
Models from Eq.~\eqref{eq:model_function_1} and Eq.~\eqref{eq:model_function_2} describe similar systems. The principal difference between these two models is in the degree of smoothness and the corresponding accuracy of the PCE approximation. We note the exponential convergence of PCE series for smooth functions. Therefore, it is expected that PCE surrogates for expected information gain have different quality for two models defined in Eq.~\eqref{eq:model_function_1} and Eq.~\eqref{eq:model_function_2}. The model defined by Eq.~\eqref{eq:model_function_1} has a discontinuity in the first order derivatives while the model defined by Eq.~\eqref{eq:model_function_2} has continuous derivatives of any order. Due to that, systems described by Eq.~\eqref{eq:model_function_1} and Eq.~\eqref{eq:model_function_2} are referred to as non-smooth and smooth, respectively.

For both models, the parameter $\mathbf{\theta}$ is a uniformly distributed random variable in $\U[0,1]$ and $d \in [0,1]$. For small values of $\sigma$ the following approximation for the KL-divergence can be derived (see Appendix~\ref{sec:approximation} for more details):
\begin{equation}
    D_{KL,a}(f_a(\theta, \mathbf{d}) + \mathbf{\eta}, \mathbf{d}) \approx D_{KL,a}(f_a(\theta, \mathbf{d}), \mathbf{d}) = \log\bigg( \frac{1}{(2\pi)^{1/2} \sigma} \frac{\partial f_a}{\partial \theta}(\mathbf{\theta}, \mathbf{d}) \bigg) - 1/2
    \label{eq:dkl_analytical}
\end{equation}
Here, we neglect the value of $\mathbf{\eta}$. Therefore, Eq.~\eqref{eq:dkl_analytical} is only valid for small values of $\sigma$.

\begin{figure}[H]
    \centering
    \begin{subfigure}[b]{0.45\textwidth}
        \includegraphics[width=0.99\linewidth]{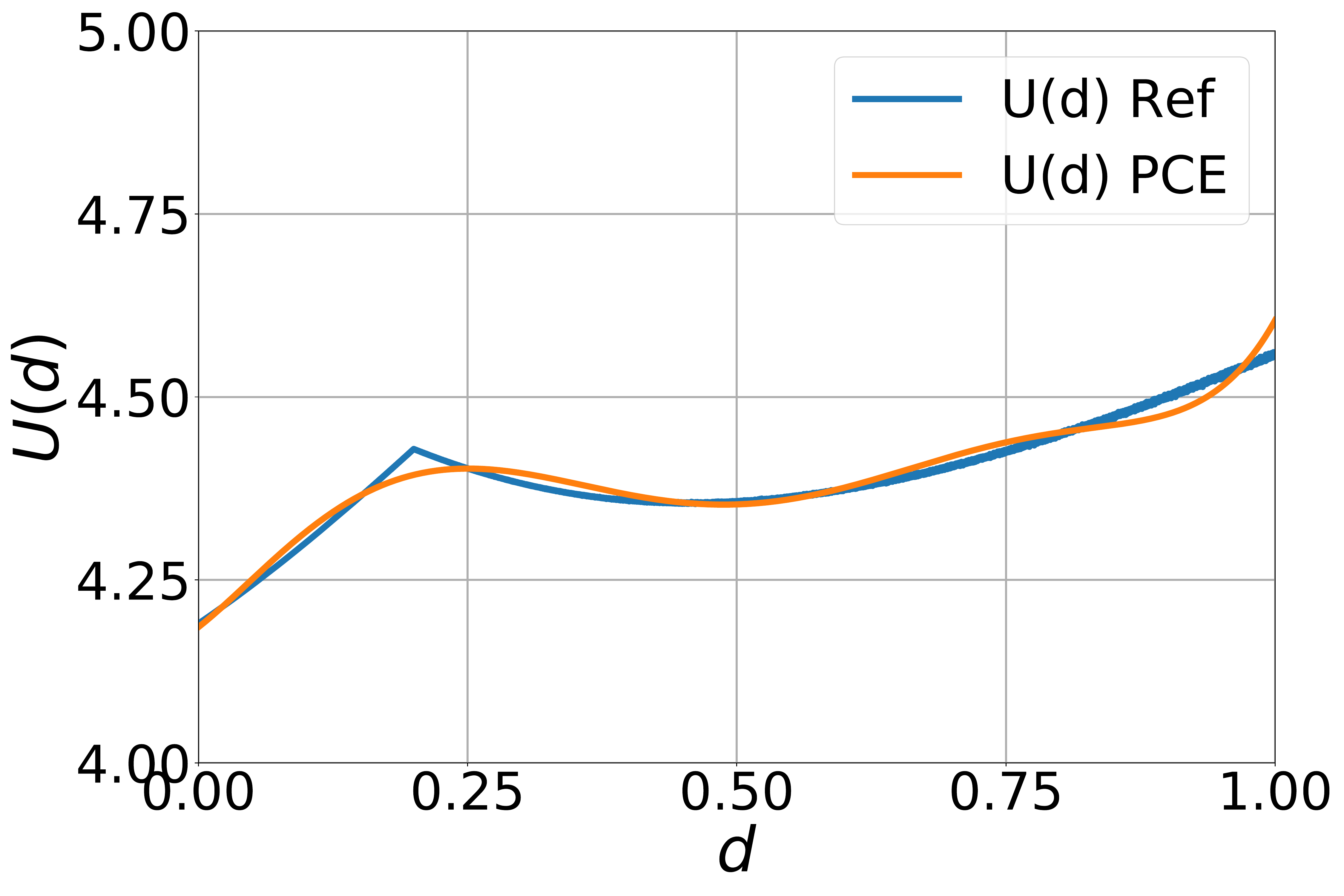}
        \caption{$\sigma_m = 3.0\times{10}^{-3}$}
        \label{fig:test_case_1_non_smooth_sigma_m_3em3}
    \end{subfigure}
    \begin{subfigure}[b]{0.45\textwidth}
        \includegraphics[width=0.99\linewidth]{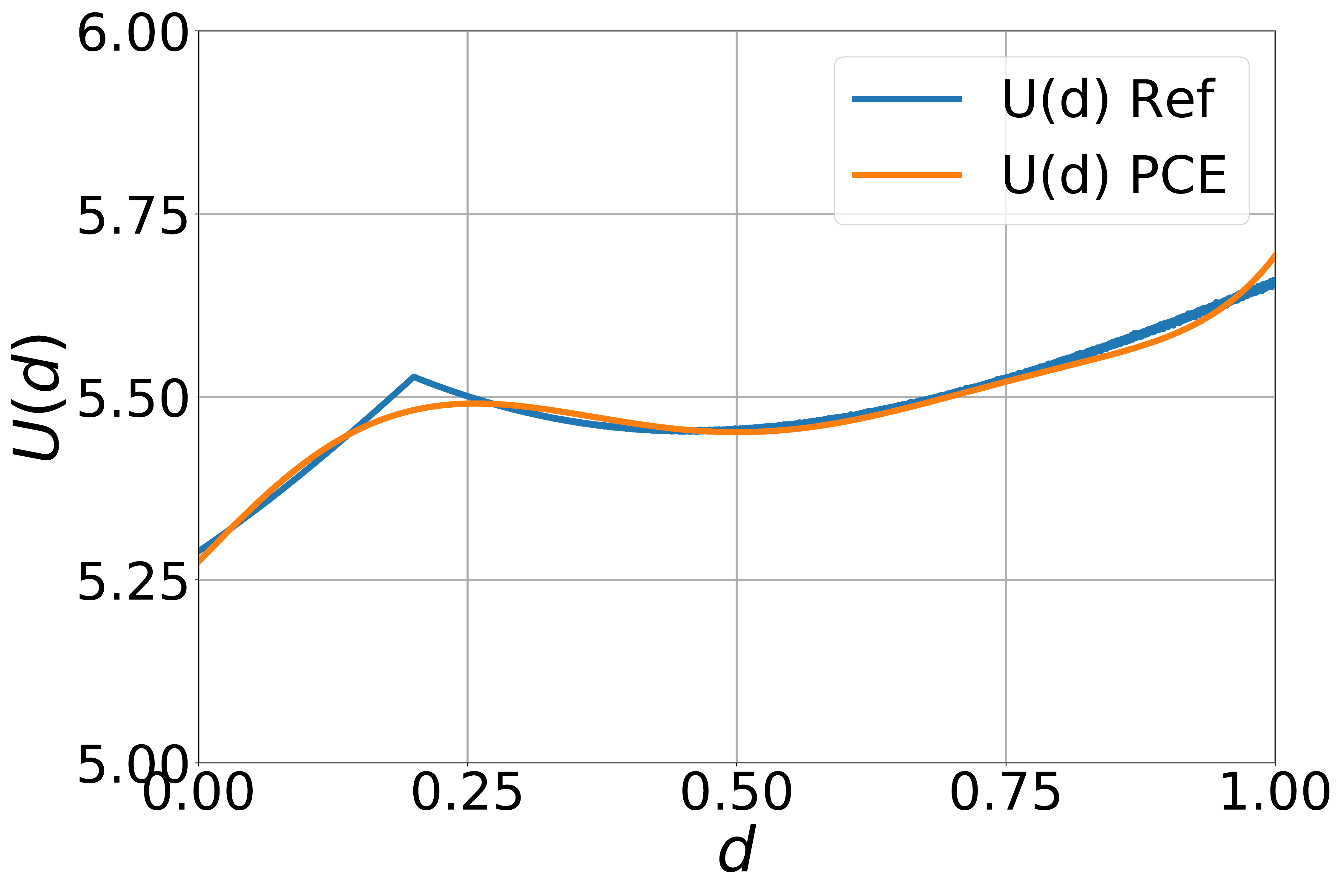}
        \caption{$\sigma_m = 1.0\times{10}^{-3}$}
        \label{fig:test_case_1_non_smooth_sigma_m_1em3}
    \end{subfigure}
    \begin{subfigure}[b]{0.45\textwidth}
        \includegraphics[width=0.99\linewidth]{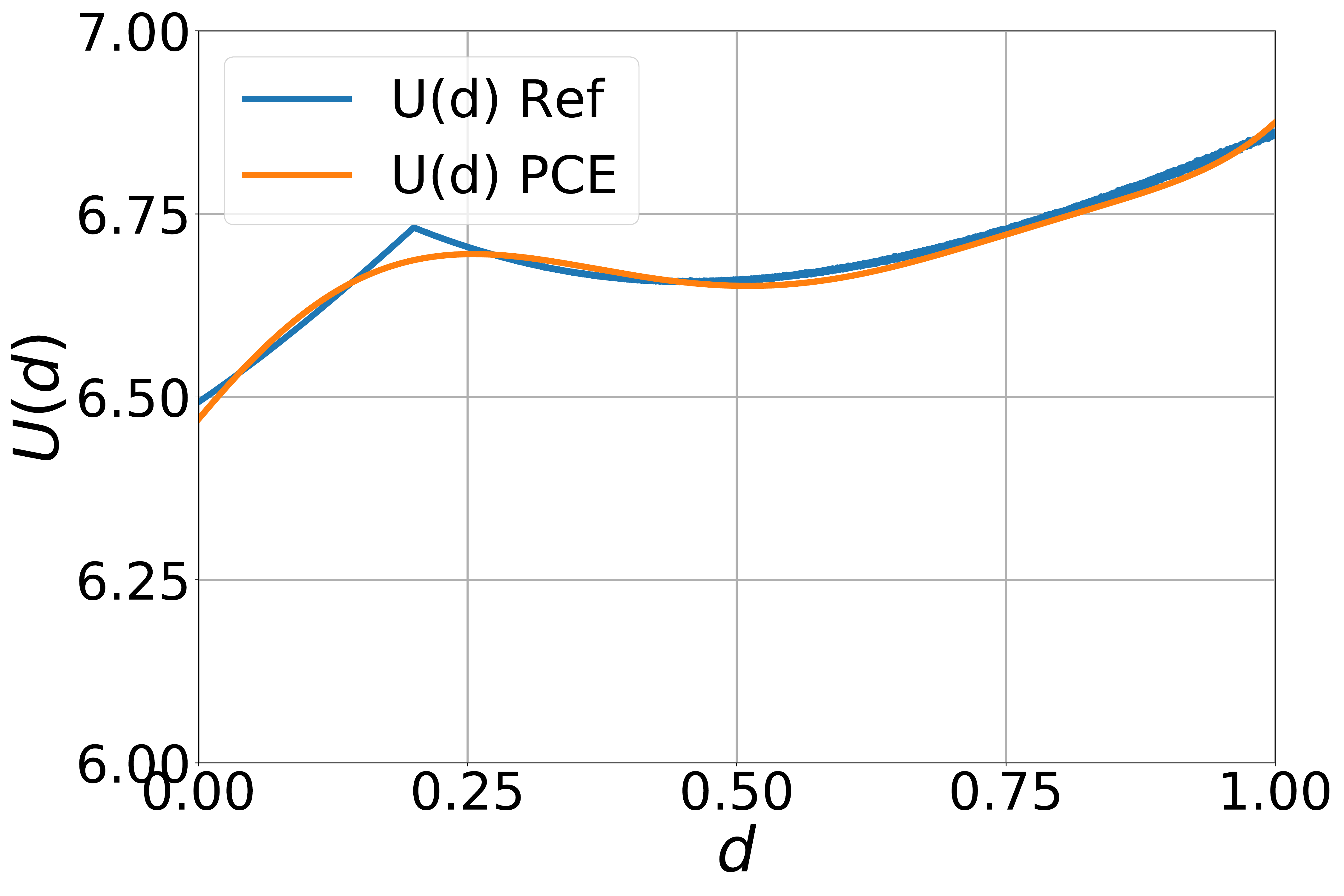}
        \caption{$\sigma_m = 3.0\times{10}^{-4}$}
        \label{fig:test_case_1_non_smooth_sigma_m_3em4}
    \end{subfigure}
    \begin{subfigure}[b]{0.45\textwidth}
        \includegraphics[width=0.99\linewidth]{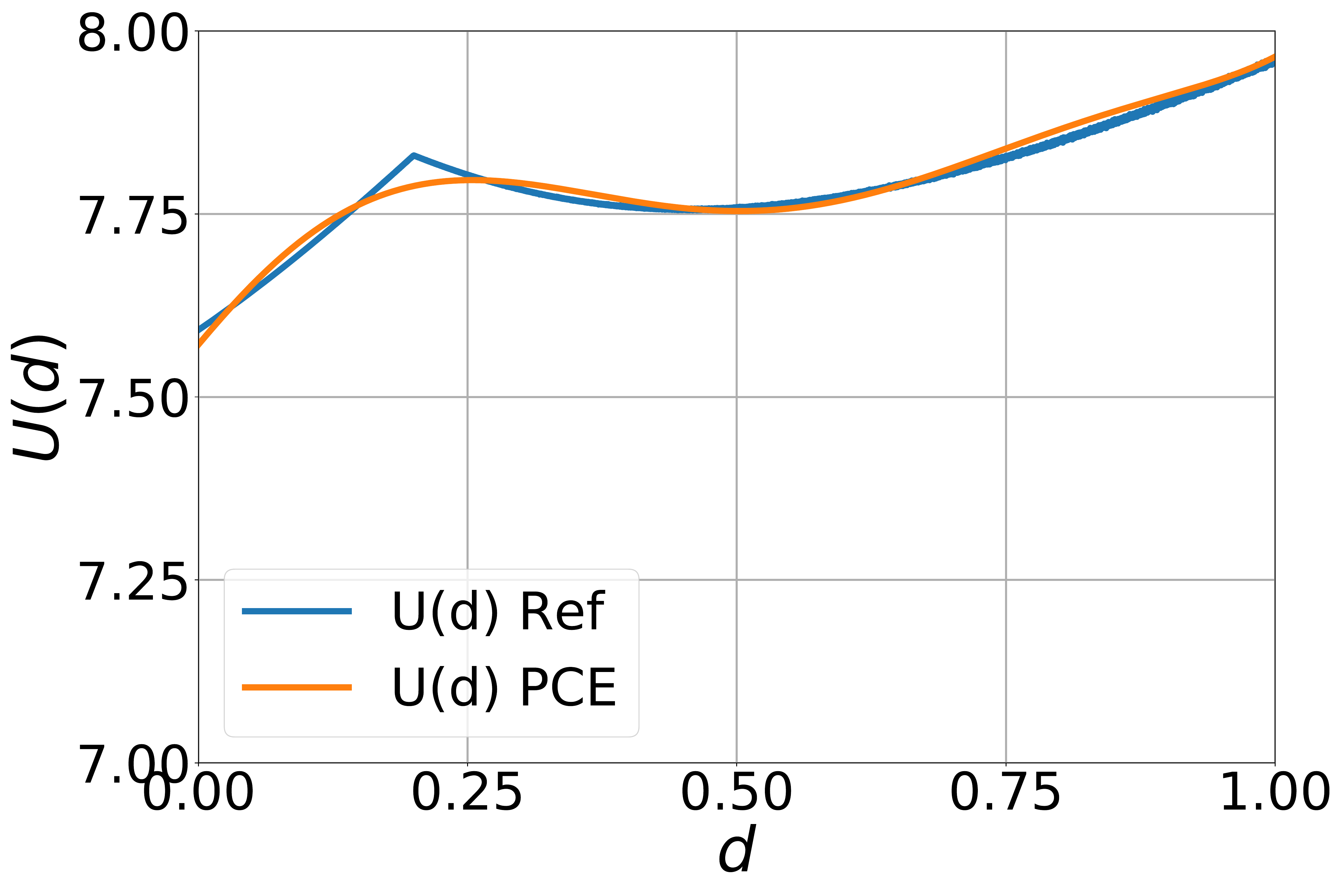}
        \caption{$\sigma_m = 1.0\times{10}^{-4}$}
        \label{fig:test_case_1_non_smooth_sigma_m_1em4}
    \end{subfigure}
    \caption{Plots of the expected information gain $U(d)$ versus the design parameter $d$ for non-smooth objective functions defined in Eq.~\eqref{eq:model_function_1} for different values of $\sigma$.}
    \label{fig:test_case_1_non_smooth}
\end{figure}

\begin{figure}[H]
    \centering
    \begin{subfigure}[b]{0.45\textwidth}
        \includegraphics[width=0.99\linewidth]{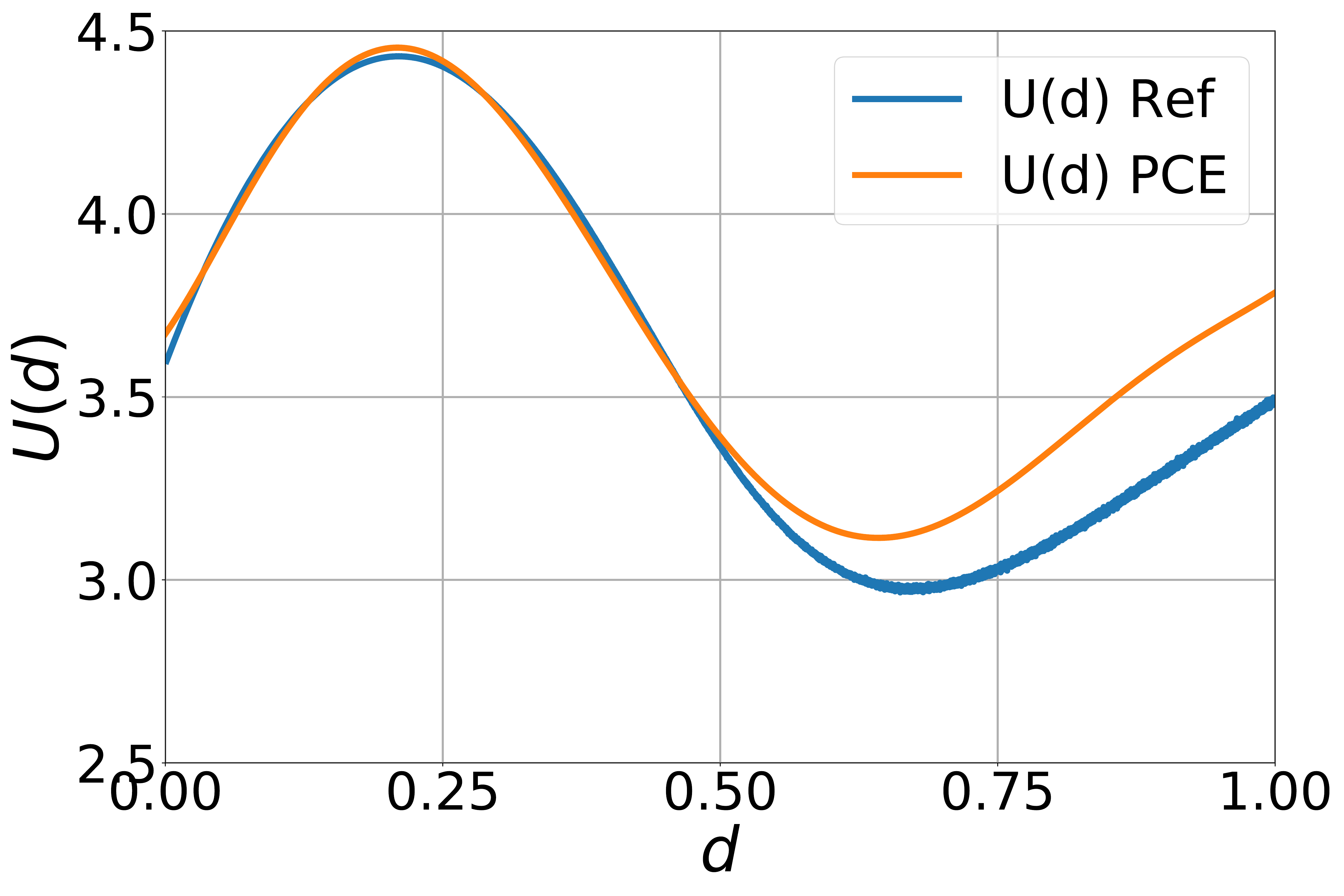}
        \caption{$U(d)$ for $\sigma_m = 3.0 \times{10}^{-3}$}
        \label{fig:test_case_1_smooth_sigma_m_3em3}
    \end{subfigure}
    \begin{subfigure}[b]{0.45\textwidth}
        \includegraphics[width=0.99\linewidth]{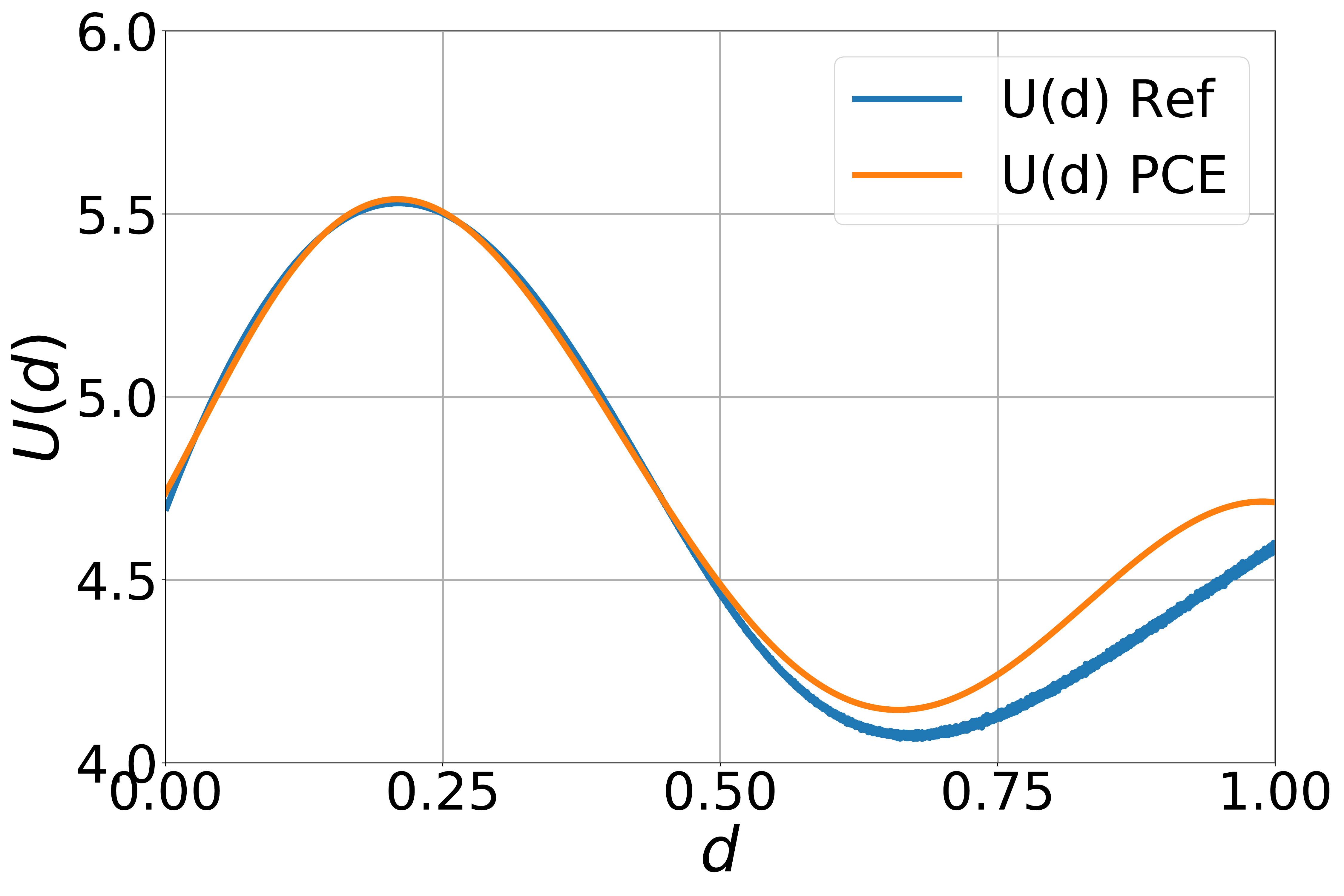}
        \caption{$\sigma_m = 1.0 \times{10}^{-3}$}
        \label{fig:test_case_1_smooth_sigma_m_1em3}
    \end{subfigure}
    \begin{subfigure}[b]{0.45\textwidth}
        \includegraphics[width=0.99\linewidth]{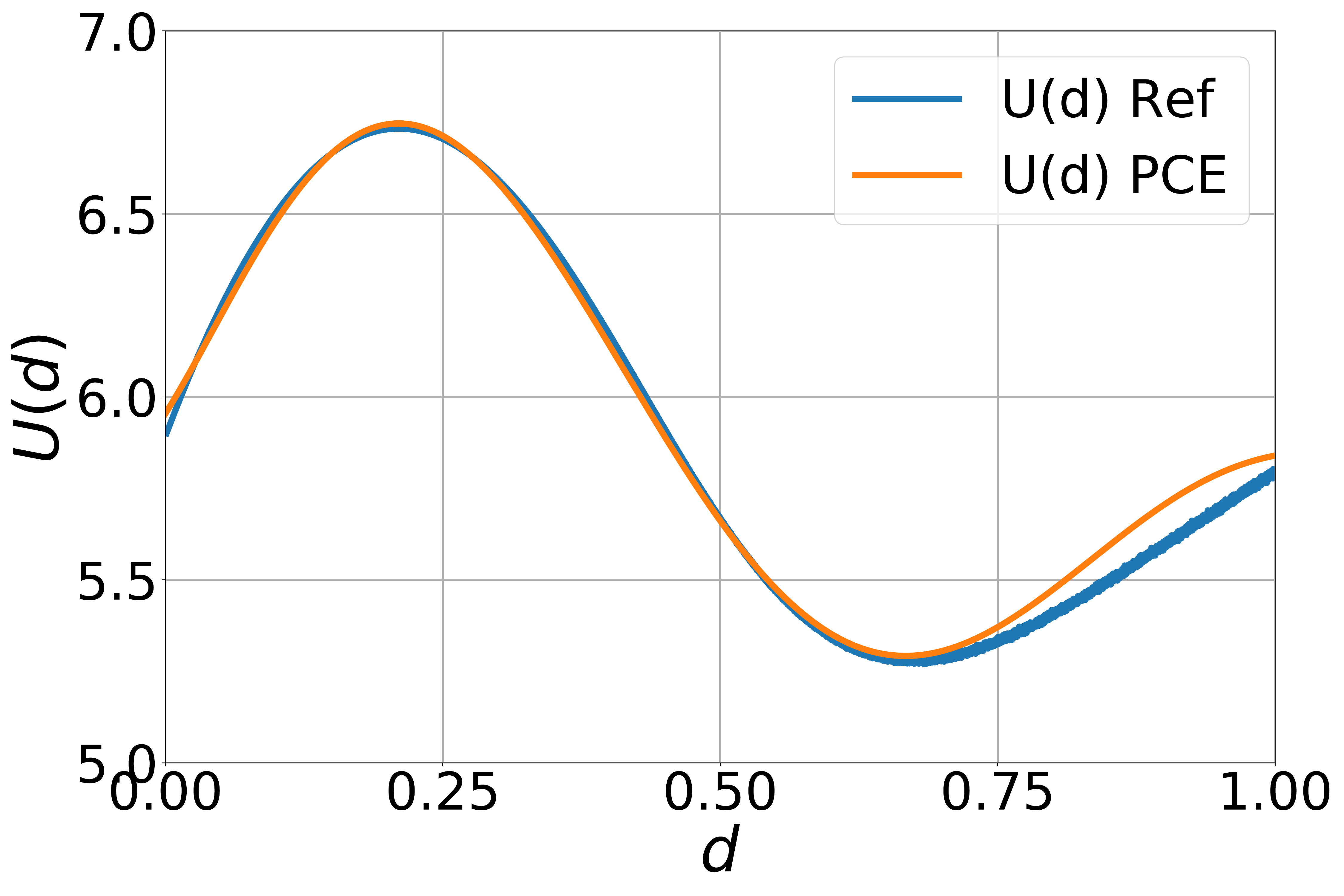}
        \caption{$\sigma_m = 3.0 \times{10}^{-4}$}
        \label{fig:test_case_1_smooth_sigma_m_3em4}
    \end{subfigure}
    \begin{subfigure}[b]{0.45\textwidth}
        \includegraphics[width=0.99\linewidth]{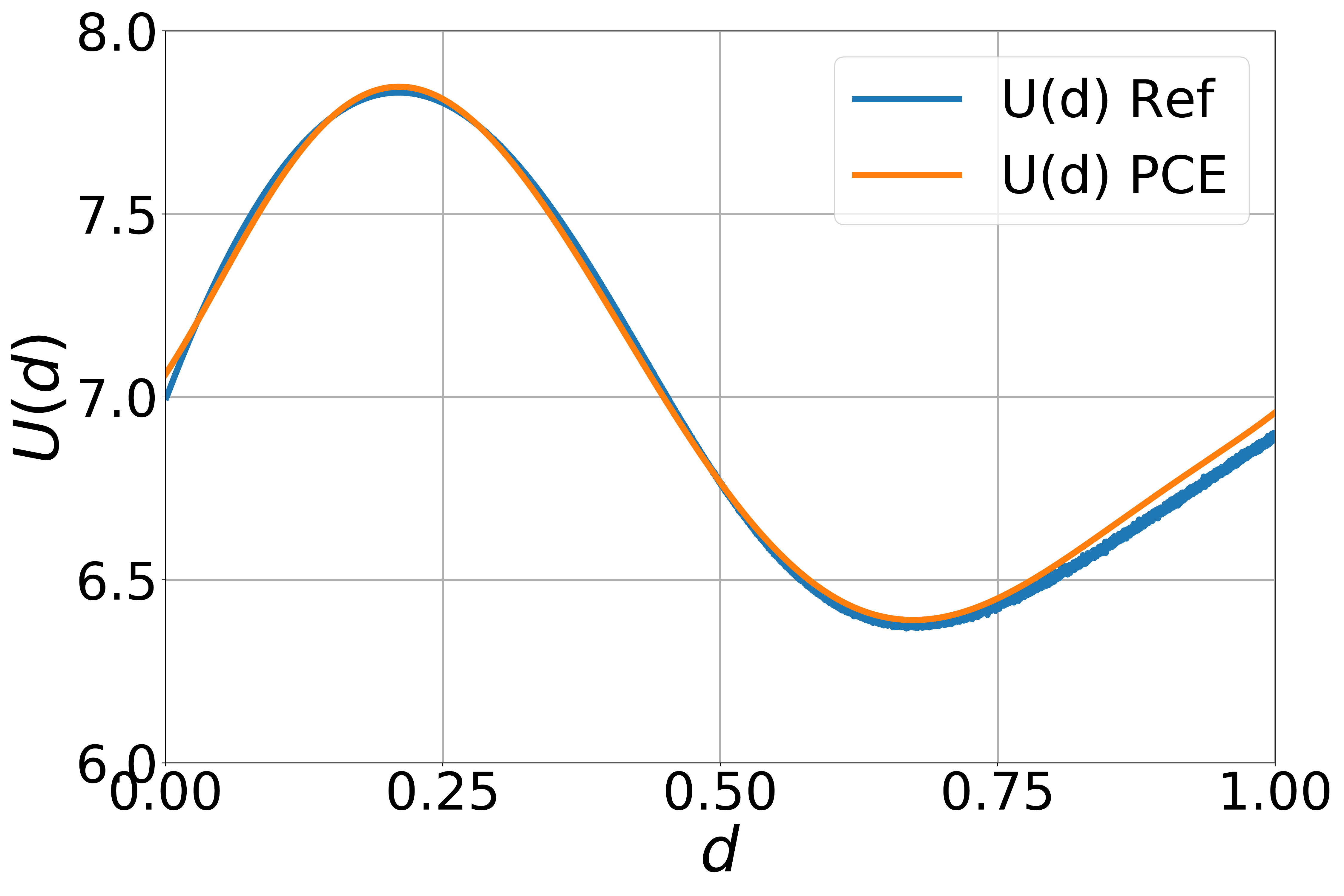}
        \caption{$\sigma_m = 1.0 \times{10}^{-4}$}
        \label{fig:test_case_1_smooth_sigma_m_1em4}
    \end{subfigure}
    \caption{Plots of the expected information gain $U(d)$ versus the design parameter $d$ for smooth objective functions defined in Eq.~\eqref{eq:model_function_2} for different values of $\sigma$.}
    \label{fig:test_case_1_smooth}
\end{figure}

In the present test case, $200,000$ values of $\mathbf{\theta}$ and $\mathbf{\eta}$ are sampled from prior distribution $\U[0,1]$ and from normal distribution with standard deviation $\sigma$, respectively. The following values of $\sigma$ are considered: $3.0\times10^{-3},~1.0\times{10}^{-3},~3.0\times{10}^{-4}$ and $1.0\times{10}^{-4}$. The range of parameter $\sigma$ is selected in such a way that assumption about small magnitude of $\sigma$ is valid for the smallest $\sigma$ considered and is violated for the highest one. In such setting, $200,000$ samples for $\mathbf{d}$ are generated for uniform distribution $\U[0,1]$. For each generated sample the KL-divergence is computed numerically using Eq.~\eqref{eq:central_equation}. The first term in Eq.~\eqref{eq:central_equation} is computed with $50,000$ MCMC-samples generated from the posterior distribution and the second term in Eq.~\eqref{eq:central_equation} is computed with $200,000$ MCMC-samples generated from the prior distribution Eq.~\eqref{eq:b_ev_sampling_from_prior}. Since the KL-divergence values have already been generated for all samples, $\mathbf{d}$ is rescaled to $\U[-1,1]$ in order to allow for using Legendre polynomials as a basis function in the PCE expansion. Polynomials up to degree eight were utilized in order to produce a response surface for $U(\mathbf{d})$ in accordance with Eq.~\eqref{eq:minimization_problem_b_ed}. For the purposes of validation, the utility function values at any given $\mathbf{d}$ is computed via Eq.~\eqref{eq:expected_kl_divergence} and Eq.~\eqref{eq:dkl_analytical}, where the integration is replaced by averaging over $200,000$ samples generated in agreement with the prior distribution $\U[0,1]$ and the likelihood defined in Eq.~\eqref{eq:likelihood_fun}.  Results of comparison of two techniques for $U(\mathbf{d})$ calculation are shown in Fig.~\ref{fig:test_case_1_non_smooth} and Fig.~\ref{fig:test_case_1_smooth}.

It can be observed that for both of the test cases the proposed PCE approach provides a relatively accurate approximation of the utility function $U(\mathbf{d})$.
Almost an exact match can be observed for the cases of small $\sigma$ where the KL-divergence approximate defined in Eq.~\eqref{eq:dkl_analytical} is supposed to be valid. For small values of $\sigma$, both methods that correspond to Eq.~\eqref{eq:expected_kl_divergence} and Eq.~\eqref{eq:dkl_analytical} provides similar estimates for $U(\mathbf{d})$. However, for high $\sigma$ values some divergence between those techniques is observed. This is expected because Eq.~\eqref{eq:dkl_analytical} is not supposed to work in those cases. In addition to that, the PCE response surface for $U(\mathbf{d})$ failed to reproduce the discontinuity in the derivative of $U(\mathbf{d})$ as it can be observed in Fig.~\ref{fig:test_case_1_non_smooth}. This is an expected behavior of PCE response surface because of the smooth basis functions. What is more important, the design value $\mathbf{d}$ corresponding to the local optimum is accurately reproduced. Therefore, the present test case demonstrates that the introduced PCE based technique for estimating the utility $U(\mathbf{d})$ function is more accurate for smooth problems. However, the proposed PCE based approach could still be used for both smooth and non-smooth cases to estimate the optimal Bayesian experimental design, because the design values $\mathbf{d}$ maximizing the utility $U(\mathbf{d})$ are accurately approximated.

\subsection{Test Case 2}

In this test case, we consider a two-phase subsurface flow related to the forecast of hydrocarbon oil production. On one hand, the accuracy of the forecast is directly related to the quantity and quality of available data used to estimate the subsurface rock properties. On the other hand, direct measurements of those properties is an expensive process. Therefore, utilizing optimal experimental design techniques to decide which data to be collected in order to produce accurate predictions of hydrocarbons production is of great practical importance. In the present numerical example we optimize the design of experiment in order to maximize the accuracy of subsurface parameters measurements, which in turn reduce the uncertainty in the oil production forecast.

In the present test case, numerical simulations of oil production enhanced by well-known water-flooding technique are studied. During this process water is injected into the reservoir via a group of wells called injection wells (aka. injectors) and displaces oil that saturates the pores of the reservoir rocks. Hydrocarbons, in turn, are produced via another group of wells called production wells (aka. producers). The fluid flow in the reservoir together with the oil production rates is controlled by the spatial distribution of reservoir properties namely, the porosity field $\phi(\mathbf{r})$ and the permeability field $k(\mathbf{r})$ where $\mathbf{r}$ is a vector of spatial coordinates of a given point in space. Typically, the porosity and permeability are known at several locations in the reservoir where rock samples have been extracted during drilling. These point values are then used within stochastic interpolation frameworks (aka. geo-statistics~\citep{Book_on_Geostatistics}) to populate the model parameters over the entire domain of interest.

In the present test case, we solve for optimal design of experiment based on Bayesian framework. We consider a five-spot injection pattern where an injection well is located at the center of a square surrounded by four production wells. Given the symmetry of this pattern, only one quarter of the domain is modeled with one producer and one injector located at the opposite corners of a square domain. The length of the edge of that square is set to $L=500 \text{m}$. The thickness of the reservoir is $h=10 \text{m}$. We do not consider discretization along the vertical direction and we only consider a two-dimensional flow problem. Further, the porosity is considered to be constant value, $\phi(\mathbf{r}) = 0.2$. Also, we assume that data is collected by drilling additional wells and results in a measurement of the permeability value at the location and a measurement of the pressure value at specified moments of time. Alternative sources of data, like seismic measurement are not considered in the present example.

In this setting, the vector of design parameters $\mathbf{d}$ is formed by the 2D coordinates additional wells and the dimension of the design parameter space can be computed as following:
\begin{equation}
    \label{eq:dim_d}
    \text{dim}(\mathbf{d}) = 2 n_s
\end{equation} 
where $n_s$ is total number of new wells. In the present example only two cases are considered: $n_s = 1, 2$.

\begin{figure}[H]
    \centering
    \begin{subfigure}[b]{0.23\textwidth}
        \includegraphics[width=0.99\linewidth]{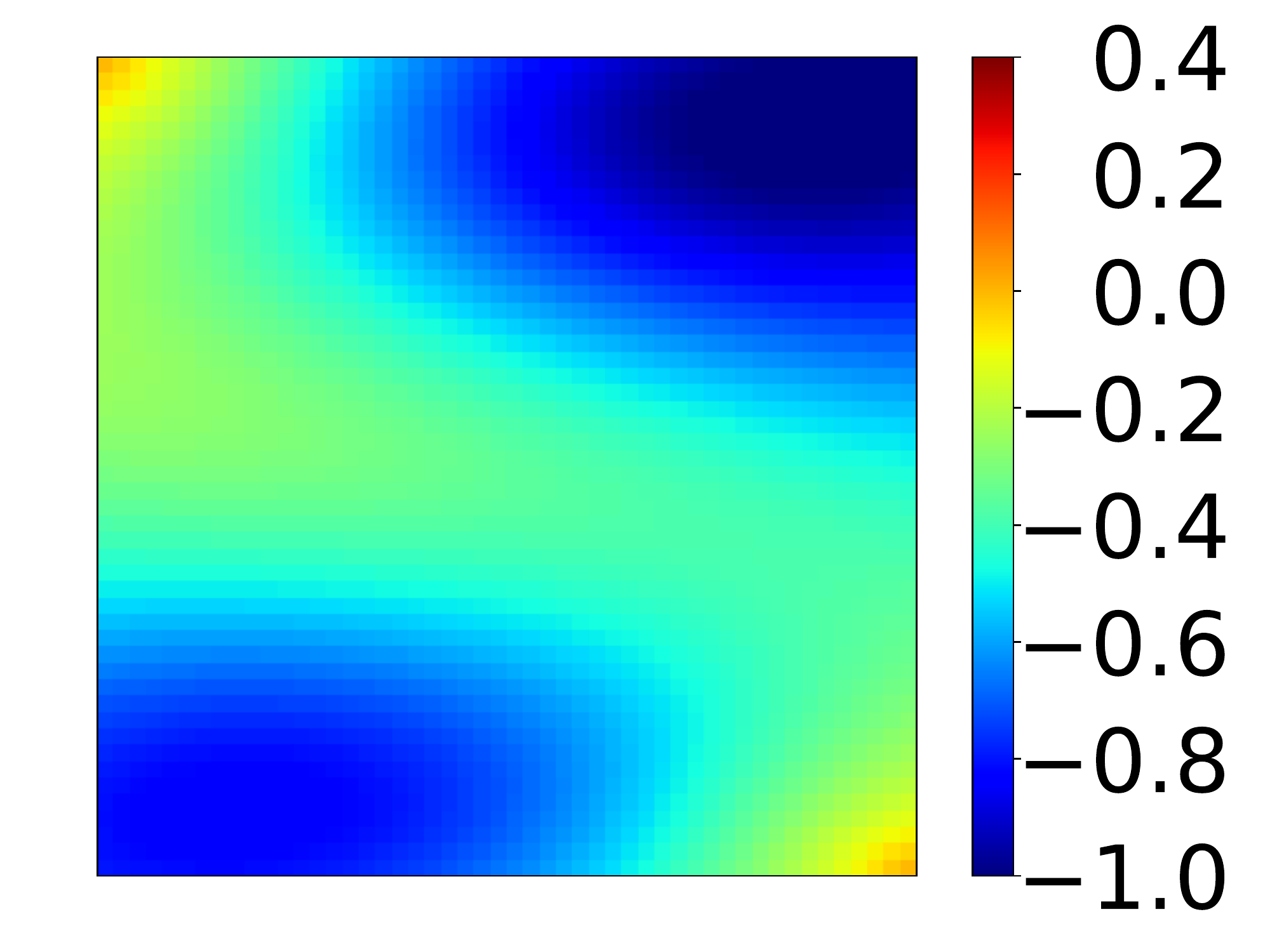}
        \caption{ }
        \label{fig:test_case_2_dlogperm_kind_0}
    \end{subfigure}
    \begin{subfigure}[b]{0.23\textwidth}
        \includegraphics[width=0.99\linewidth]{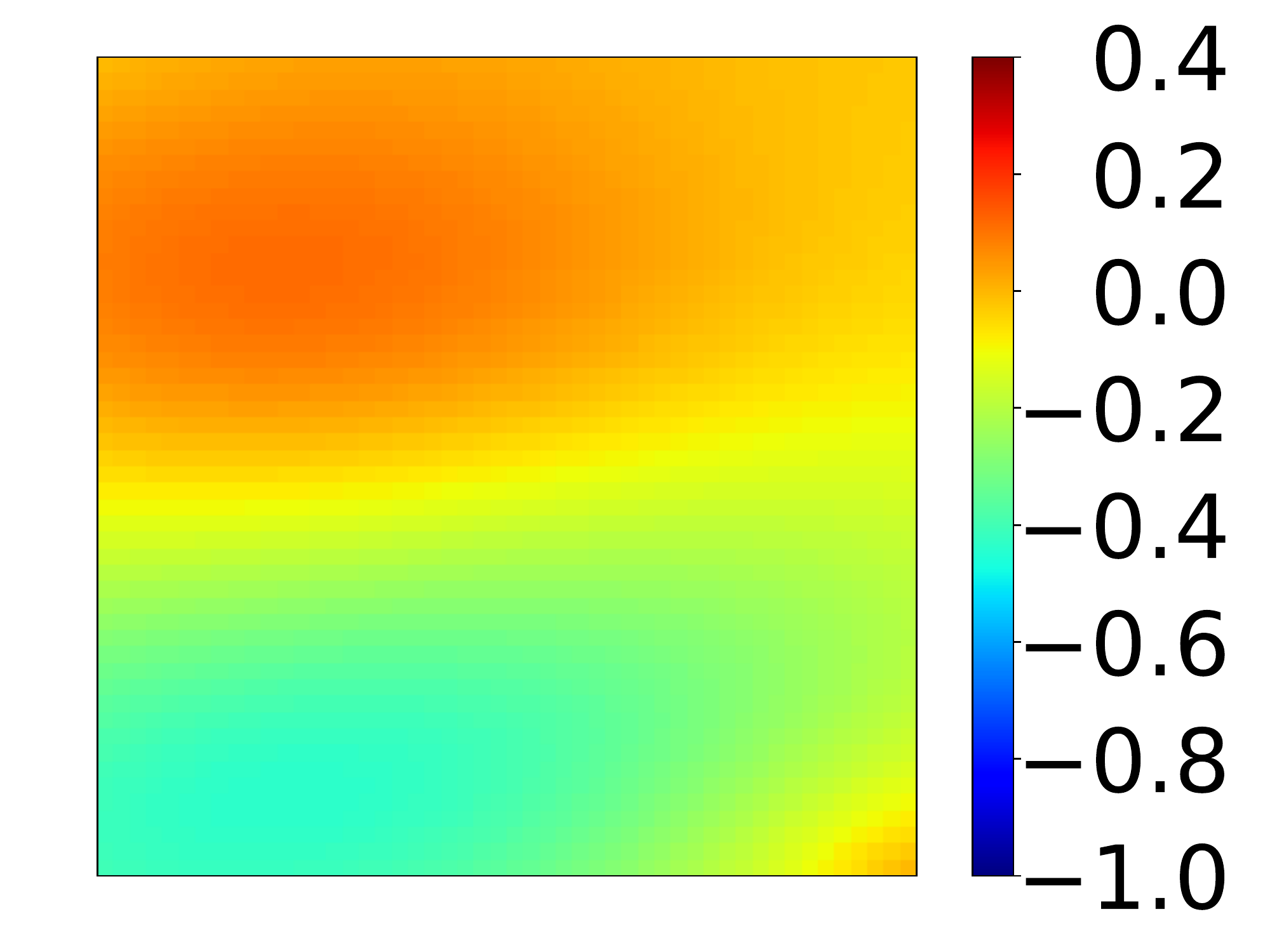}
        \caption{ }
        \label{fig:test_case_2_dlogperm_kind_1}
    \end{subfigure}
    \begin{subfigure}[b]{0.23\textwidth}
        \includegraphics[width=0.99\linewidth]{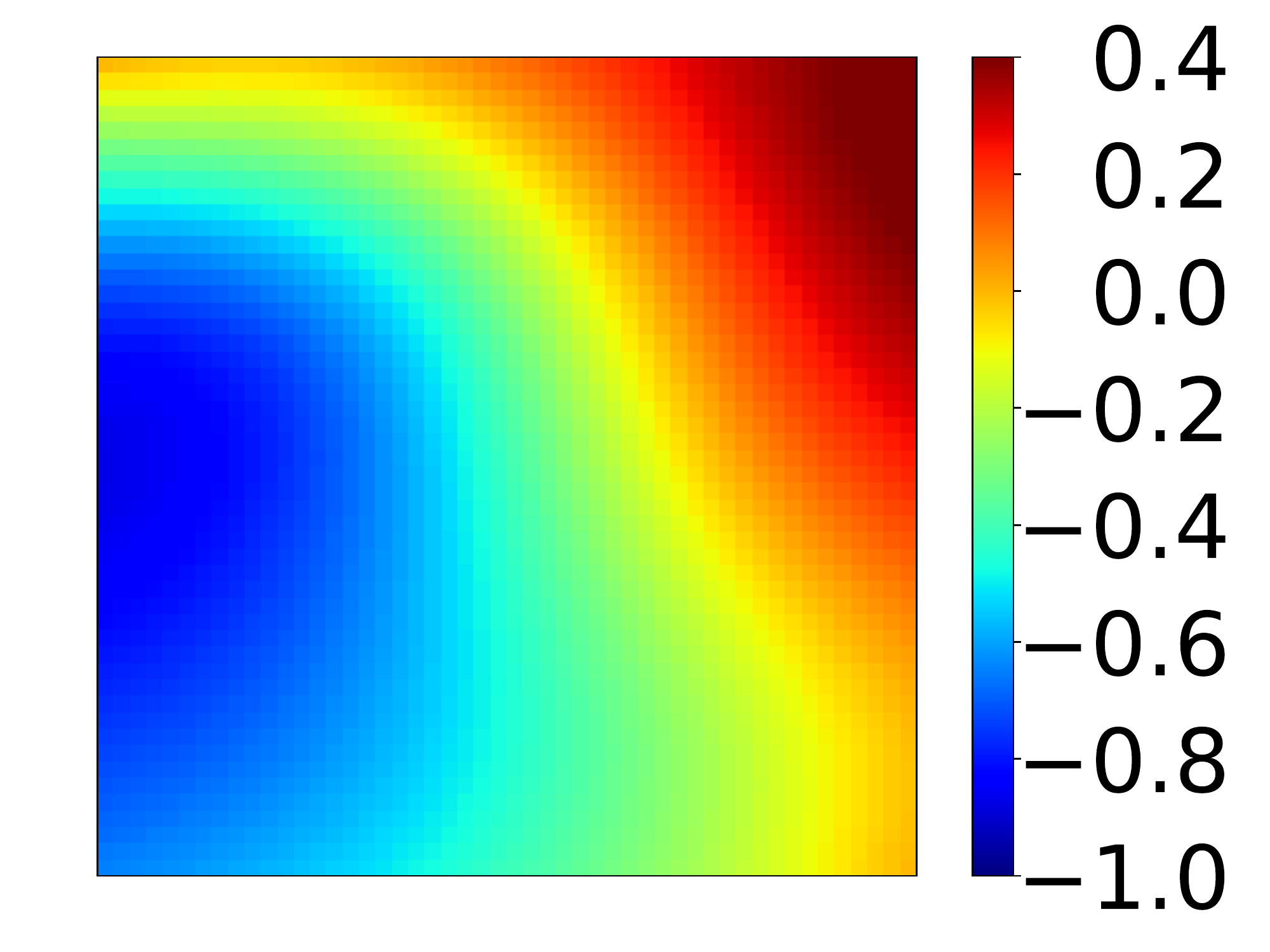}
        \caption{ }
        \label{fig:test_case_2_dlogperm_kind_2}
    \end{subfigure}
    \begin{subfigure}[b]{0.23\textwidth}
        \includegraphics[width=0.99\linewidth]{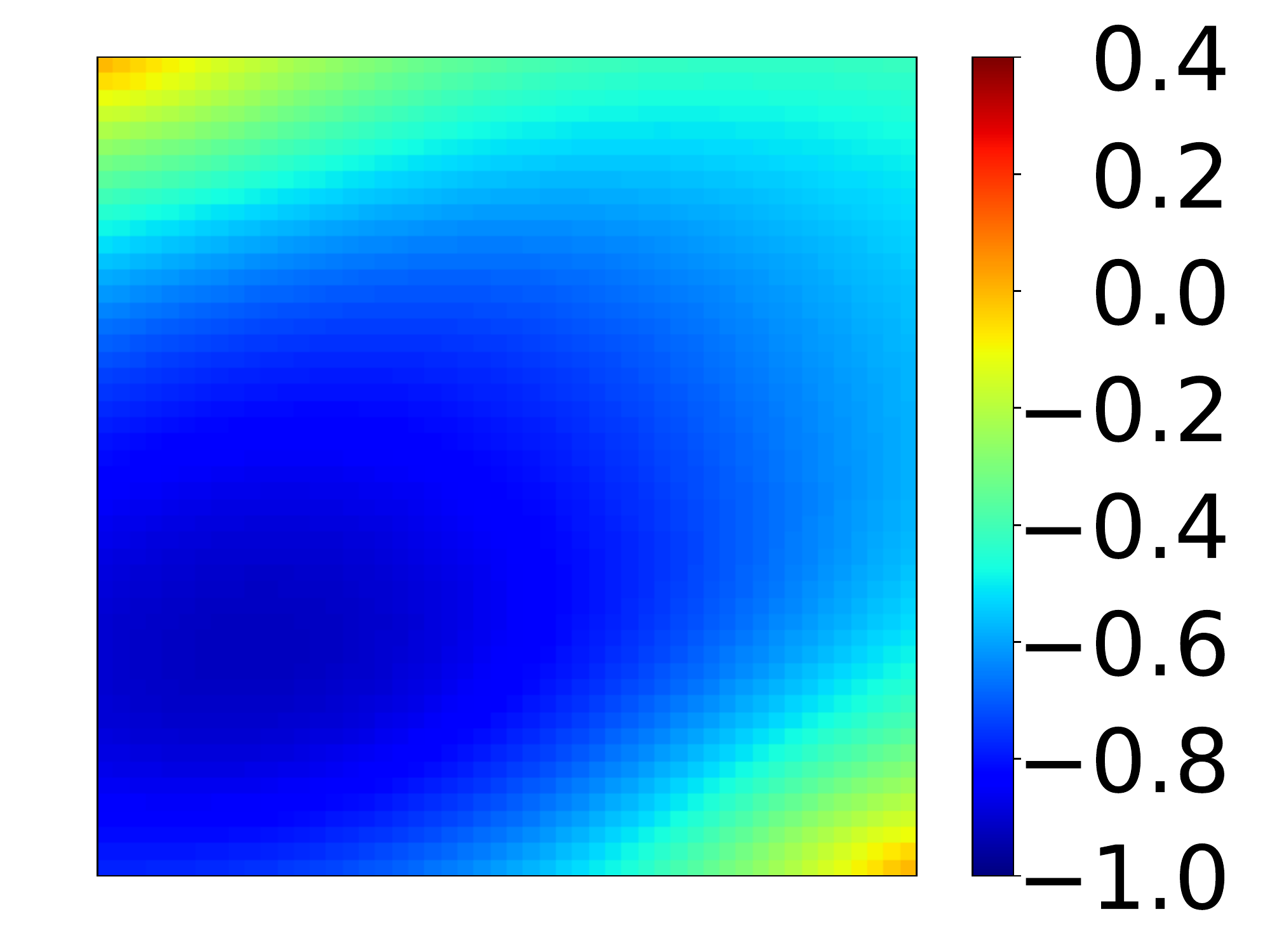}
        \caption{ }
        \label{fig:test_case_2_dlogperm_kind_3}
    \end{subfigure}

    \begin{subfigure}[b]{0.23\textwidth}
        \includegraphics[width=0.99\linewidth]{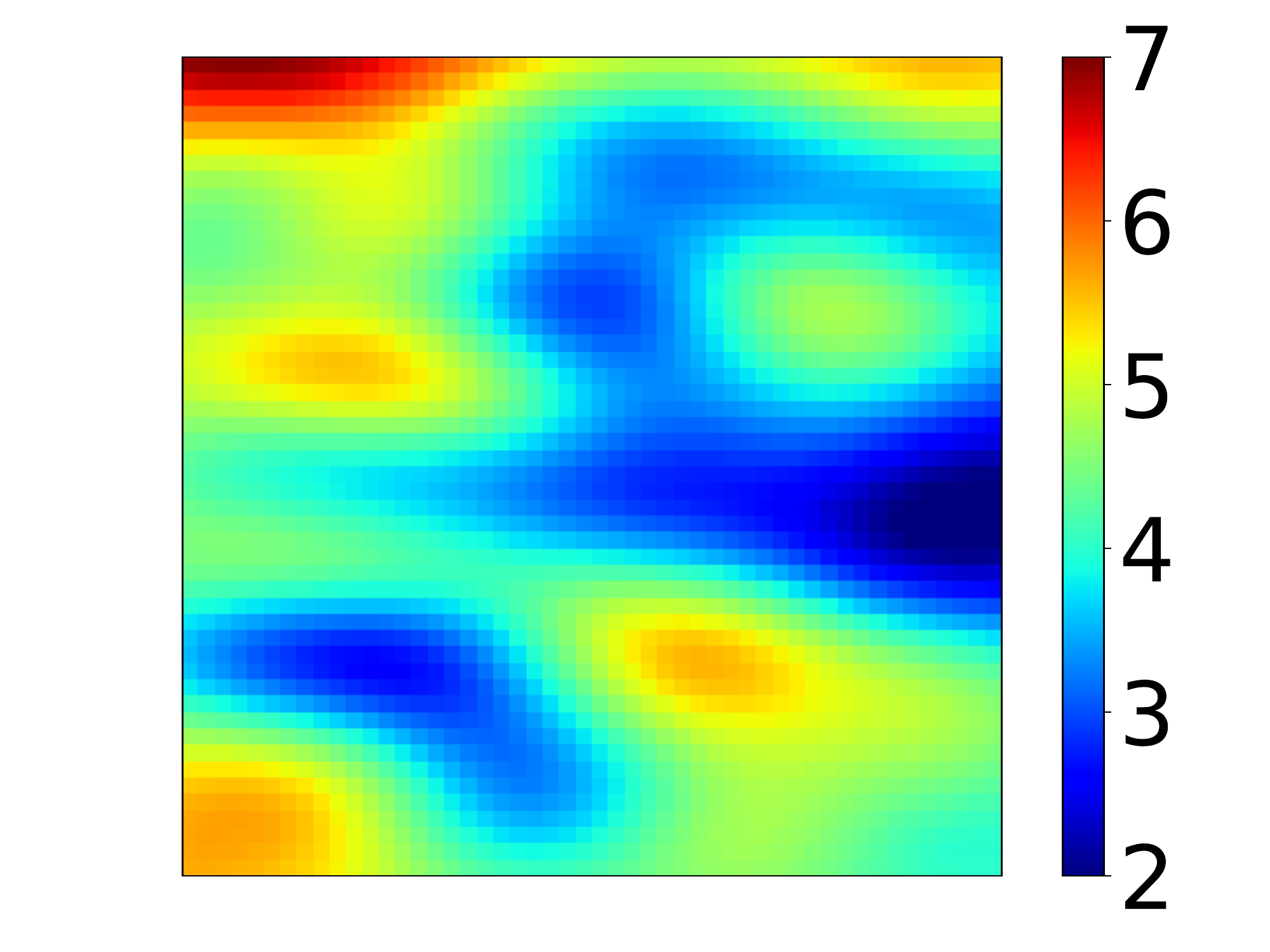}
        \caption{ }
        \label{fig:test_case_2_logperm_kind_0}
    \end{subfigure}
    \begin{subfigure}[b]{0.23\textwidth}
        \includegraphics[width=0.99\linewidth]{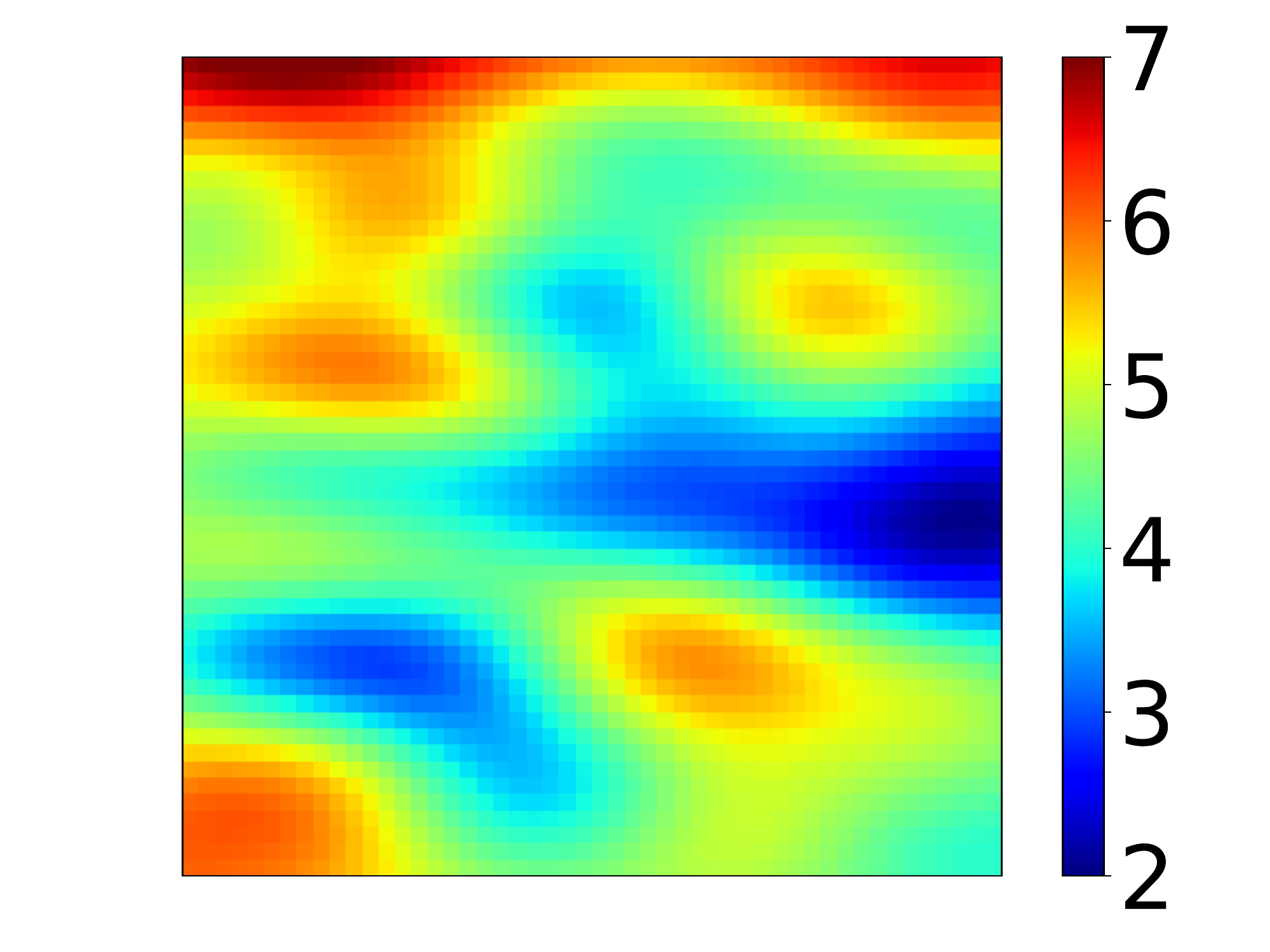}
        \caption{ }
        \label{fig:test_case_2_logperm_kind_1}
    \end{subfigure}
    \begin{subfigure}[b]{0.23\textwidth}
        \includegraphics[width=0.99\linewidth]{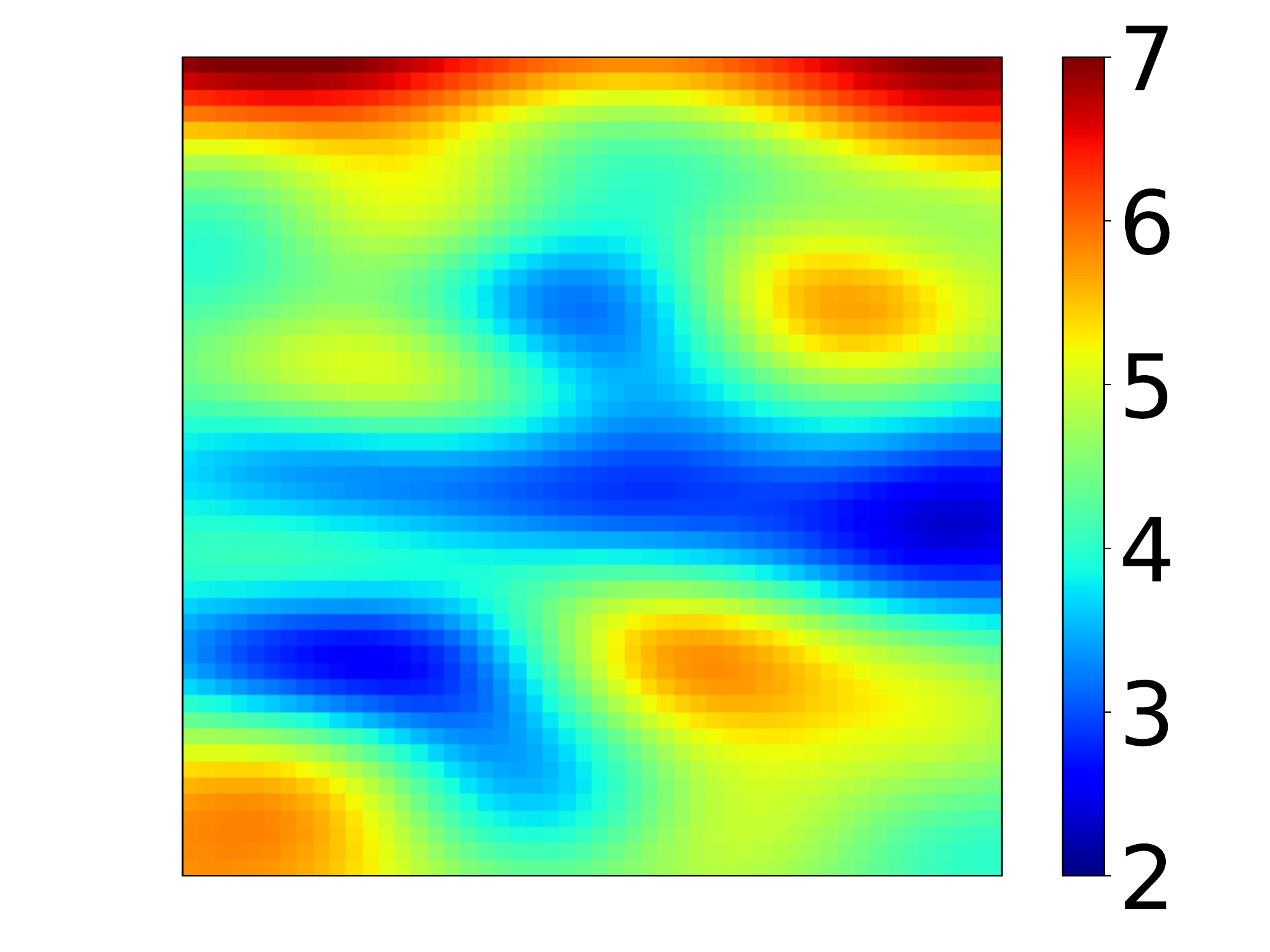}
        \caption{ }
        \label{fig:test_case_2_logperm_kind_2}
    \end{subfigure}
    \begin{subfigure}[b]{0.23\textwidth}
        \includegraphics[width=0.99\linewidth]{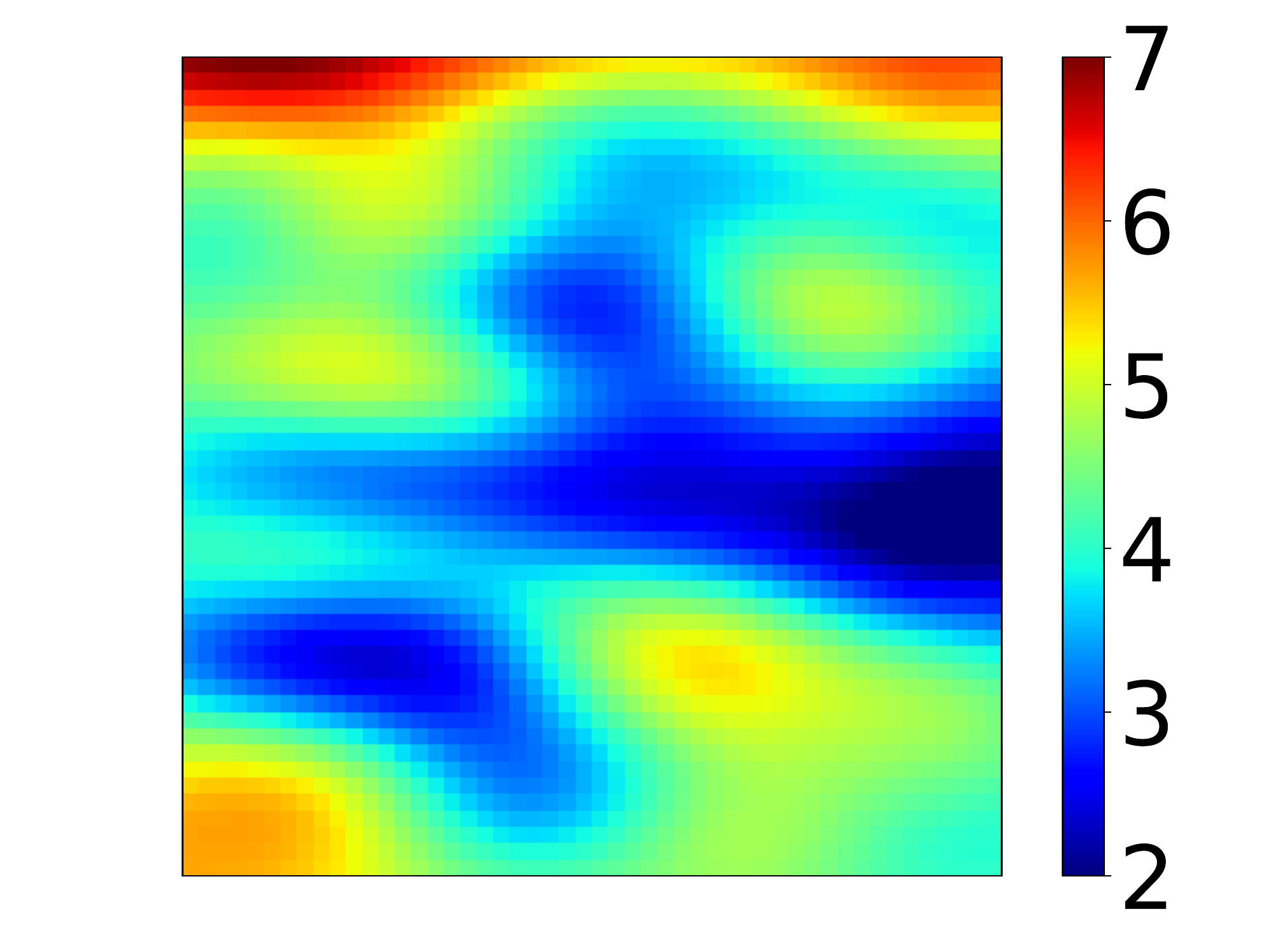}
        \caption{ }
        \label{fig:test_case_2_logperm_kind_3}
    \end{subfigure}

    \caption{Realizations of the perturbations $\zeta(\mathbf{r})$ shown in panels (a, b, c, d) and the corresponding permeability field $\log(k(\mathbf{r}))$ shown in the panels (e, f, g, h), respectively.}
    \label{fig:test_case_2_logperm}
\end{figure}

The vector of model parameters $\mathbf{\theta}$ is introduced via Karhunen-Lo\`eve (KL) expansion for the spatially discretized log-permeability field $\log(k(\mathbf{r}))$.
The log-permeability distribution is assumed to be a linear combination of the reference permeability field representing the general trend of the field and random perturbation that is defined stochastically:
\begin{equation}
    \label{eq:perm}
    \log(k(\mathbf{r})) = \log(k_{\text{ref}}(\mathbf{r})) + \zeta(\mathbf{r})
\end{equation}
where $k_{\text{ref}}(\mathbf{r})$ is the reference permeability field at the point $\mathbf{r}$, $k(\mathbf{r})$ is a value of permeability field at $\mathbf{r}$ and $\zeta(\mathbf{r})$ represents the perturbations to the logarithm of the reference permeability. The perturbation $\zeta(\mathbf{r})$ is set as zero at the locations of injector and producer wells as the permeability is known at those grid blocks. For generating multiple realizations of the  $\log(k_{\text{ref}}(\mathbf{r}))$, KL expansion is applied to spatially discretized permeability field. In more details, it is assumed that values of $\log(k_{\text{ref}}(\mathbf{r}))$ at grid-blocks are exponentially correlated:
\begin{equation}
\begin{gathered}
    \label{eq:klog_correlation}
    \langle \log(k_\text{ref}(\mathbf{r}_1)) , \log(k_\text{ref}(\mathbf{r}_2)) \rangle = \exp\bigg( -\frac{|\mathbf{r}_1-\mathbf{r}_2|}{L_\text{ref}} \bigg ) \\
    \langle \zeta(\mathbf{r}_1) , \zeta(\mathbf{r}_2) \rangle = \exp\bigg( -\frac{|\mathbf{r}_1-\mathbf{r}_2|}{L_{\text{p}}} \bigg )
\end{gathered}
\end{equation}
where $L_\text{ref} = 0.3 L$ and $L_{\text{p}} = 0.1 L$ is the correlation length for reference permeability and perturbation and $L = 500 \text{m}$ is the side of the square reservoir. 
For both $\log(k(\mathbf{r}))$ and $\zeta(\mathbf{r})$ KL expansion is performed:
\begin{equation}
\begin{gathered}
    \label{eq:KL_expansion}
    \log(k_{\text{ref}}(\mathbf{r})) = \sum_j \lambda_{\text{ref}, j} \chi_{\text{ref}, j} \xi_{\text{ref}, j}(\mathbf{r}) \\
    \zeta(\mathbf{r}) = \sum_j \lambda_{\text{p}, j} \chi_{\text{p}, j} \xi_{\text{p}, j}(\mathbf{r})
\end{gathered}
\end{equation}
where $\lambda_{\text{ref}, j}$ and $\xi_{\text{ref}, j}(\mathbf{r})$ are eigenvalues and eigenfunctions, respectively for KL expansion for the random field with correlation length $L_{\text{ref}}$, $\lambda_{\text{p}, j}$ and $\xi_{\text{p}, j}(\mathbf{r})$ are eigenvalues and eigenfunctions, respectively for KL expansion for the random field with correlation length $L_{\text{p}}$ respectively, $\chi_{\text{ref}, j}$ and $\chi_{\text{p}, j}$ are uncorrelated normally distributed random variables with zero means and standard deviations $\sigma_{\text{ref}} = 2.0$ and $\sigma_{\text{p}} = 0.5$. In the present work permeability field is normalized in such a way that zero values of $\chi_{\text{ref}, j}$ correspond to permeability of $1 \text{mD}$ ($1~\textrm{milliDarcy} = 9.869233 \times 10^{-16} \textrm{m}^2$). Both reference permeability distribution and perturbation are generated stochastically by sampling of random values from appropriate normal distribution for $\chi_{\text{ref}, j}$ and $\chi_{\text{p}, j}$.
In the present example, the KL expansion for the reference permeability distribution is truncated and only first $50$ eigenfunctions are considered. Then a single realization of reference permeability distribution is selected and utilized in all calculations in the present test case. The perturbation $\zeta(\mathbf{r})$ to the $\log(k_{\text{ref}}(\mathbf{r}))$ is constructed in almost the same fashion. However, different values of correlation length and variance are utilized to perform the KL expansion. Moreover, only first five terms of the KL expansion are considered. In addition to that, two linear constraints on $\zeta(\mathbf{r})$ are utilized, because the perturbation vanishes at the injector and production wells. Therefore, $\zeta(\mathbf{r})$ is effectively parametrized with five coefficients of the KL expansion: $\chi_{\text{p}, 1}, ... ,  \chi_{\text{p}, 5}$. It is clear that for a fixed $k_{\text{ref}}(\mathbf{r})$, the permeability distribution $k(\mathbf{r})$ is fully parametrized by the same parameters due to Eq.~\eqref{eq:perm}. Therefore, the dimension of the model parameter space $\text{dim}(\mathbf{\theta})$ is simply $5$. Figure~\ref{fig:test_case_2_logperm} shows examples of the permeability field generated with the described approach above.

The observations vector $\mathbf{m}$ is formed by values of the permeability and pressure at selected locations. The pressure is measured at early stage of water-flooding at four different moments of time that correspond to four different values of total injected volume of water measured as a fraction of total reservoir pore volume or PVI. Two scenarios of pressure measurements are considered. In the first scenario only value of pressure for $\text{PVI} = 1\%$ is measured. In the second scenario, the pressure is measured for $\text{PVI} = 1\%,\  2\%,\  3\%,\  4\%$

In the current numerical example, the observations vector is computed numerically by solving a coupled system of PDEs namely the conservation of mass and conservation of momentum coupled via Darcy law~\citep{doi:10.1002/0470092718.ch2}:
\begin{equation}
    \label{eq:mass_conservation}
    \frac{\partial \phi s_a \rho_a}{\partial t} - \sum_{\gamma = 1}^{3}  \frac{\partial}{\partial r^{\gamma}} \bigg ( \frac{\rho_a k k_a}{\mu_a} \frac{\partial P} {\partial r^{\gamma}} \bigg ) = Q_{a}
\end{equation}
where $s_a = s_a(\mathbf{r})$ is a volumetric fraction or saturation of fluid with index $a$ ($a = \text{w}$ and $a = \text{o}$ correspond to water and oil, respectively) at the point $\mathbf{r}$, $r^\gamma$ for $\gamma = 1, 2, 3$ are coordinates of the point in space, $\phi = \phi(\mathbf{r})$ is rock porosity at the point $\mathbf{r}$, $k = k(\mathbf{r})$ is permeability at the point $\mathbf{r}$, $k_a = k_a(\mathbf{s})$ is the relative phase permeability that depends only on the fluid saturations $\mathbf{s}$ at the point $\mathbf{r}$, $P = P(\mathbf{r})$ is the pressure at point $\mathbf{r}$, $\rho_a$ is the density of fluid $a$, $\mu_a$ is the  viscosity of fluid $a$, $Q_a = Q_a(\mathbf{r})$ is source term at the point $\mathbf{r}$. In addition to Eq.~\eqref{eq:mass_conservation}, we assume that the pore space is fully saturated with fluids as defined by the following equation:
\begin{equation}
    \label{eq:saturation_constraint}
    \sum_a s_a = s_{\text{w}} +  s_{\text{o}} = 1.
\end{equation}
Given that constraint on saturations, Eq.~\eqref{eq:mass_conservation} determines the time evolution of saturation and pressure distributions for a given porosity and permeability fields and distribution of the source terms. In the present test case, the observation vectors only depend on the parameters of KL expansion for permeability field and on the position of sensors in space.

In the case of incompressible flow, Eq.~\eqref{eq:mass_conservation} admits the following simplification:
\begin{equation}
    \label{eq:mass_conservation_volumetric}
    \phi \frac{\partial s_a}{\partial t} - \sum_{\gamma = 1}^{3} \frac{\partial}{\partial r^{\gamma}} \bigg ( \frac{k k_a}{\mu_a} \frac{\partial P} {\partial r^{\gamma}} \bigg ) = q_{a}
\end{equation}
where $q_a = Q_a/\rho_a$ is the source term for fluid $a$ normalized to the density of corresponding fluid. For calculating the relative phase permeabilities, Brooks-Corey model~\citep{Relative_Phase_Permeability} is utilized:
\begin{equation}
    \label{eq:Corey_model}
    \begin{gathered}
    k_{w}(S_{\text{wn}}) = k_w^{(0)} S_{\text{wn}} ^ {p_w} \\
    k_{w}(S_{\text{wn}}) = k_o^{(0)} (1-S_{\text{wn}}) ^ {p_o}
    \end{gathered}
\end{equation}
where $k_w$ and $k_o$ are the values of the relative phase permeability for water and oil, respectively and $k_w^{(0)}$ and $k_o^{(0)}$ are the maximum values of the relative phase permeability for water and oil, respectively. The values $p_w$ and $p_o$ are dimensionless parameters of the model and $S_{\text{wn}}$ is the normalized water saturation defined as:
\begin{equation}
    \label{eq:normalized_water_saturation}
    S_{\text{wn}} = \frac{s - S_{\text{wir}}} {1 - S_{\text{wir}} -  S_{\text{owr}}}
\end{equation}
where $S_\text{wir}$ and $S_\text{owr}$ are the irreducible water and oil saturations, respectively. For the purposes of simplicity, incompressible immiscible fluids is considered while neglecting gravity effects.

A uniform square grid is used for simulations and the dimensions of each grid-block is $10 \text{m}$ by $10 \text{m}$ by $10 \text{m}$. In other words, a $50$ by $50$ by $1$ mesh is used for discretization. Pressures at injection and production wells are considered to be constant and equal to $200~\text{Bar}$ and $100~\text{Bar}$ respectively. The fluid properties and parameters of Corey model are essentially the same as in~\citep{TARAKANOV2019108909} and are summarized in the Table~\ref{tab:parameters}.
\begin{table}
\begin{center}
\begin{tabular}{ |c|c|c|c|c|c| }\hline
 $\mu_o$, cP & $\mu_w, cP$ & $p_o$ & $p_w$ &  $k_o^{(0)}$ &  $k_w^{(0)}$ \\ \hline
 $10.0$ & $1.0$ & $2.0$ & $2.0$ & $1.0$ & $1.0$  \\ \hline
\end{tabular}
\caption{Fluid properties and parameters of the model for relative-phase permeability.}
\label{tab:parameters}
\end{center}
\end{table}

The evolution of incompressible flow is fully determined by the pressure differences between the injection well and the production wells and  does not depend on the absolute values of those pressures. Therefore, the pressure distribution is rescaled in the following way:
\begin{equation}
    \label{eq:pressure_rescale}
    P_*(t, \mathbf{r}) = \frac{P(t, \mathbf{r}) - P_0} {P_1 - P_0}
\end{equation}
where $P_0$ and $P_1$ are the pressures at the injection well and production wells and $P_*(t, \mathbf{r})$ is a normalized pressure. In the present test case, normalized pressure $P_*(t, \mathbf{r})$ is utilized for construction of observations vector Eq.~\eqref{eq:system_bed}.

\begin{figure}[H]
    \centering
    \begin{subfigure}[b]{0.23\textwidth}
        \includegraphics[width=0.99\linewidth]{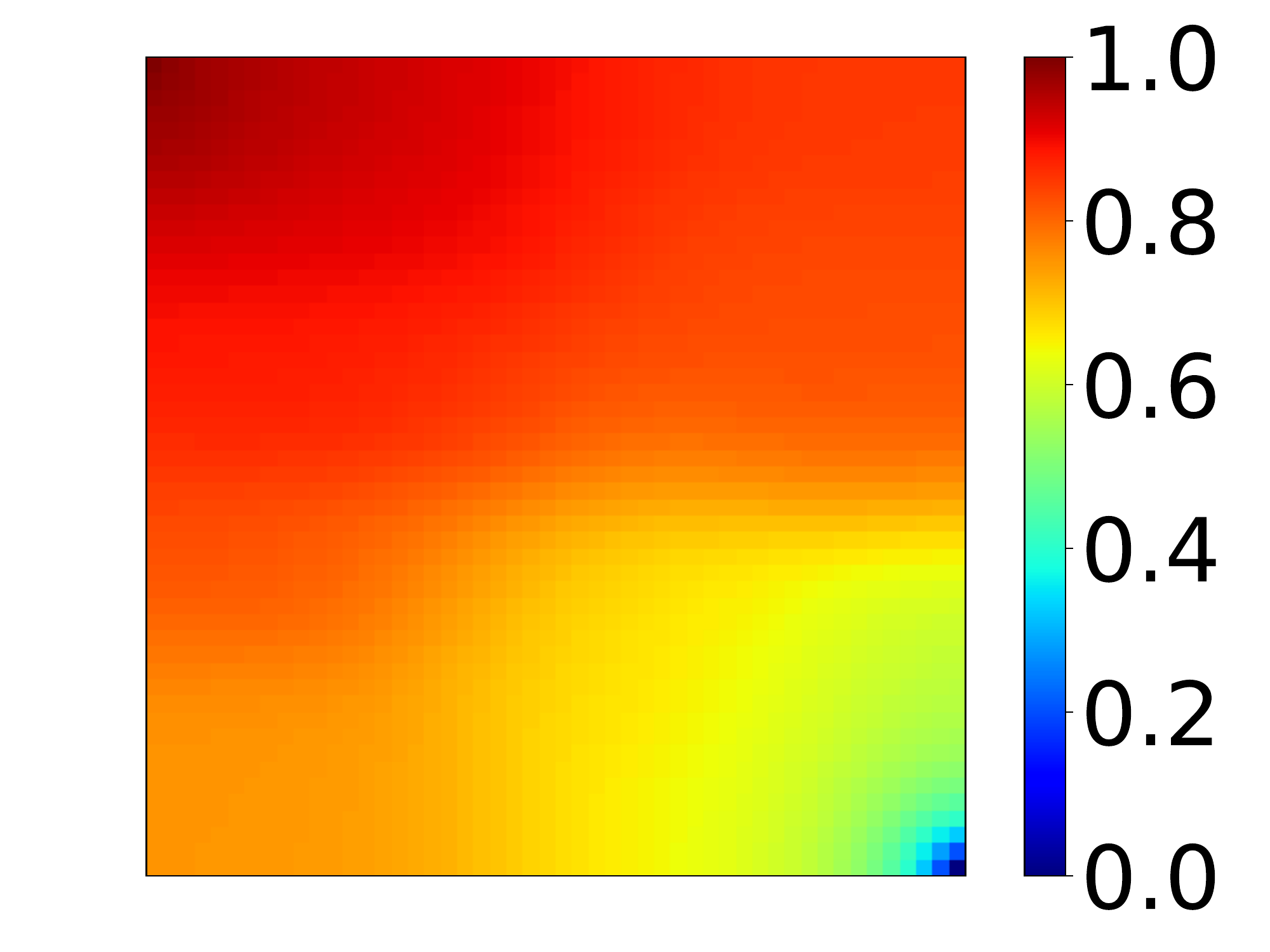}
        \caption{$\text{PVI} = 1\%$}
        \label{fig:test_case_2_pref_kt_0}
    \end{subfigure}
    \begin{subfigure}[b]{0.23\textwidth}
        \includegraphics[width=0.99\linewidth]{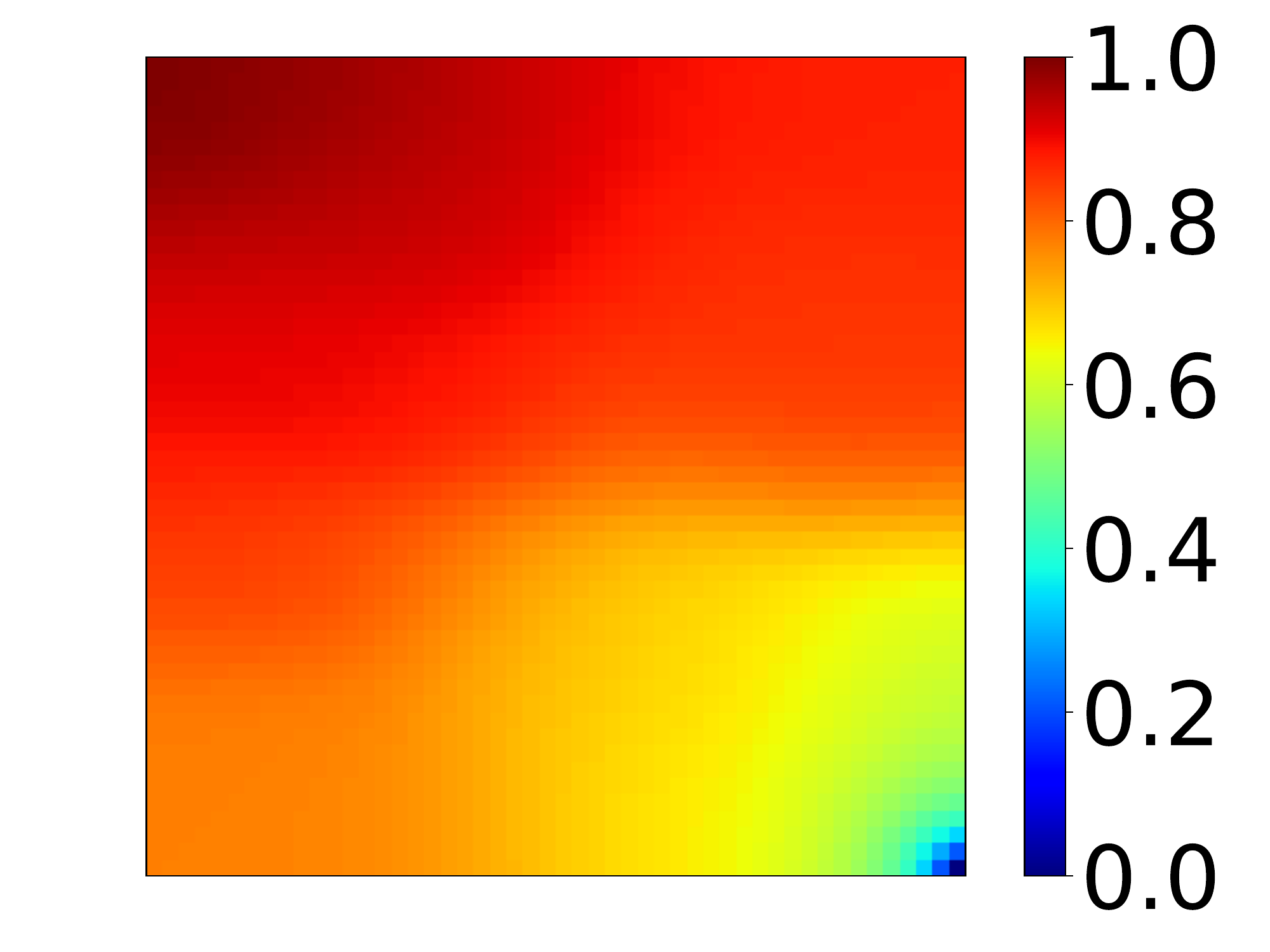}
        \caption{$\text{PVI} = 2\%$}
        \label{fig:test_case_2_pref_kt_1}
    \end{subfigure}
    \begin{subfigure}[b]{0.23\textwidth}
        \includegraphics[width=0.99\linewidth]{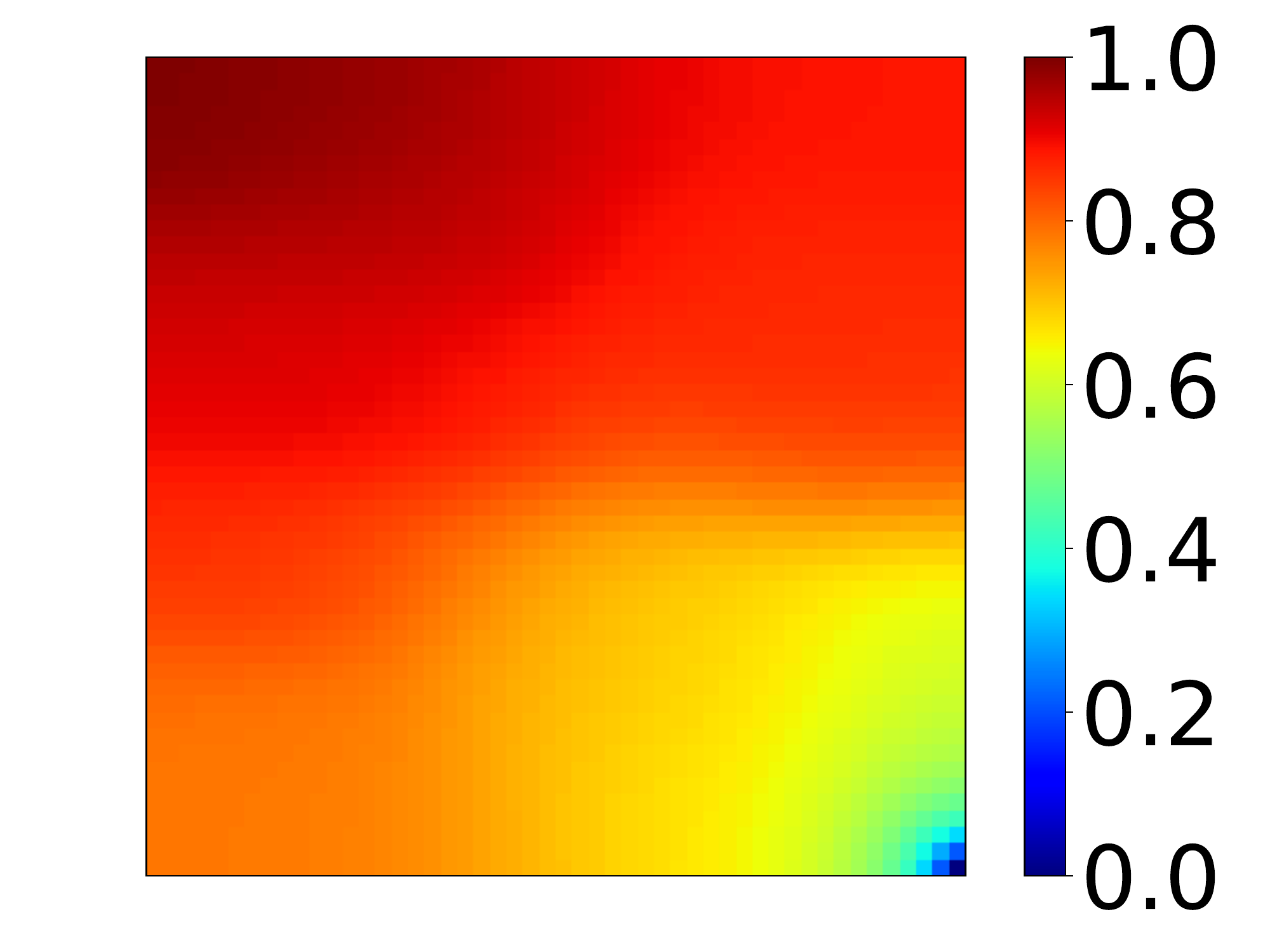}
        \caption{$\text{PVI} = 3\%$}
        \label{fig:test_case_2_pref_kt_2}
    \end{subfigure}
    \begin{subfigure}[b]{0.23\textwidth}
        \includegraphics[width=0.99\linewidth]{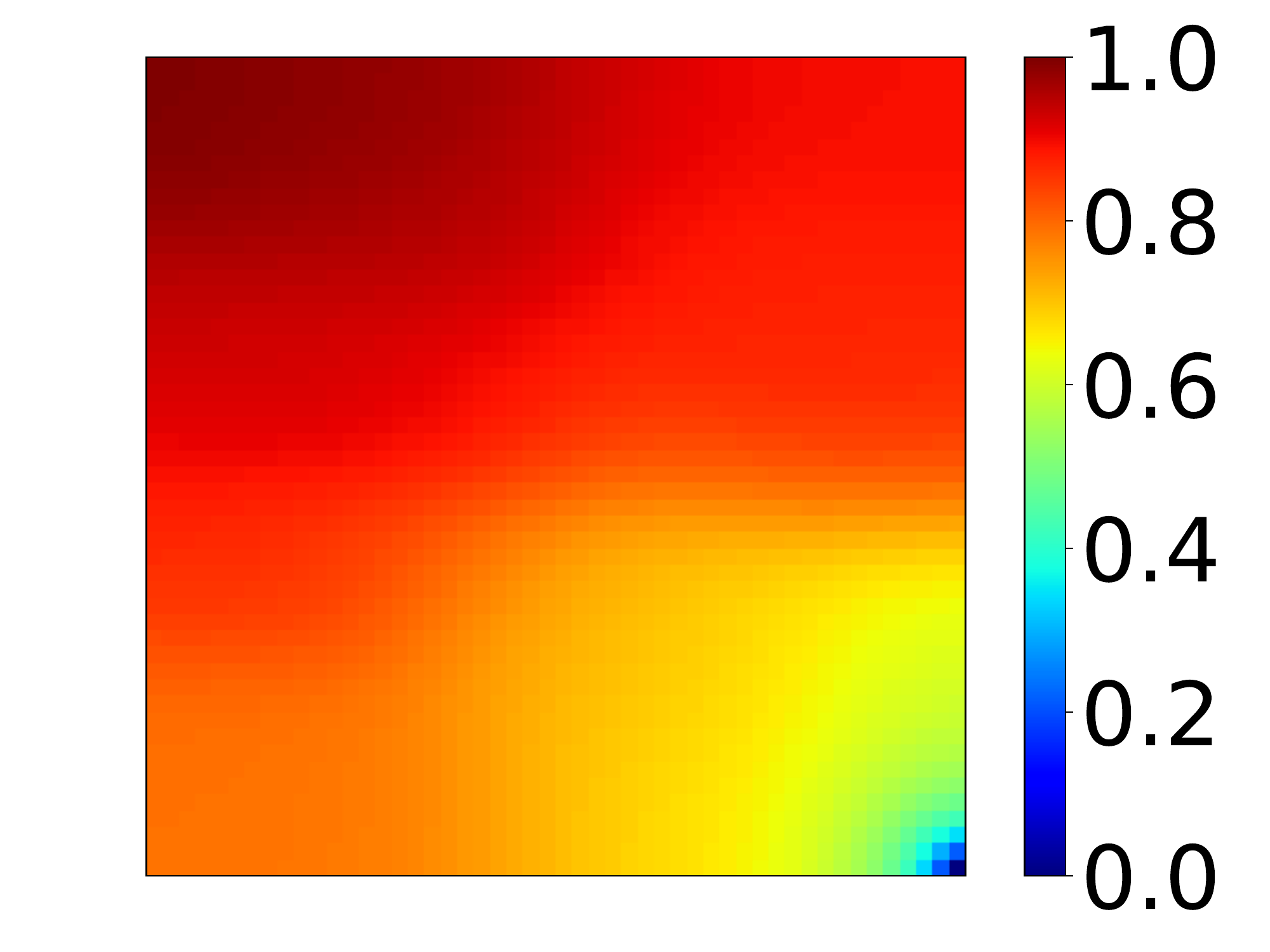}
        \caption{$\text{PVI} = 4\%$}
        \label{fig:test_case_2_pref_kt_3}
    \end{subfigure}

    \begin{subfigure}[b]{0.23\textwidth}
        \includegraphics[width=0.99\linewidth]{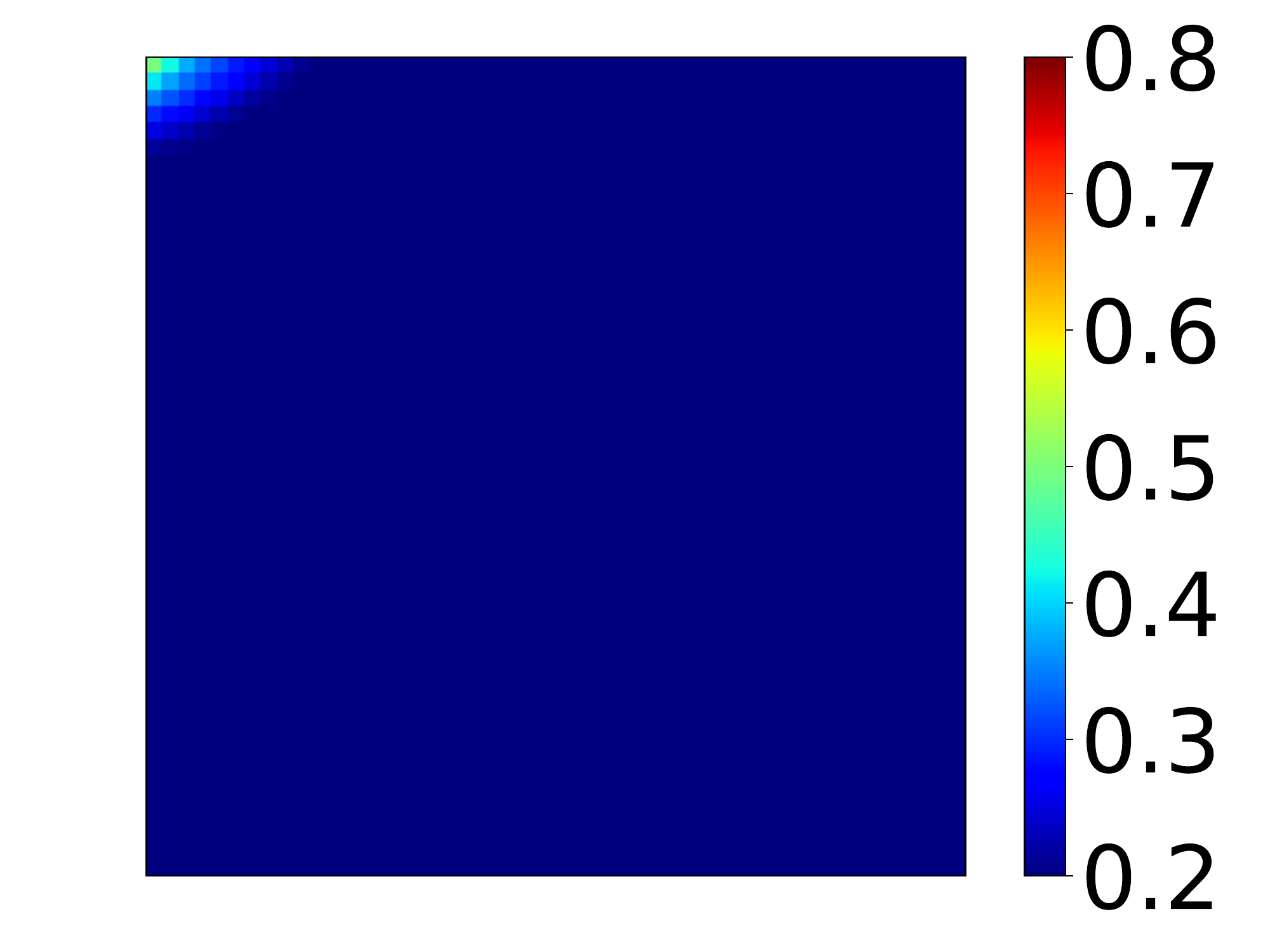}
        \caption{$\text{PVI} = 1\%$}
        \label{fig:test_case_2_sref_kt_0}
    \end{subfigure}
    \begin{subfigure}[b]{0.23\textwidth}
        \includegraphics[width=0.99\linewidth]{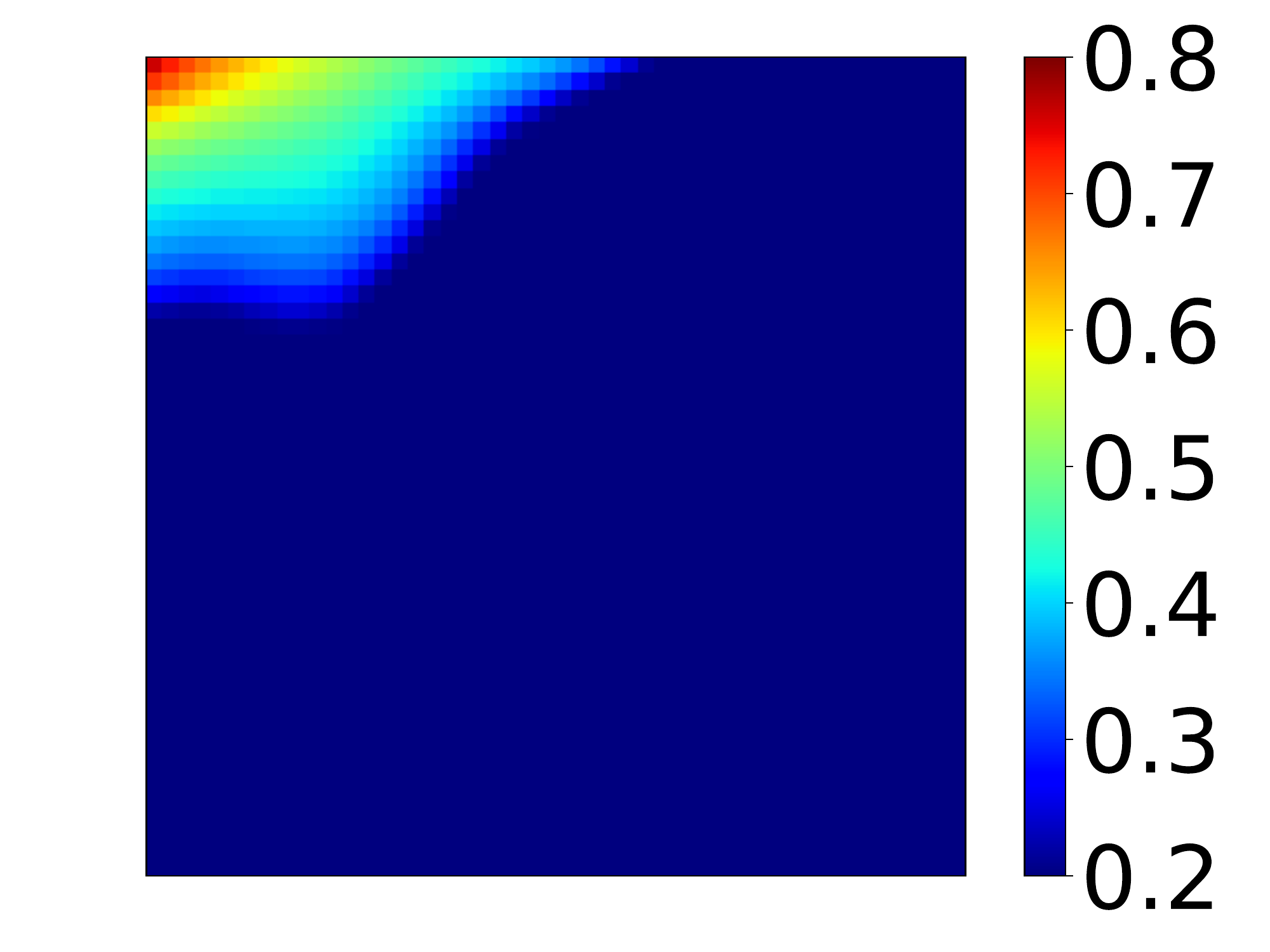}
        \caption{$\text{PVI} = 2\%$}
        \label{fig:test_case_2_sref_kt_1}
    \end{subfigure}
    \begin{subfigure}[b]{0.23\textwidth}
        \includegraphics[width=0.99\linewidth]{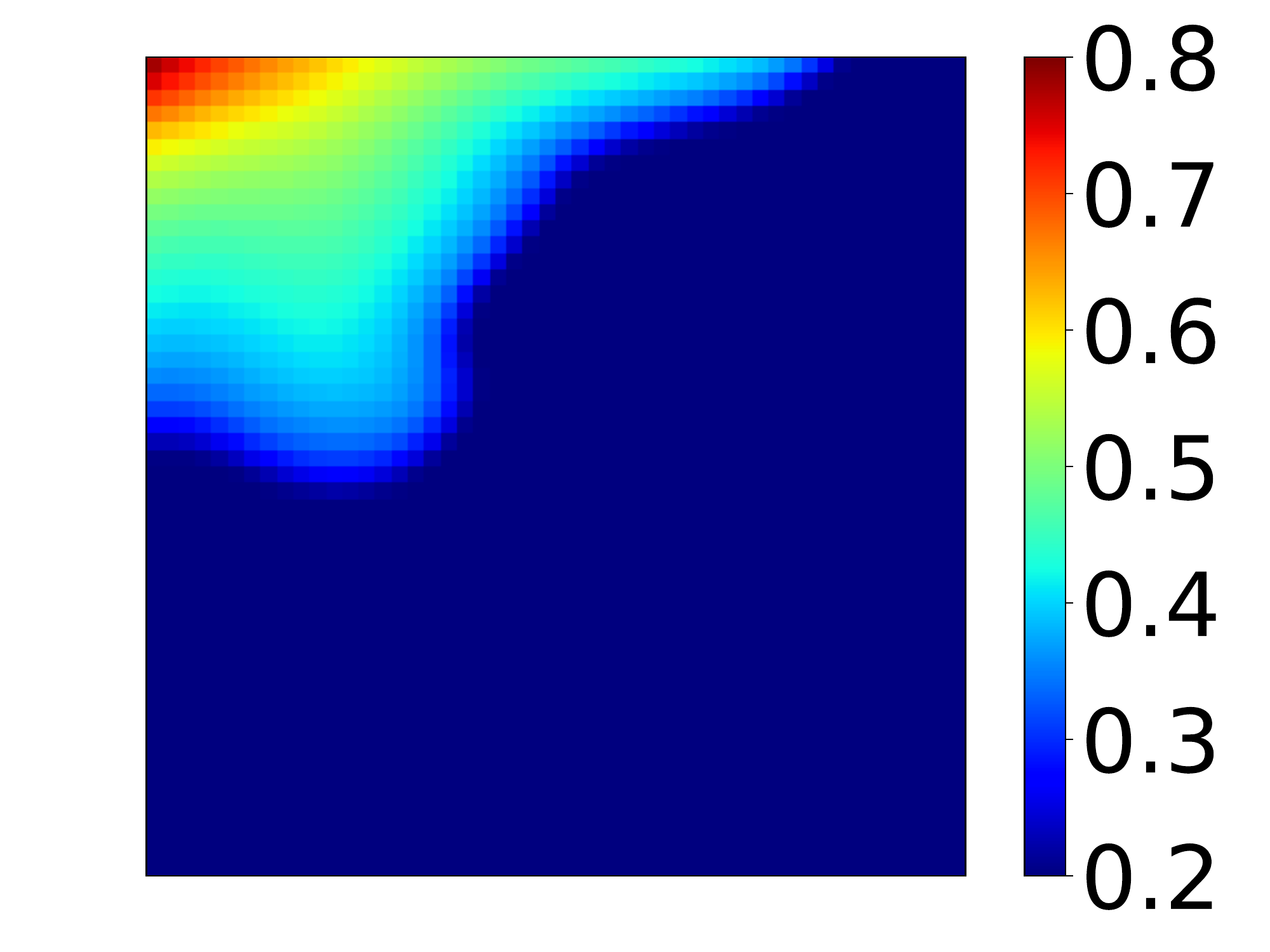}
        \caption{$\text{PVI} = 3\%$}
        \label{fig:test_case_2_sref_kt_2}
    \end{subfigure}
    \begin{subfigure}[b]{0.23\textwidth}
        \includegraphics[width=0.99\linewidth]{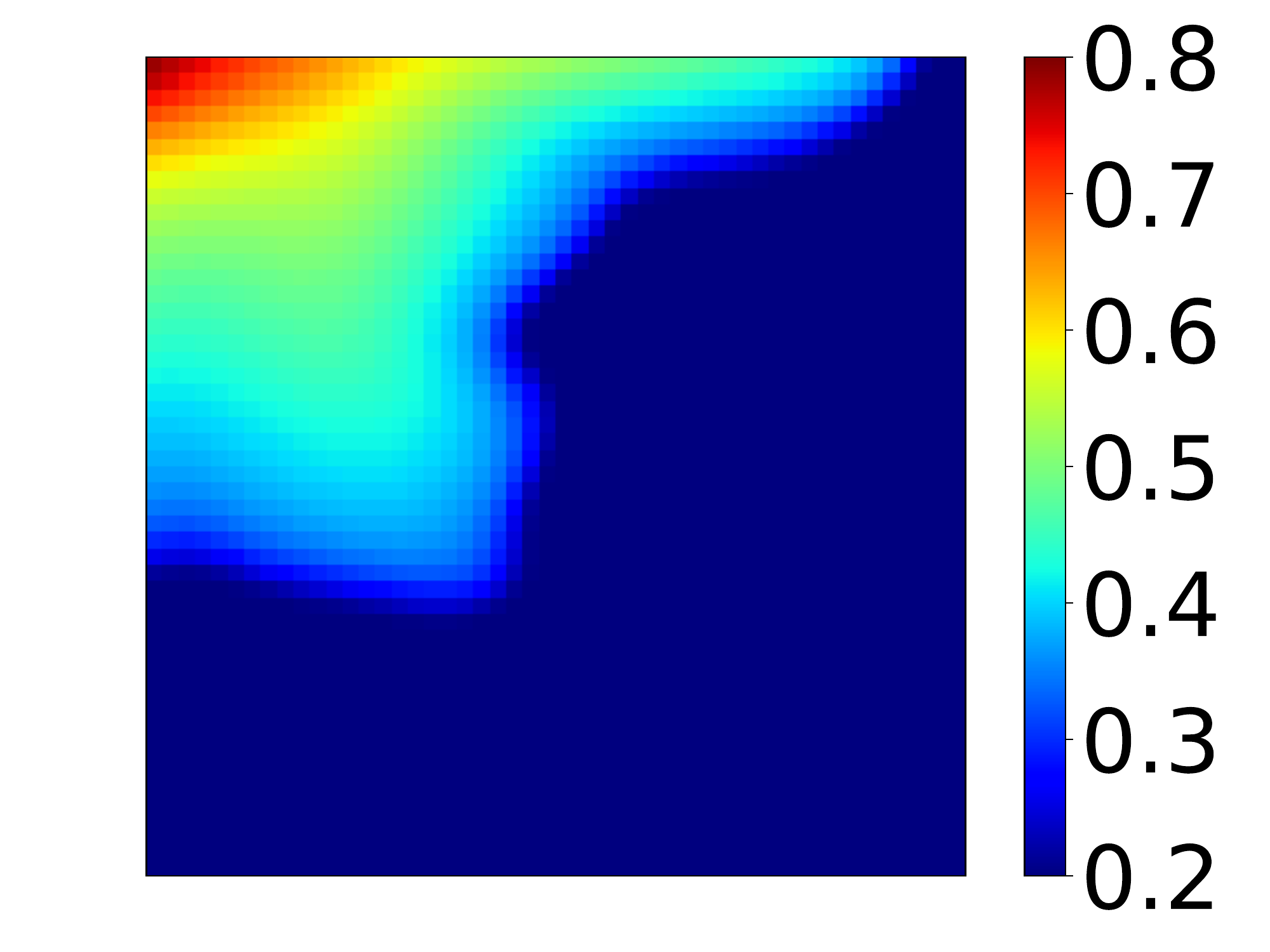}
        \caption{$\text{PVI} = 4\%$}
        \label{fig:test_case_2_sref_kt_3}
    \end{subfigure}
    \caption{Snapshots of pressure (a, b, c, d) and saturation (e, f, g, h) distributions for $\text{PVI} = 1\%, 2\%, 3\%$ and $4\%$ respectively computed for the reference permeability field.}
    \label{fig:test_case_2_ref}
\end{figure}

Figure.~\eqref{fig:test_case_2_ref} shows the pressure and saturation distributions for the reference permeability field $k_{\text{ref}}(\mathbf{r})$ at different PVI values. The plots demonstrate that reference permeability field is highly heterogeneous, leading to a highly heterogeneous distribution of the saturation field.

\begin{figure}[H]
    \centering
    \begin{subfigure}[b]{0.18\textwidth}
        \includegraphics[width=0.99\linewidth]{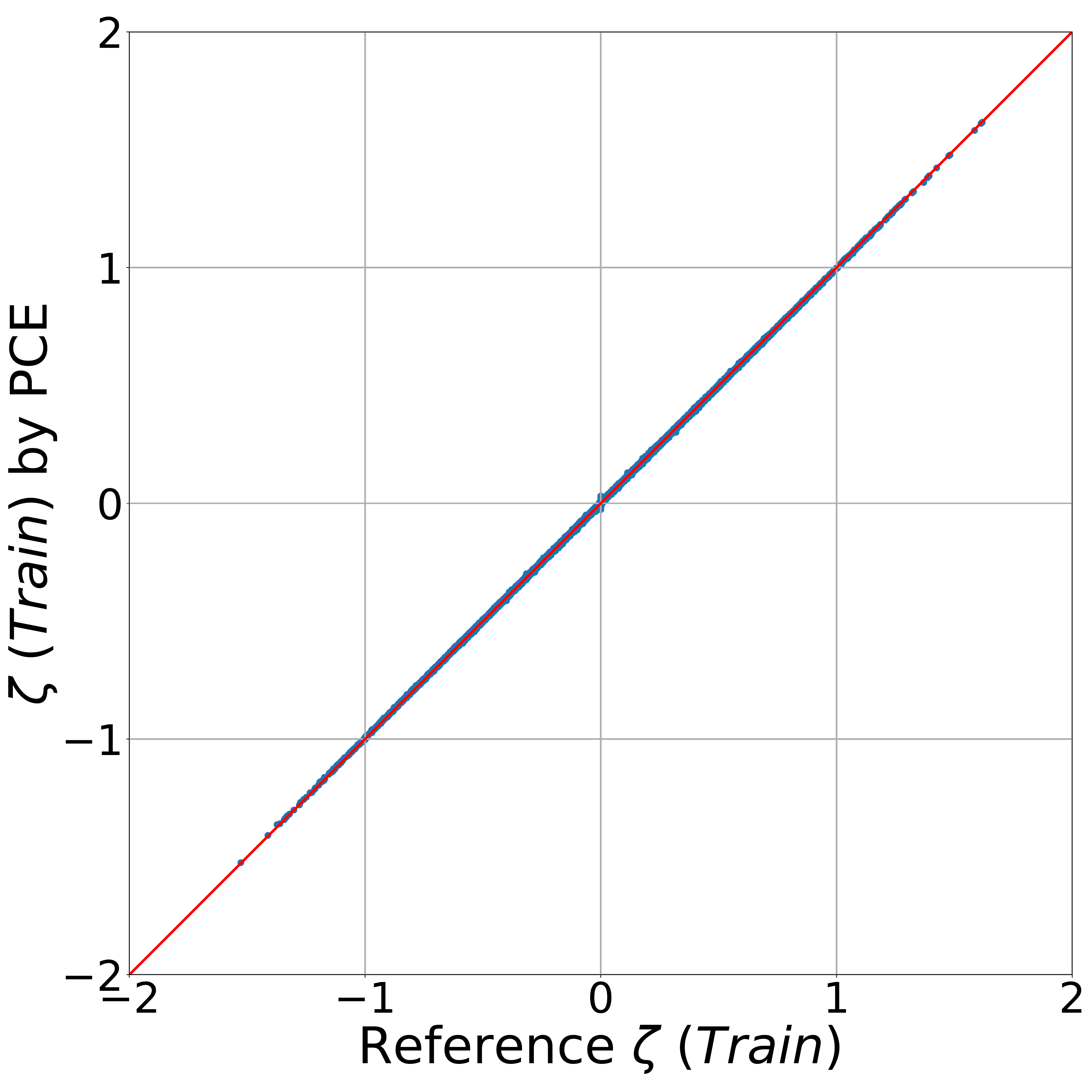}
        \caption{$\zeta$}
        \label{fig:test_case_2_pce_train_0}
    \end{subfigure}
    \begin{subfigure}[b]{0.18\textwidth}
        \includegraphics[width=0.99\linewidth]{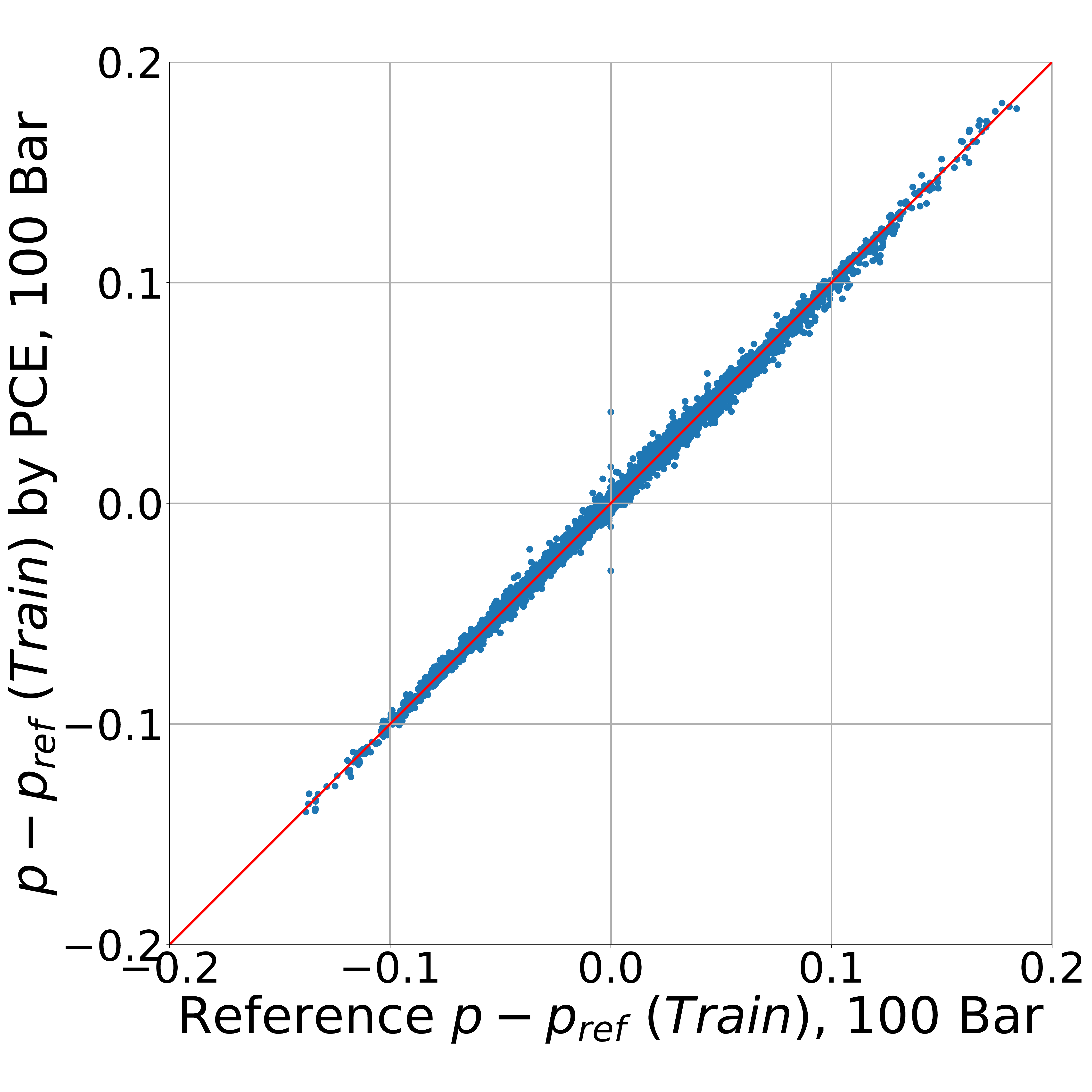}
        \caption{$\text{PVI} = 1\%$}
        \label{fig:test_case_2_pce_train_1}
    \end{subfigure}
    \begin{subfigure}[b]{0.18\textwidth}
        \includegraphics[width=0.99\linewidth]{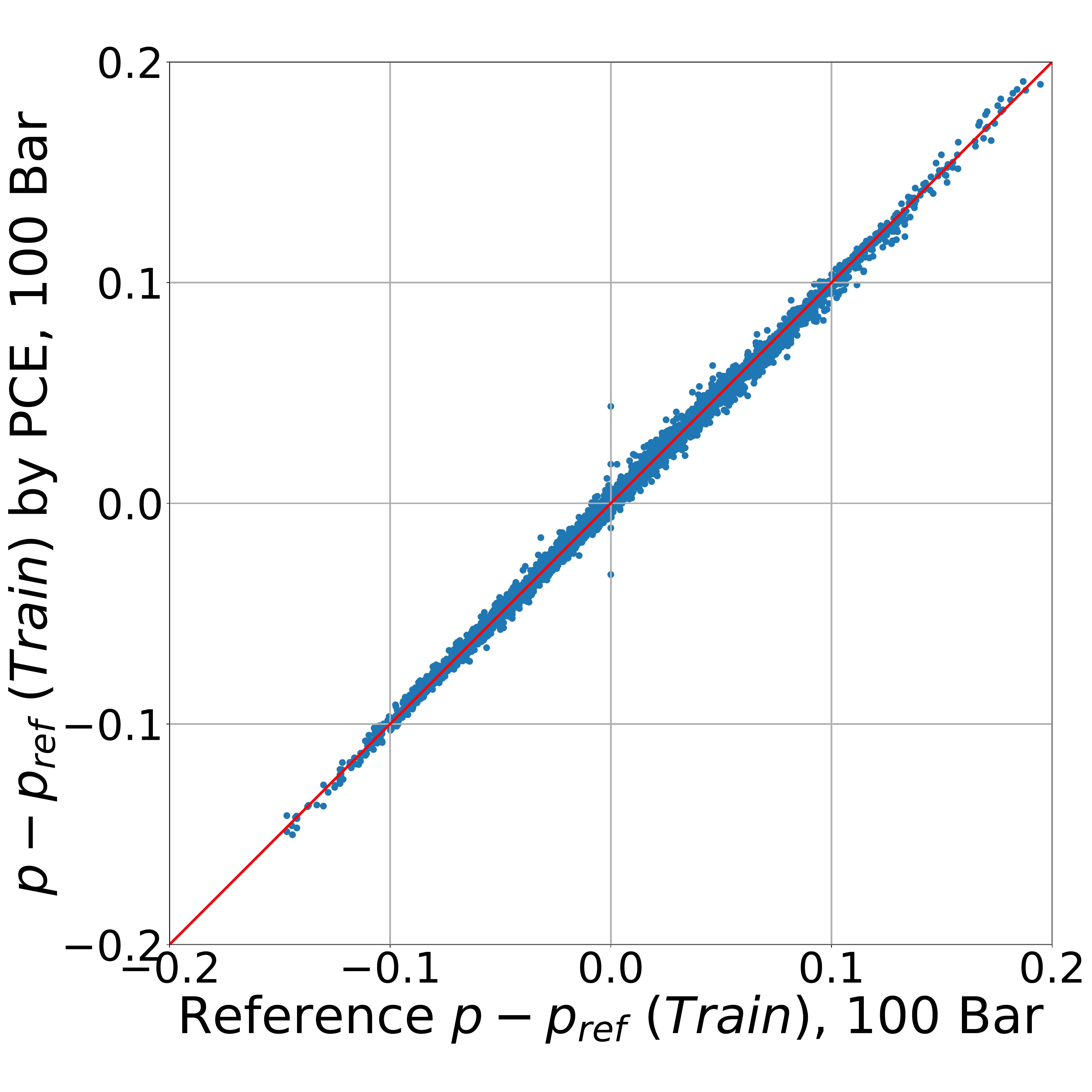}
        \caption{$\text{PVI} = 2\%$}
        \label{fig:test_case_2_pce_train_2}
    \end{subfigure}
    \begin{subfigure}[b]{0.18\textwidth}
        \includegraphics[width=0.99\linewidth]{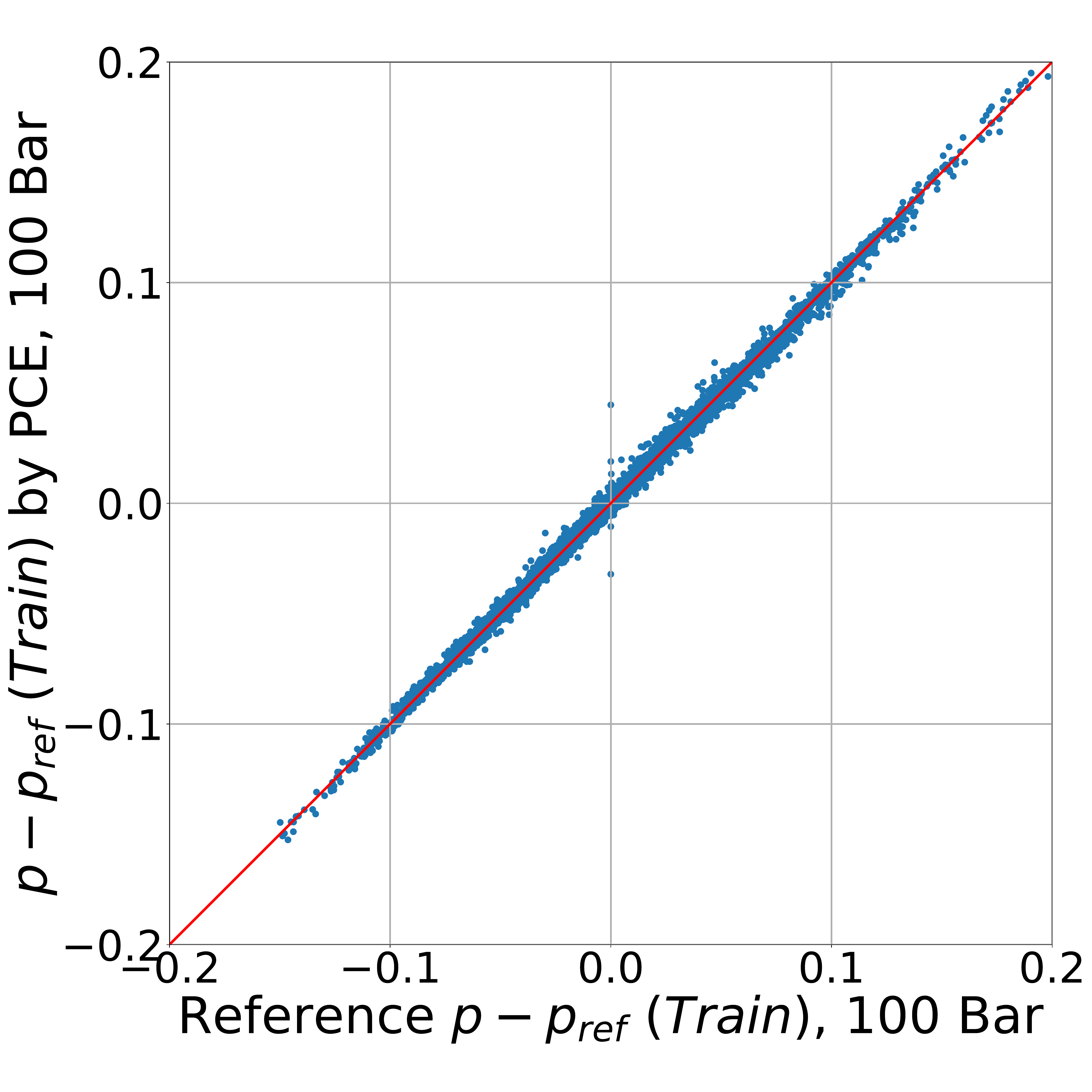}
        \caption{$\text{PVI} = 3\%$}
        \label{fig:test_case_2_pce_train_3}
    \end{subfigure}
    \begin{subfigure}[b]{0.18\textwidth}
        \includegraphics[width=0.99\linewidth]{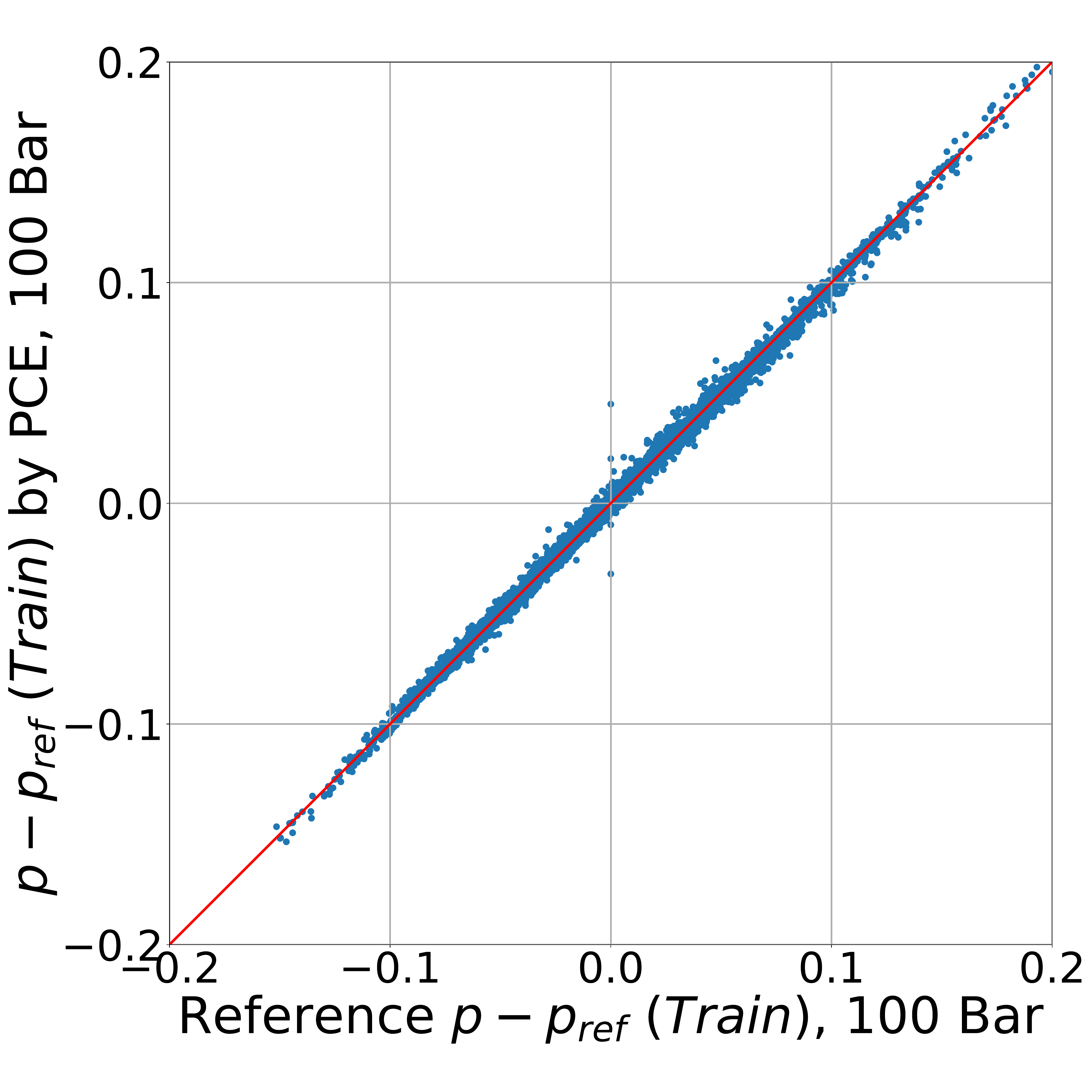}
        \caption{$\text{PVI} = 4\%$}
        \label{fig:test_case_2_pce_train_4}
    \end{subfigure}

    \begin{subfigure}[b]{0.18\textwidth}
        \includegraphics[width=0.99\linewidth]{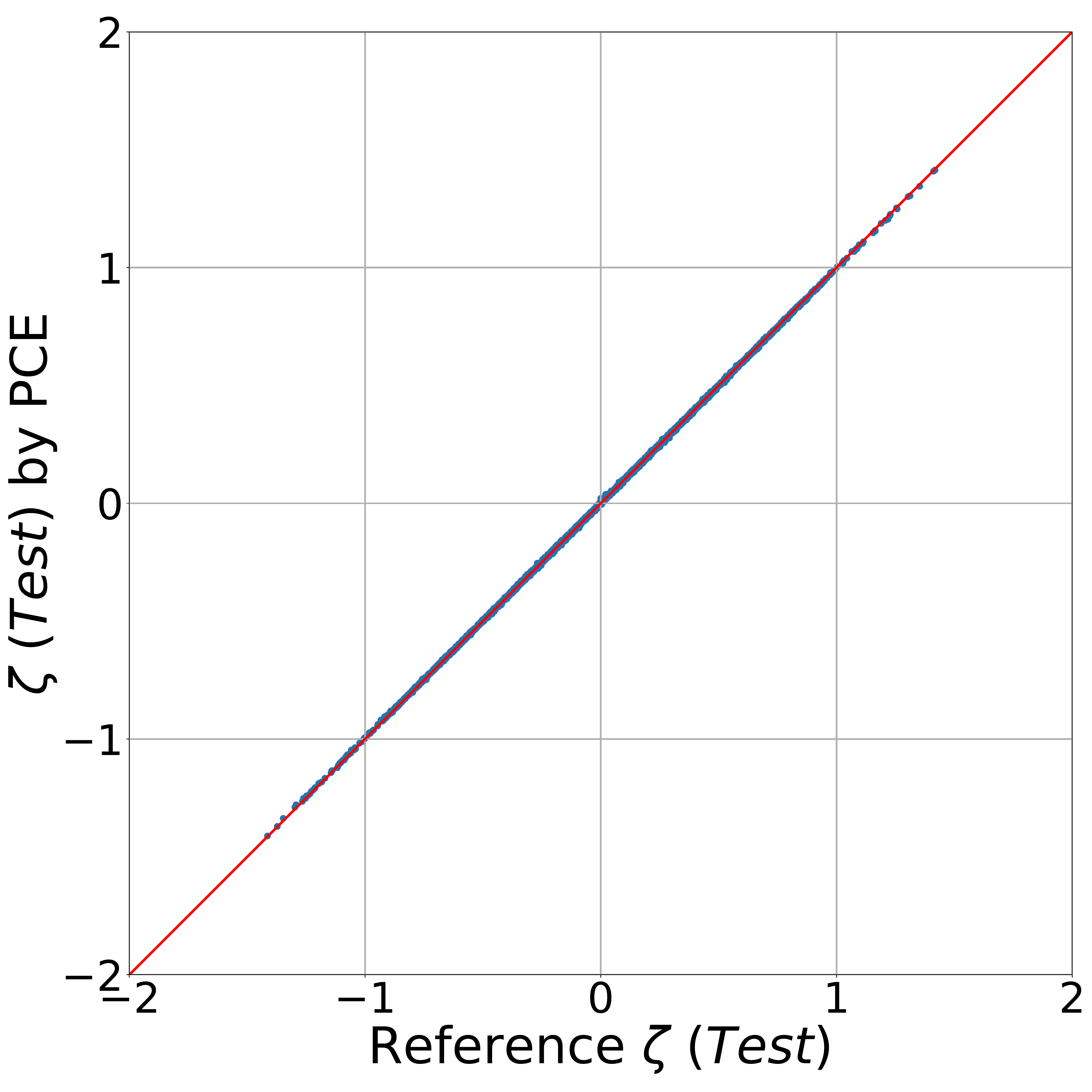}
        \caption{$\zeta$}
        \label{fig:test_case_2_pce_test_0}
    \end{subfigure}
    \begin{subfigure}[b]{0.18\textwidth}
        \includegraphics[width=0.99\linewidth]{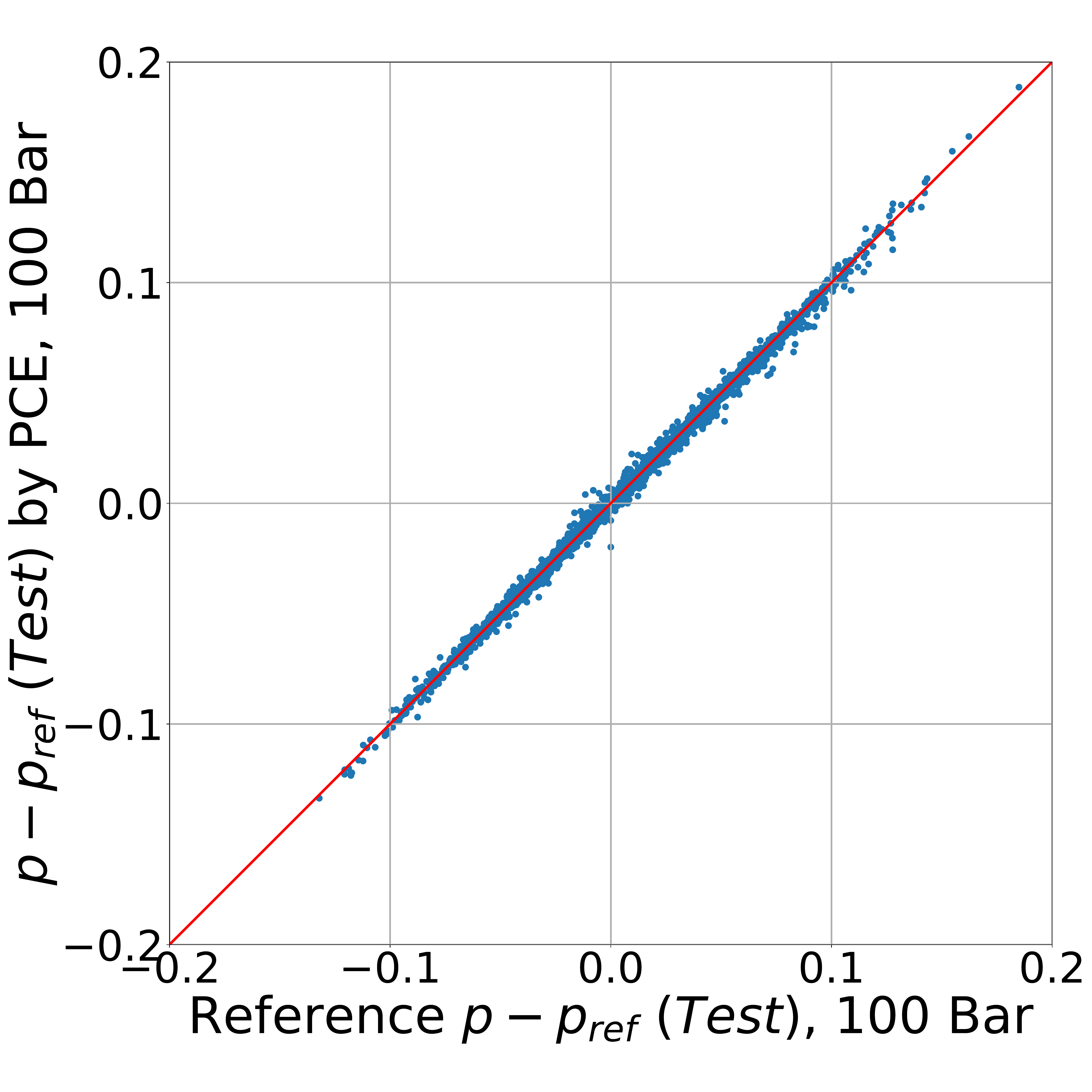}
        \caption{$\text{PVI} = 1\%$}
        \label{fig:test_case_2_pce_test_1}
    \end{subfigure}
    \begin{subfigure}[b]{0.18\textwidth}
        \includegraphics[width=0.99\linewidth]{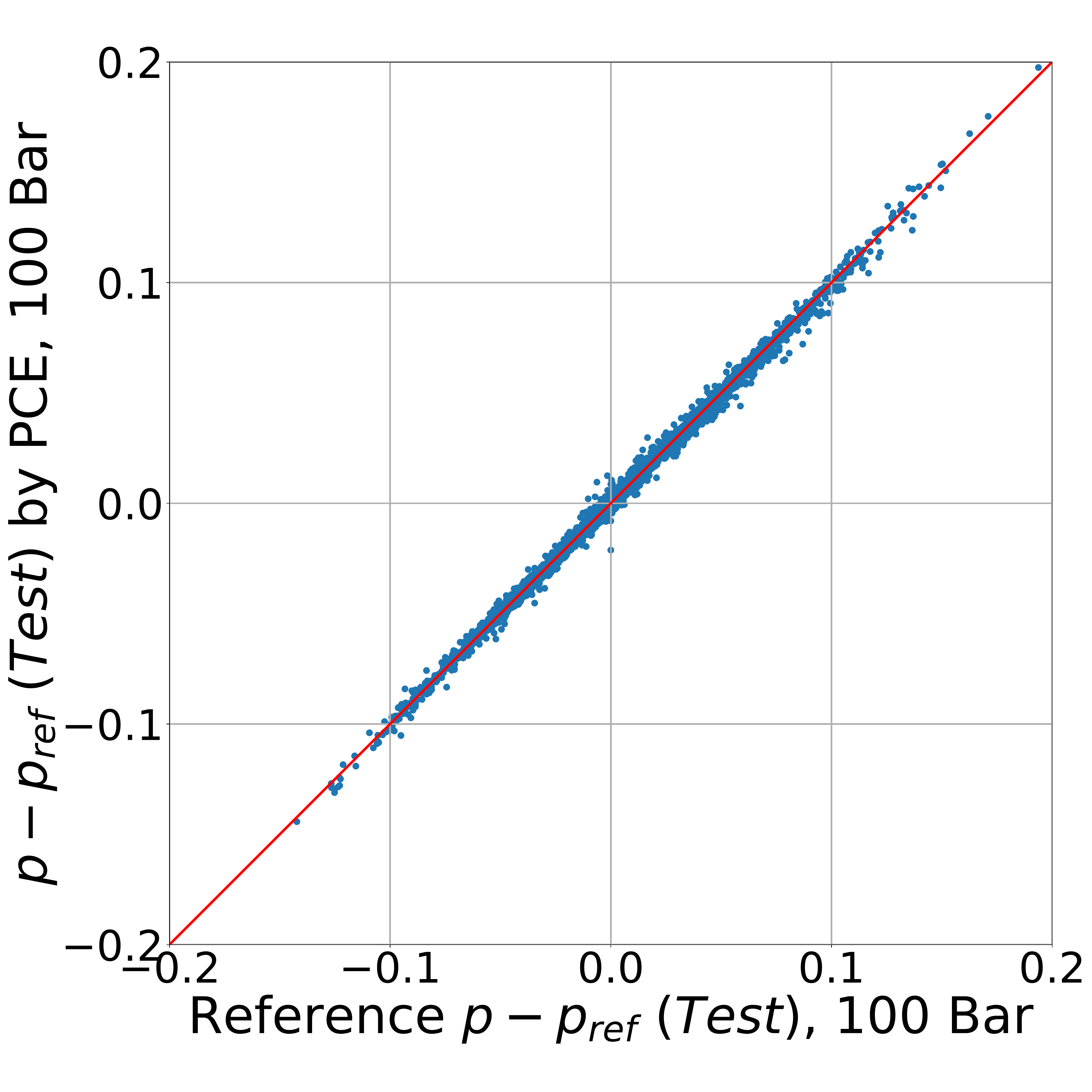}
        \caption{$\text{PVI} = 2\%$}
        \label{fig:test_case_2_pce_test_2}
    \end{subfigure}
    \begin{subfigure}[b]{0.18\textwidth}
        \includegraphics[width=0.99\linewidth]{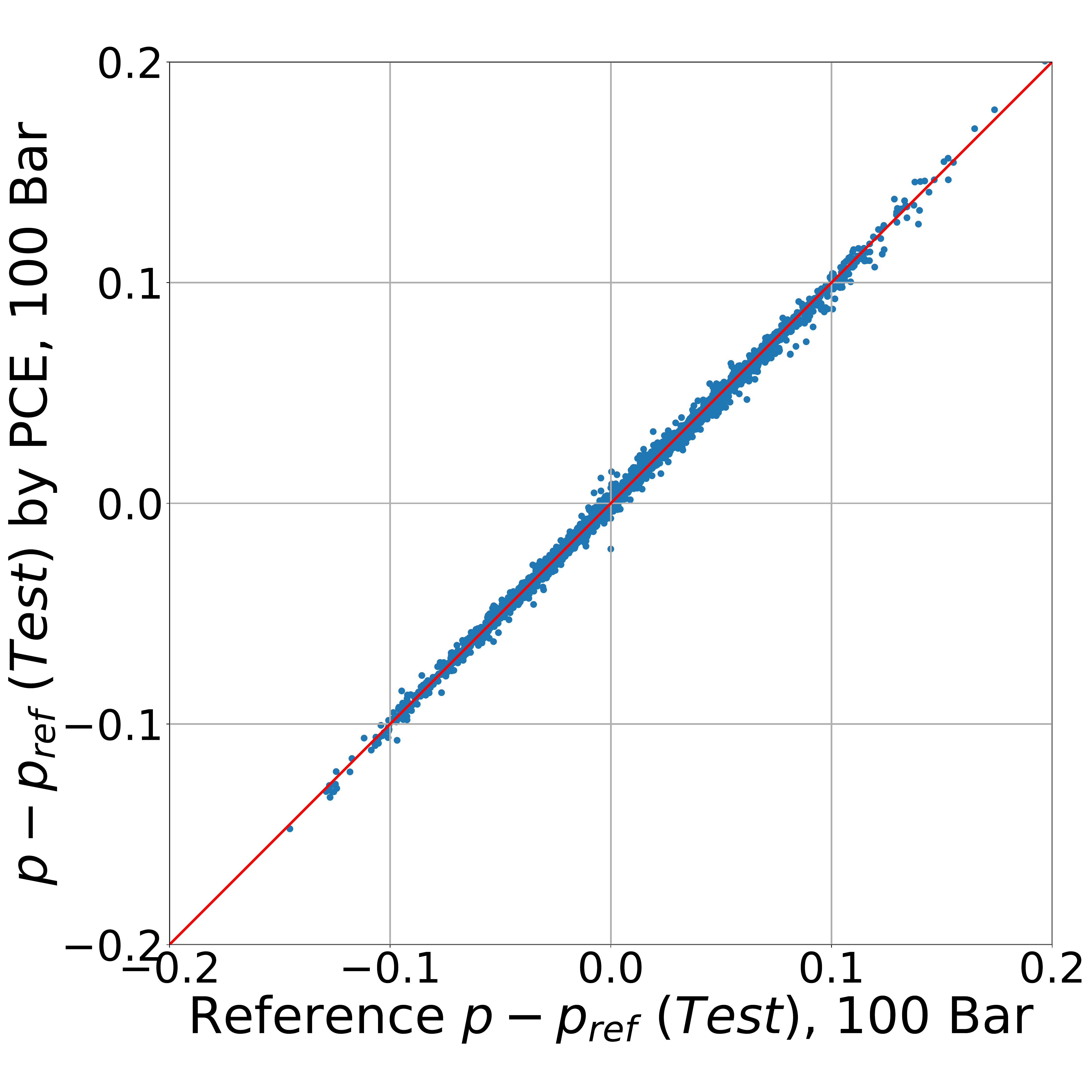}
        \caption{$\text{PVI} = 3\%$}
        \label{fig:test_case_2_pce_test_3}
    \end{subfigure}
    \begin{subfigure}[b]{0.18\textwidth}
        \includegraphics[width=0.99\linewidth]{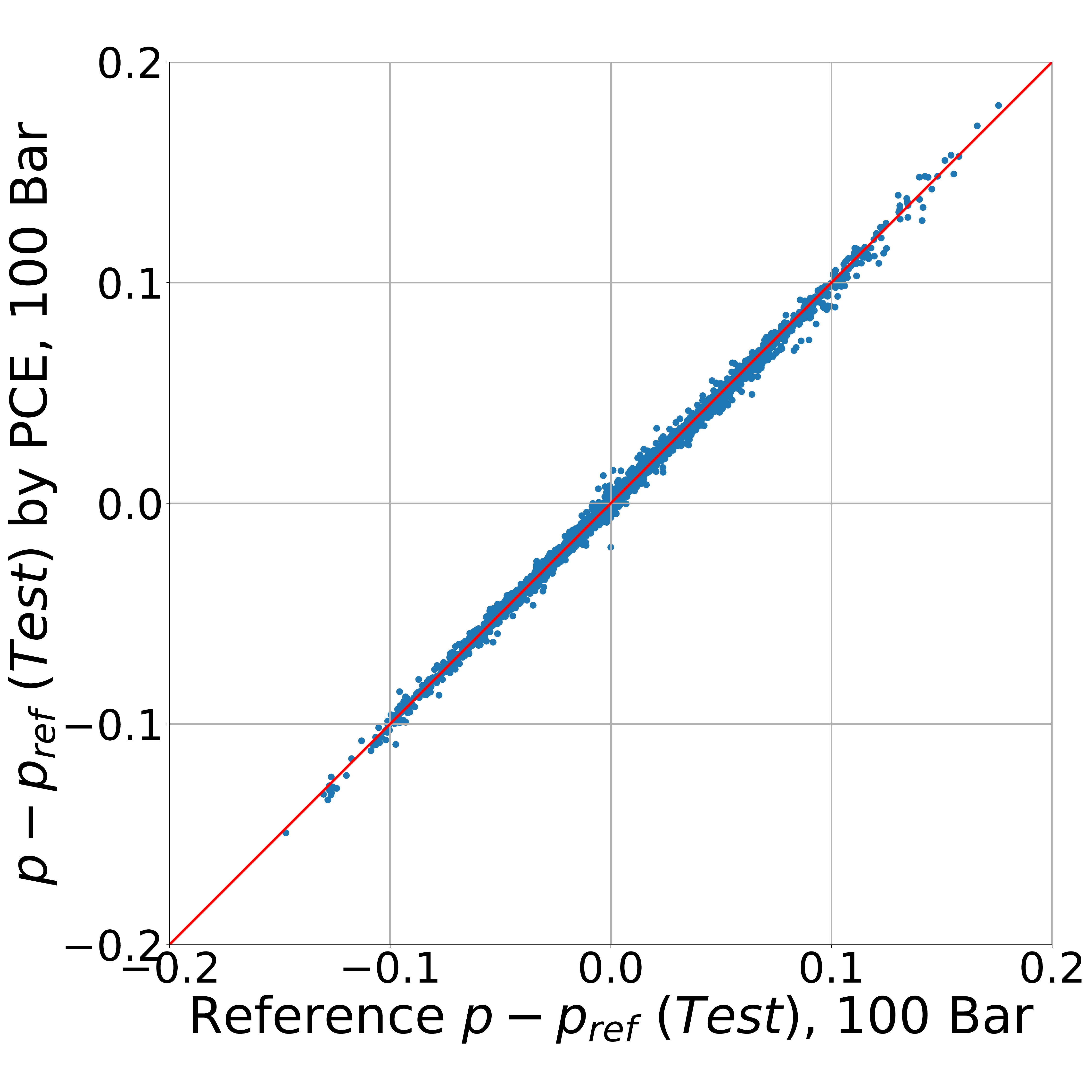}
        \caption{$\text{PVI} = 4\%$}
        \label{fig:test_case_2_pce_test_4}
    \end{subfigure}
    \caption{Cross-plots of reference values and predictions of PCE surrogate model for training (a) - (e) and test (f) - (j) data respectively. Figures (a) and (f) correspond to permeability perturbation and remaining figures correspond to deviation from the reference pressure for different values of ${\text{PVI}}$.}
    \label{fig:test_case_2_pce}
\end{figure}

Direct utilization of flow simulations in Bayesian experimental design is not feasible due to the high computational cost of estimating the utility function $U(\mathbf{d})$. Therefore, PCE-based response surface for both $\zeta(\mathbf{r})$ and $P_*(t,\mathbf{r})$ has been developed. For that purpose, numerical simulations on $5,000$ different realizations of perturbation to the reference permeability field have been performed. A total of $4,000$ of those simulations are utilized for building (aka. training) the PCE-based response surface and the remaining $1,000$ model runs are used for validation and hyper-parameters optimization. For each simulation from the training set, a $20$ grid blocks are randomly sampled and the values of $\zeta(\mathbf{r})$ and $P_*(t, \mathbf{r})$ are added to the training dataset. Finally, the PCE surrogate for permeability perturbation and pressure has been developed as a function of two spatial coordinates and five coefficients of KL expansion. Both spatial coordinates and parameters of perturbation are rescaled in such a way that classical families of orthogonal polynomials can be utilized. Namely, Legendre and Hermite probabilistic polynomials are utilized for the spatial variables and parameters of KL expansion, respectively. Basis polynomials of degree up to eight with respect to all variables are considered in PCE. Additional constraint is imposed on the Hermite polynomials, where only basis functions of degree up to four are utilized. The PCE coefficients are computed via minimization of mean-square error functional with Elastic-Net regularization terms~\citep{Regularized_Linear_Regression}. The accuracy of the response surface on the validation data is around $3\%$. The cross-plots shown in Figure~\ref{fig:test_case_2_pce} demonstrate quality of the response surface on both the training and validation dataset.

\begin{figure}[H]
    \centering
    \begin{subfigure}[b]{0.45\textwidth}
        \includegraphics[width=0.99\linewidth]{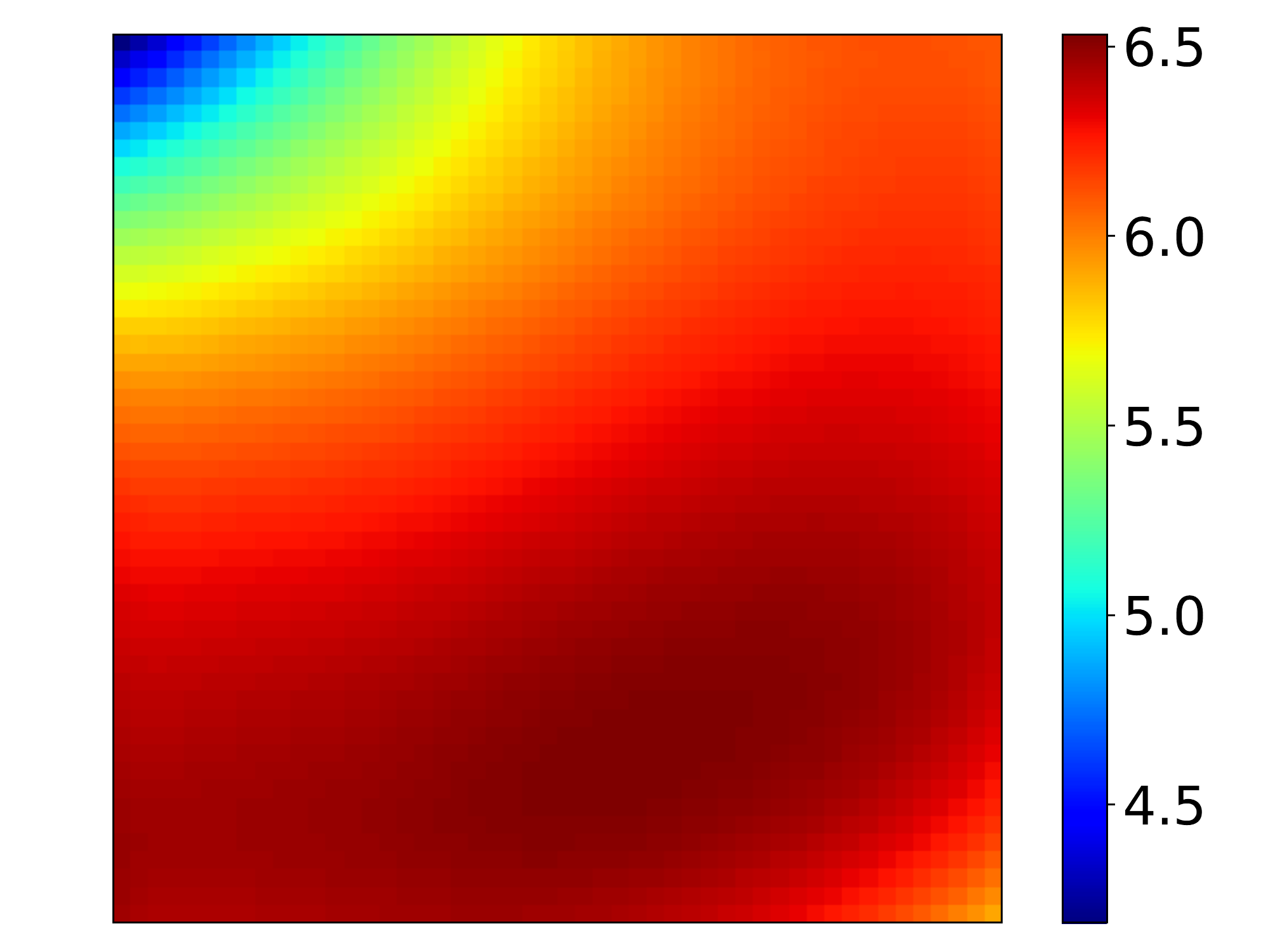}
        \caption{ }
        \label{fig:test_case_2_ufun_nwell_1_nt_2}
    \end{subfigure}
    \begin{subfigure}[b]{0.45\textwidth}
        \includegraphics[width=0.99\linewidth]{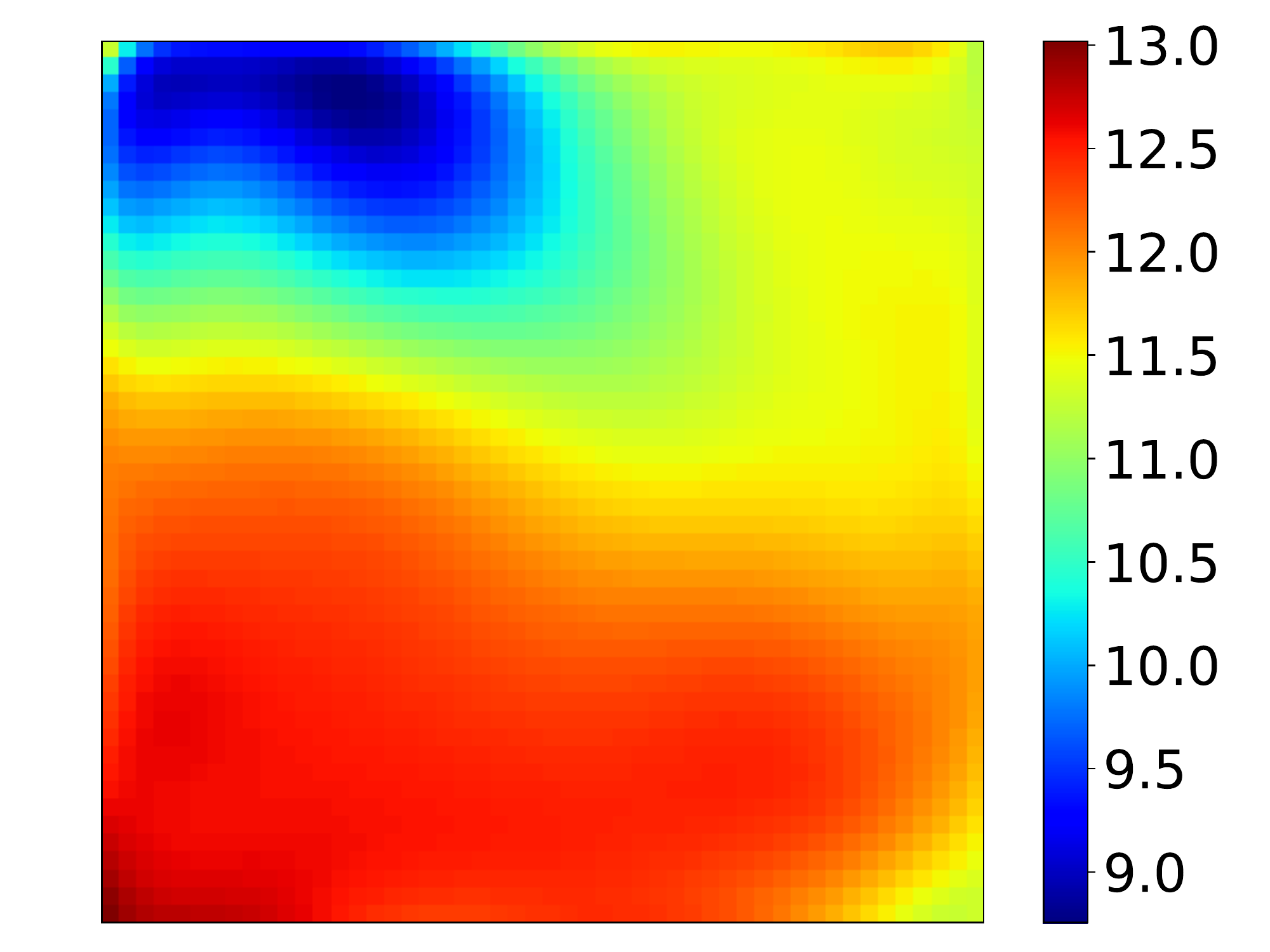}
        \caption{ }
        \label{fig:test_case_2_ufun_nwell_1_nt_5}
    \end{subfigure}
    \caption{PCE response surface for expected information gain for experiments with one (a) and four (b) pressure measurements.}
    \label{fig:test_case_2_ufun_nwell_1}
\end{figure}

In the proposed method, the model function $f(\mathbf{\theta}, \mathbf{d})$ is calculated via PCE-based response surface and the differences between the model predictions and observations are assumed to follow a normal distribution and the standard deviation for that difference $\sigma$ is assumed to be the same for all the components of vector of observables. This is generally true, as long as $\zeta(\mathbf{r})$ and $P_*(t, \mathbf{r})$ are dimensionless quantities. In this test case, the standard deviation is set to be $\sigma = 1.0 \times \textrm {10}^{-3}$. A total of $200,000$ realizations of design parameters $\mathbf{d}$ are sampled in both cases with $n_s = 1$ and $n_s = 2$. All model parameters are rescaled linearly in order to be uniformly distributed in the interval $[-1,1]$. For each of the samples KL-divergence is computed with MCMC chain of length $50,000$. The computed data is then fitted with Legendre polynomials on a rescaled design parameters only. PCE is truncated by the total polynomial degree, which is set to $5$. According to Eq.~\eqref{eq:minimization_problem_b_ed}, the surrogate model developed represents $U(\mathbf{d})$ directly. The response surfaces for $U(\mathbf{d})$ are visualized for the case of a single new well, for the two different number of pressure measurements as shown in Fig.~\ref{fig:test_case_2_ufun_nwell_1}.
\begin{figure}[H]
    \centering
    \begin{subfigure}[b]{0.45\textwidth}
        \includegraphics[width=0.99\linewidth]{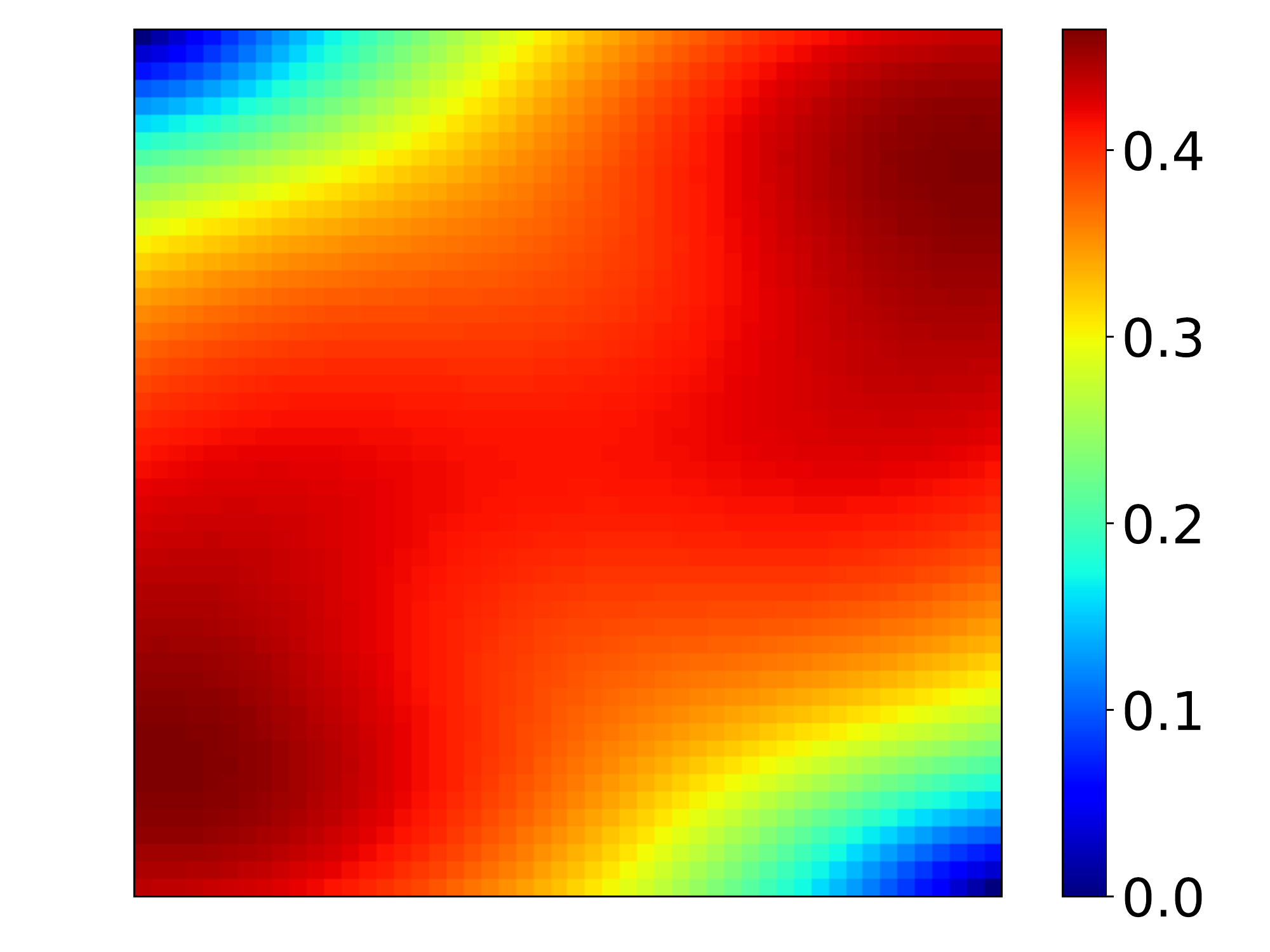}
        \caption{ }
        \label{fig:test_case_2_logperm_sigma}
    \end{subfigure}
    \begin{subfigure}[b]{0.45\textwidth}
        \includegraphics[width=0.99\linewidth]{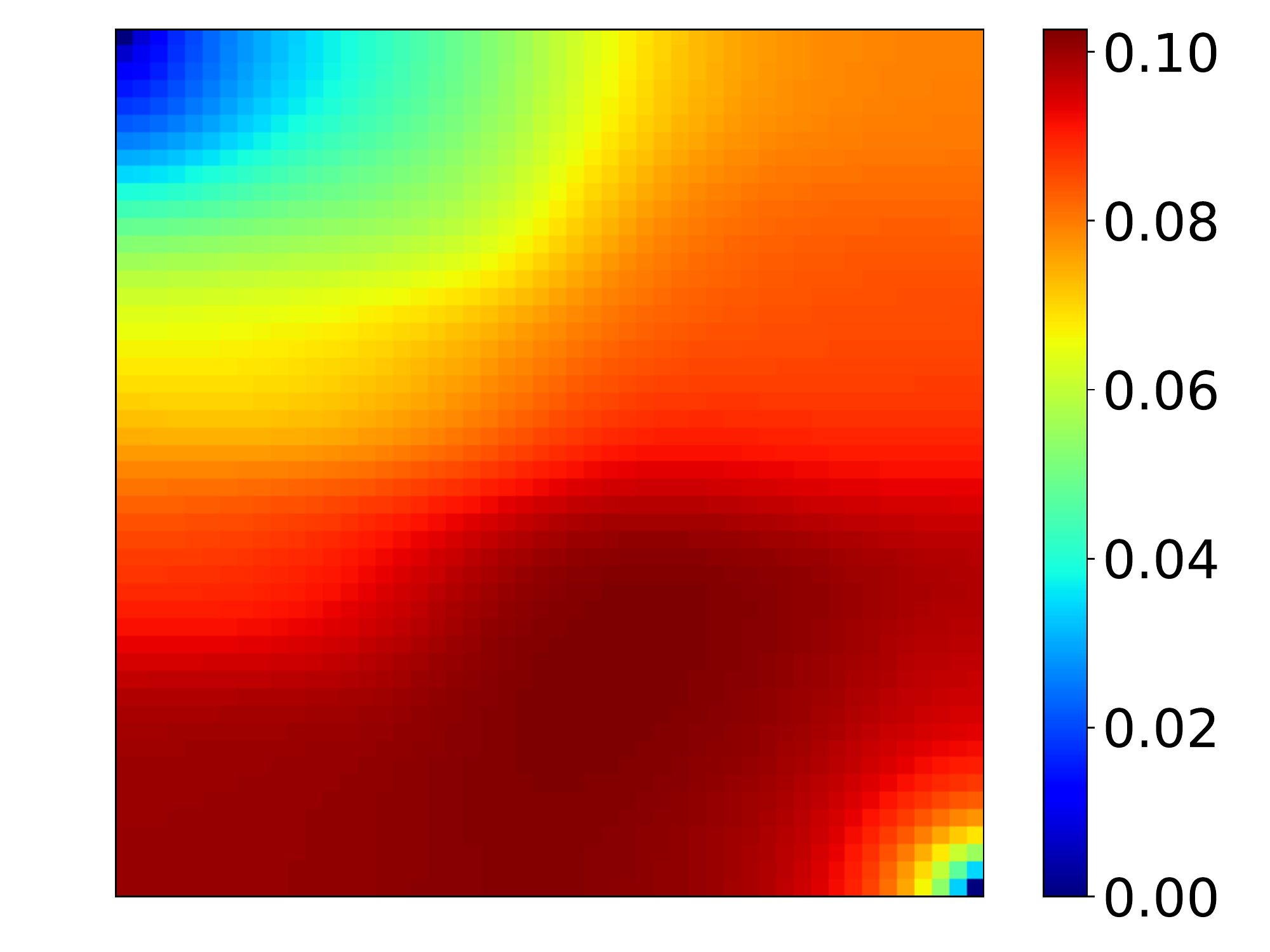}
        \caption{ }
        \label{fig:test_case_2_pressure_sigma}
    \end{subfigure}

    \caption{Variance of $\log(k)$ (a) and variance of normalized pressure (b) computed from training data.}
    \label{fig:perm_press_std}
\end{figure}
According to the color maps of expected information gain for two scenarios of pressure measurements and for single additional well (shown in  Fig.~\ref{fig:test_case_2_ufun_nwell_1}), there are two peaks of the utility function located at the corners of the model domain opposite to the injection and production wells as demonstrated in Fig.~\ref{fig:test_case_2_ufun_nwell_1_nt_2}. In the second scenario when additional measurements of pressure are added, only one maximum is observed {--} at the lower left corner of the domain {--} as shown in Fig.~\ref{fig:test_case_2_ufun_nwell_1_nt_5}. This observation is in agreement with variance of perturbation of permeability and pressure as shown in Fig.~\ref{fig:test_case_2_logperm_sigma} and Fig.~\ref{fig:test_case_2_pressure_sigma} respectively. In the case of a single measurement of pressure, the variance of permeability perturbation determines the shape of the utility function. In the second scenario, when extra pressure measurements are added the contribution of pressure variance becomes more significant. Therefore, the maximum of utility function is shifted towards the maximum of pressure variance. In other words, optimal parameters of experiment according to the Bayesian technique are in the proximity to point where the sensitivity of the model predictions to model parameters is the highest in terms of standard deviations. The latter observation is in agreement with common sense of experimental design.

\begin{figure}[H]
    \centering
    \begin{subfigure}[b]{0.19\textwidth}
        \includegraphics[width=0.99\linewidth]{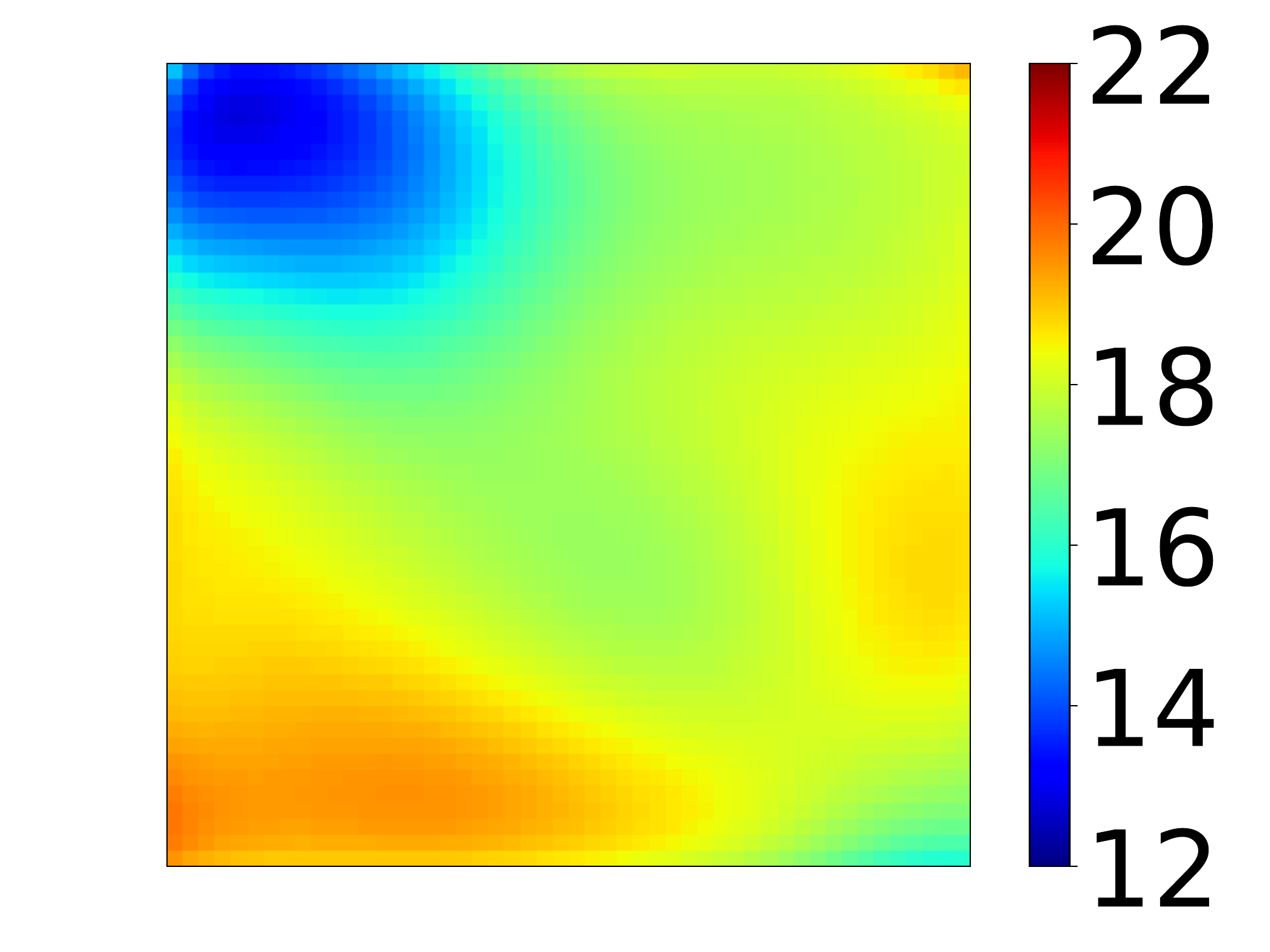}
        \caption{ }
        \label{fig:test_case_2_nwell_2_nt_2_kx0_0_ky0_0}
    \end{subfigure}
    \begin{subfigure}[b]{0.19\textwidth}
        \includegraphics[width=0.99\linewidth]{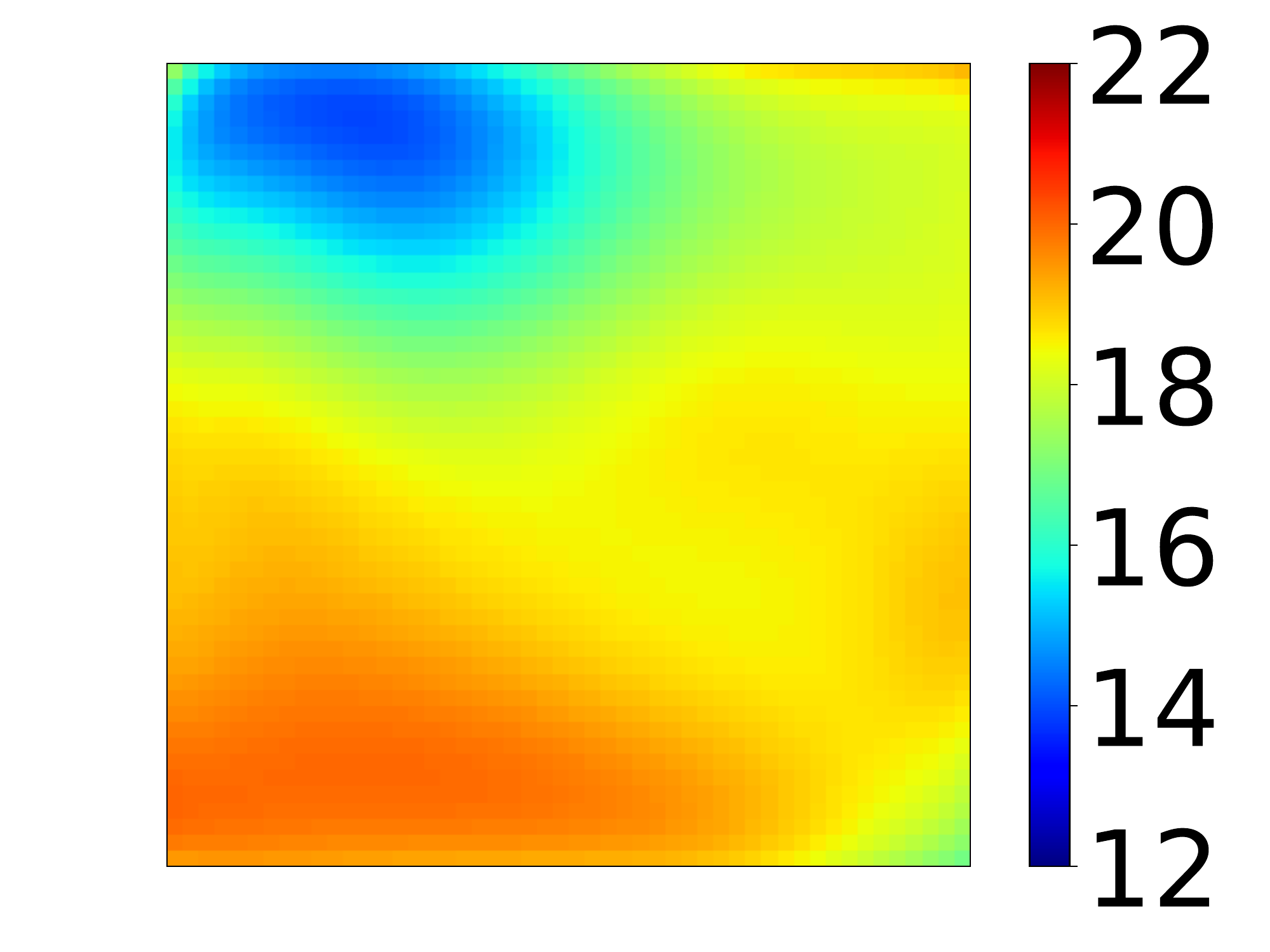}
        \caption{ }
        \label{fig:test_case_2_nwell_2_nt_2_kx0_0_ky0_1}
    \end{subfigure}
    \begin{subfigure}[b]{0.19\textwidth}
        \includegraphics[width=0.99\linewidth]{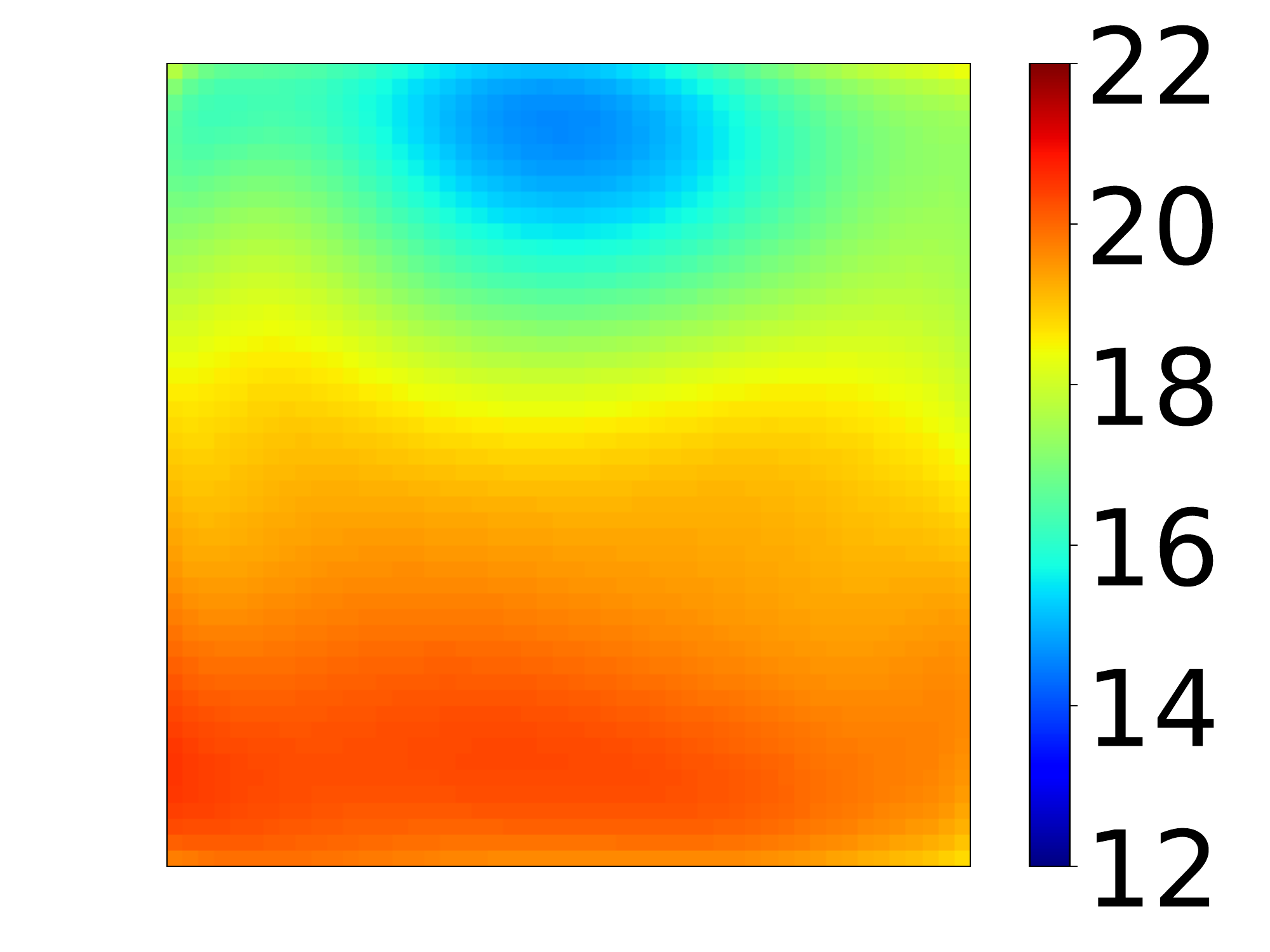}
        \caption{ }
        \label{fig:test_case_2_nwell_2_nt_2_kx0_0_ky0_2}
    \end{subfigure}
    \begin{subfigure}[b]{0.19\textwidth}
        \includegraphics[width=0.99\linewidth]{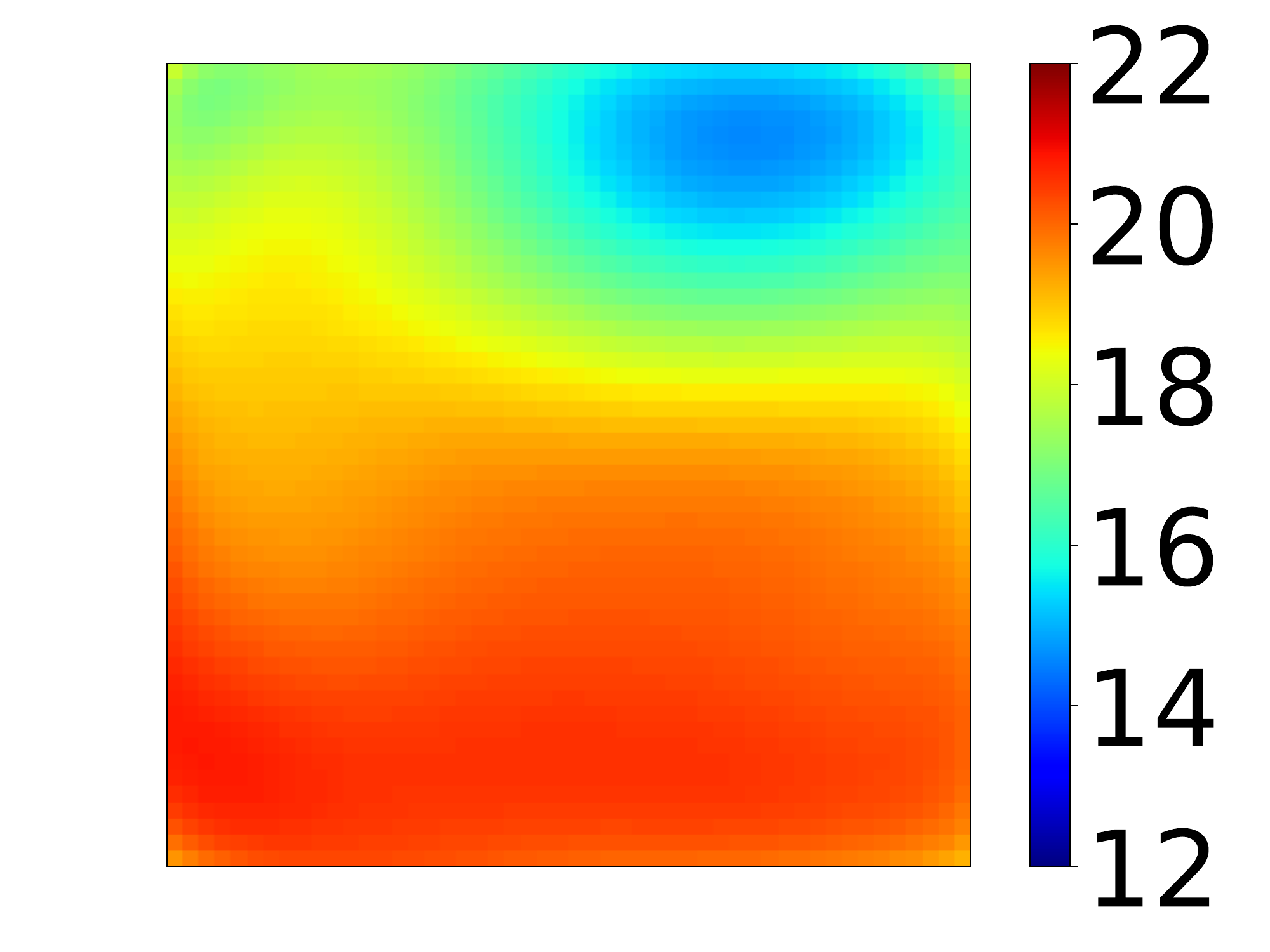}
        \caption{ }
        \label{fig:test_case_2_nwell_2_nt_2_kx0_0_ky0_3}
    \end{subfigure}
    \begin{subfigure}[b]{0.19\textwidth}
        \includegraphics[width=0.99\linewidth]{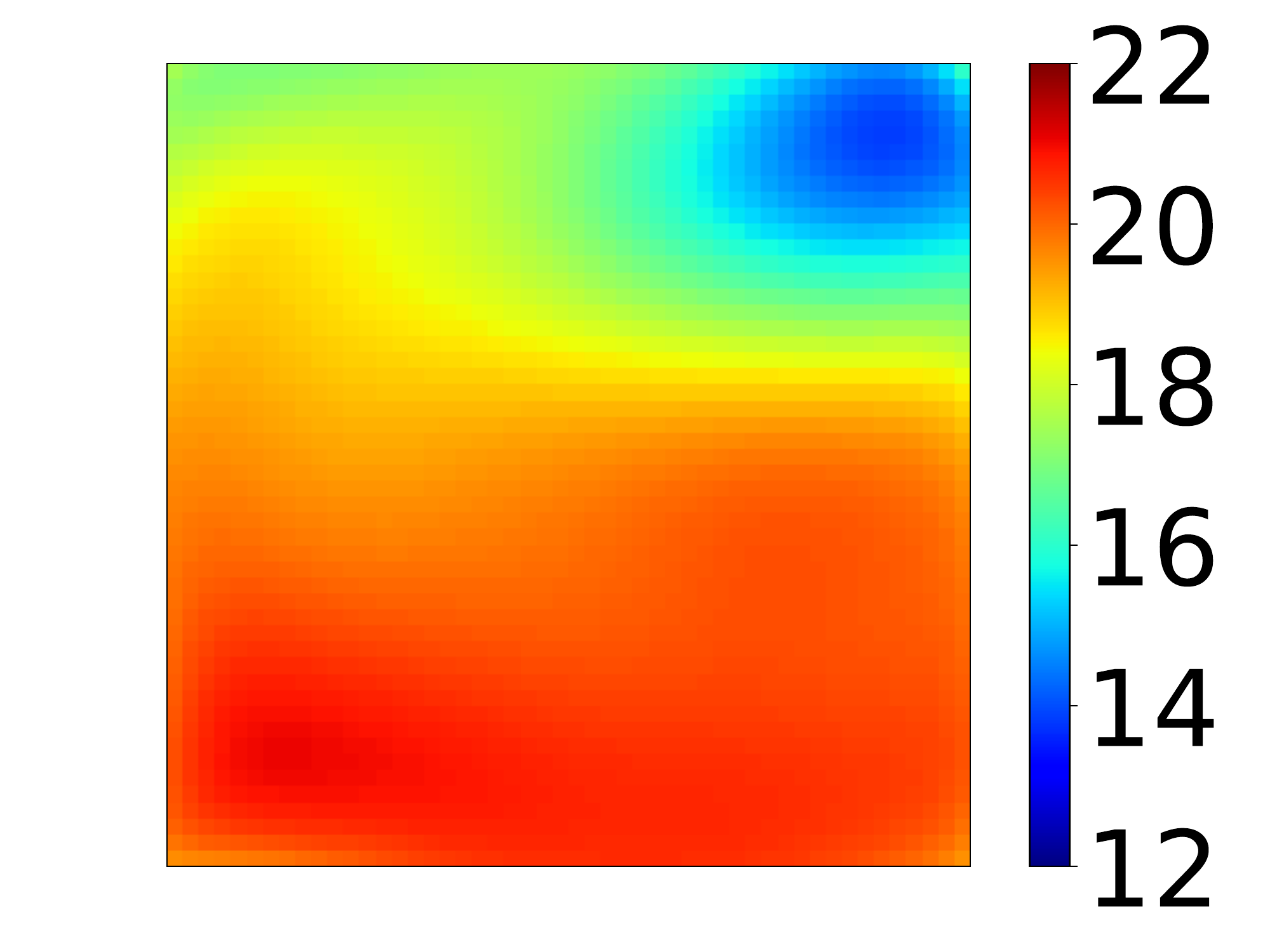}
        \caption{ }
        \label{fig:test_case_2_nwell_2_nt_2_kx0_0_ky0_4}
    \end{subfigure}

    \begin{subfigure}[b]{0.19\textwidth}
        \includegraphics[width=0.99\linewidth]{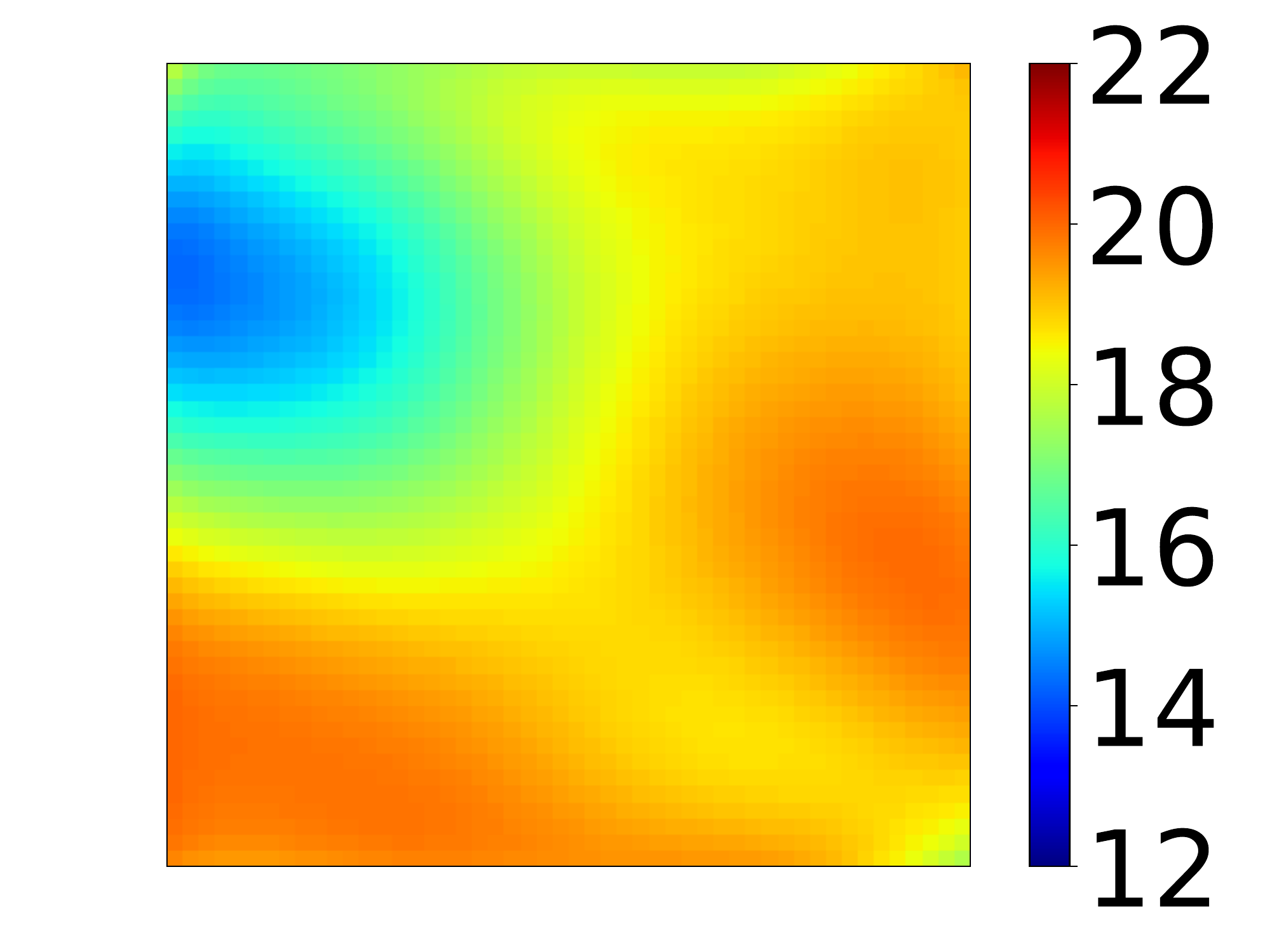}
        \caption{ }
        \label{fig:test_case_2_nwell_2_nt_2_kx0_1_ky0_0}
    \end{subfigure}
    \begin{subfigure}[b]{0.19\textwidth}
        \includegraphics[width=0.99\linewidth]{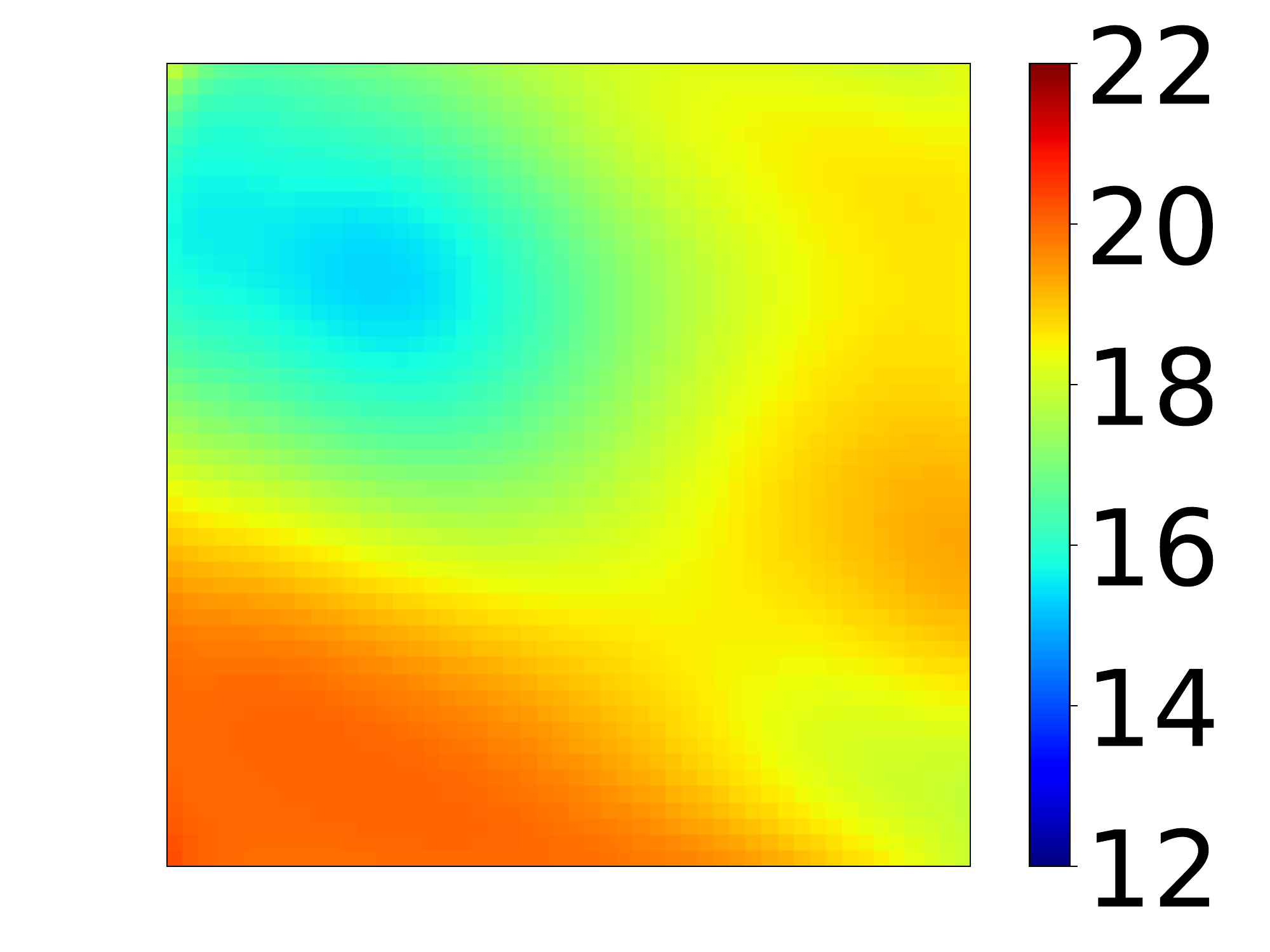}
        \caption{ }
        \label{fig:test_case_2_nwell_2_nt_2_kx0_1_ky0_1}
    \end{subfigure}
    \begin{subfigure}[b]{0.19\textwidth}
        \includegraphics[width=0.99\linewidth]{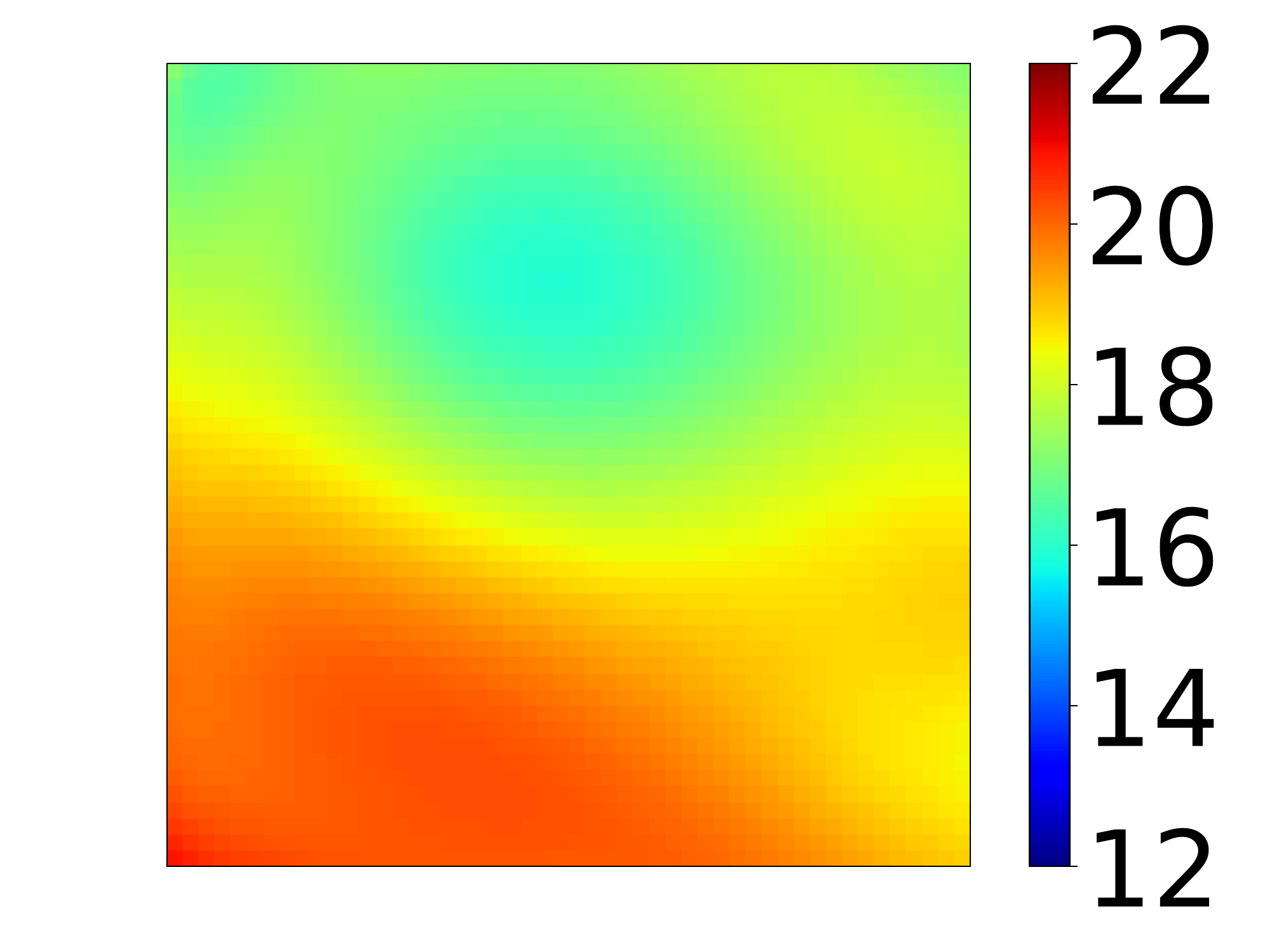}
        \caption{ }
        \label{fig:test_case_2_nwell_2_nt_2_kx0_1_ky0_2}
    \end{subfigure}
    \begin{subfigure}[b]{0.19\textwidth}
        \includegraphics[width=0.99\linewidth]{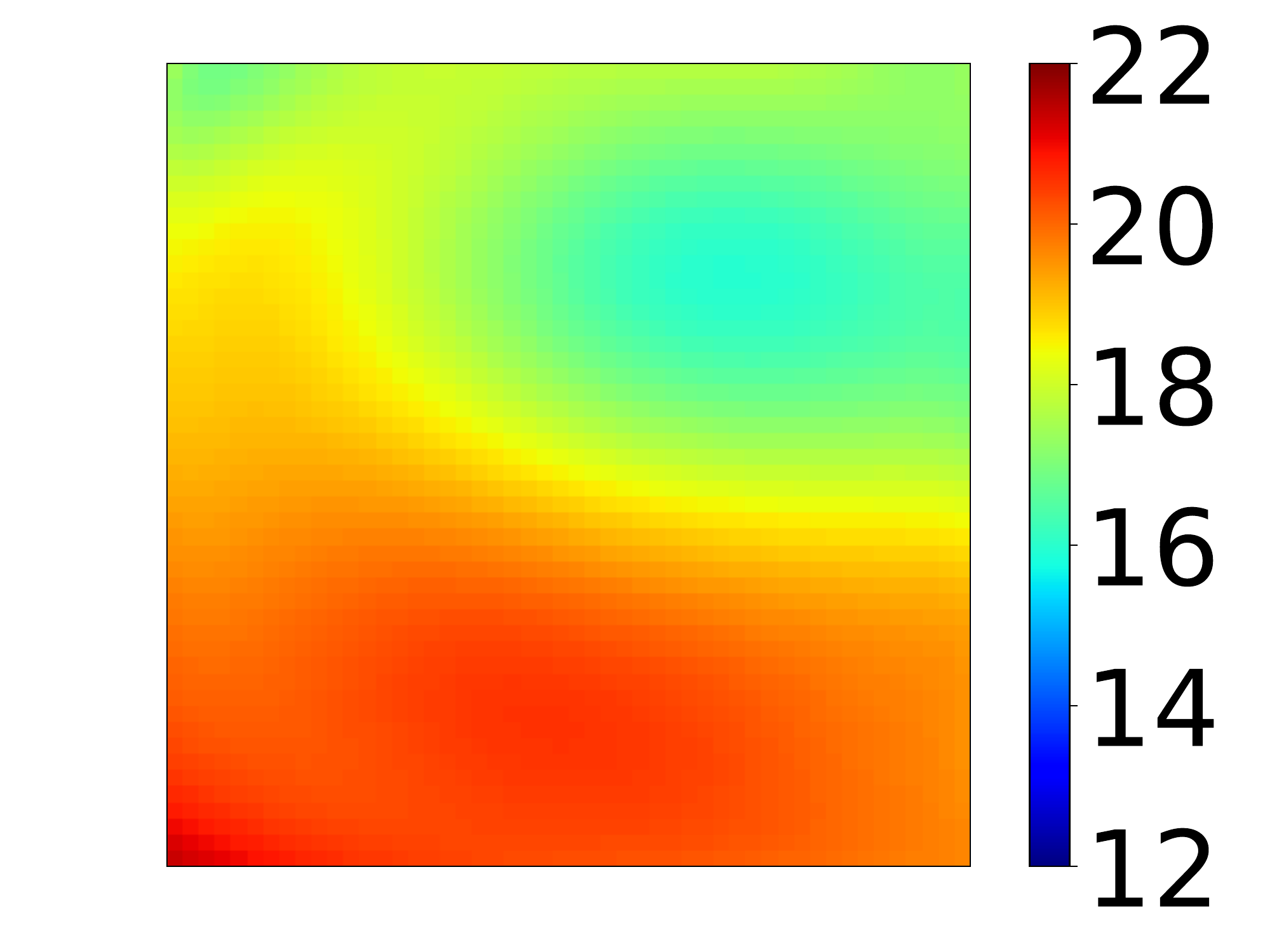}
        \caption{ }
        \label{fig:test_case_2_nwell_2_nt_2_kx0_1_ky0_3}
    \end{subfigure}
    \begin{subfigure}[b]{0.19\textwidth}
        \includegraphics[width=0.99\linewidth]{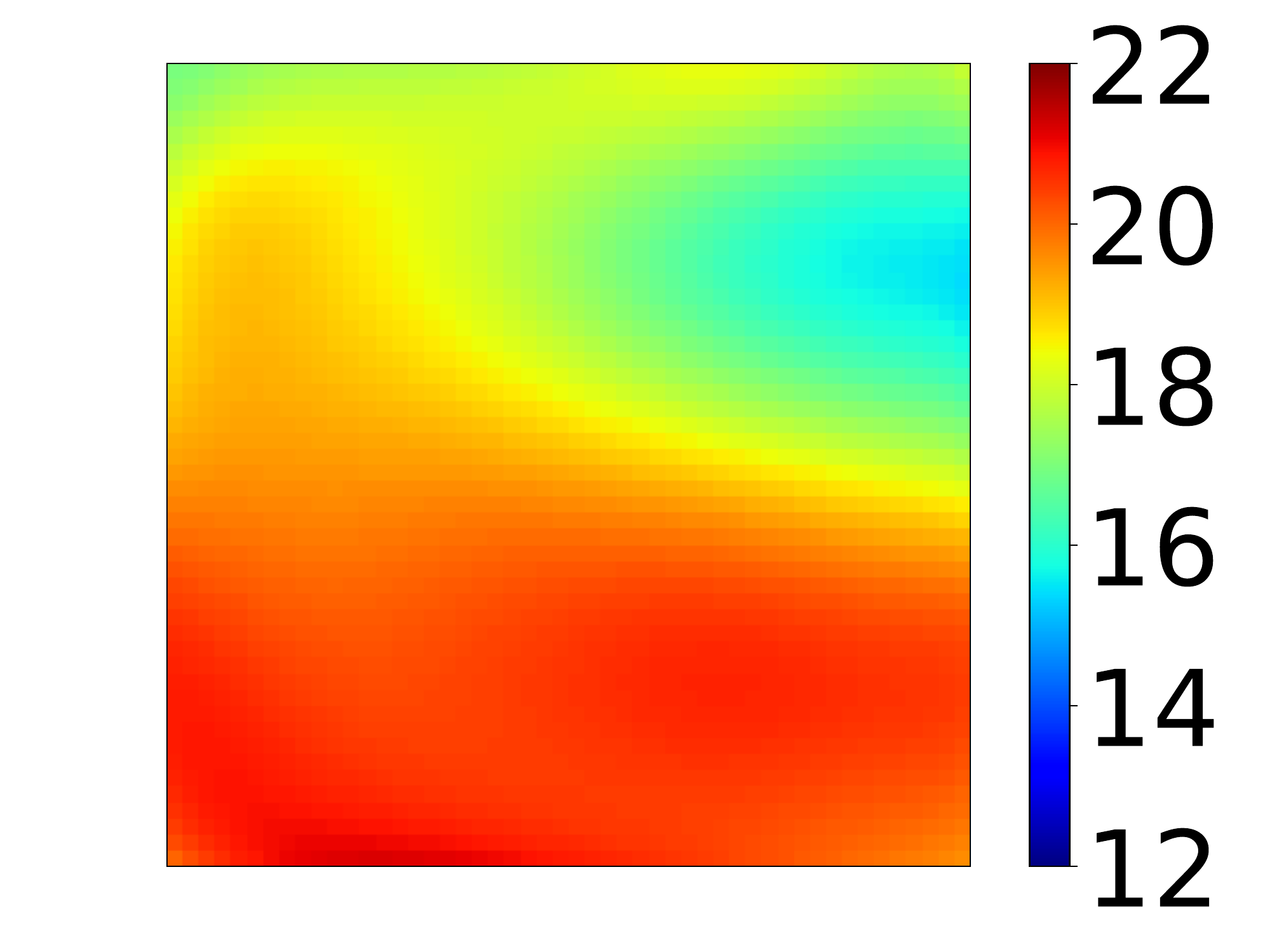}
        \caption{ }
        \label{fig:test_case_2_nwell_2_nt_2_kx0_1_ky0_4}
    \end{subfigure}

    \begin{subfigure}[b]{0.19\textwidth}
        \includegraphics[width=0.99\linewidth]{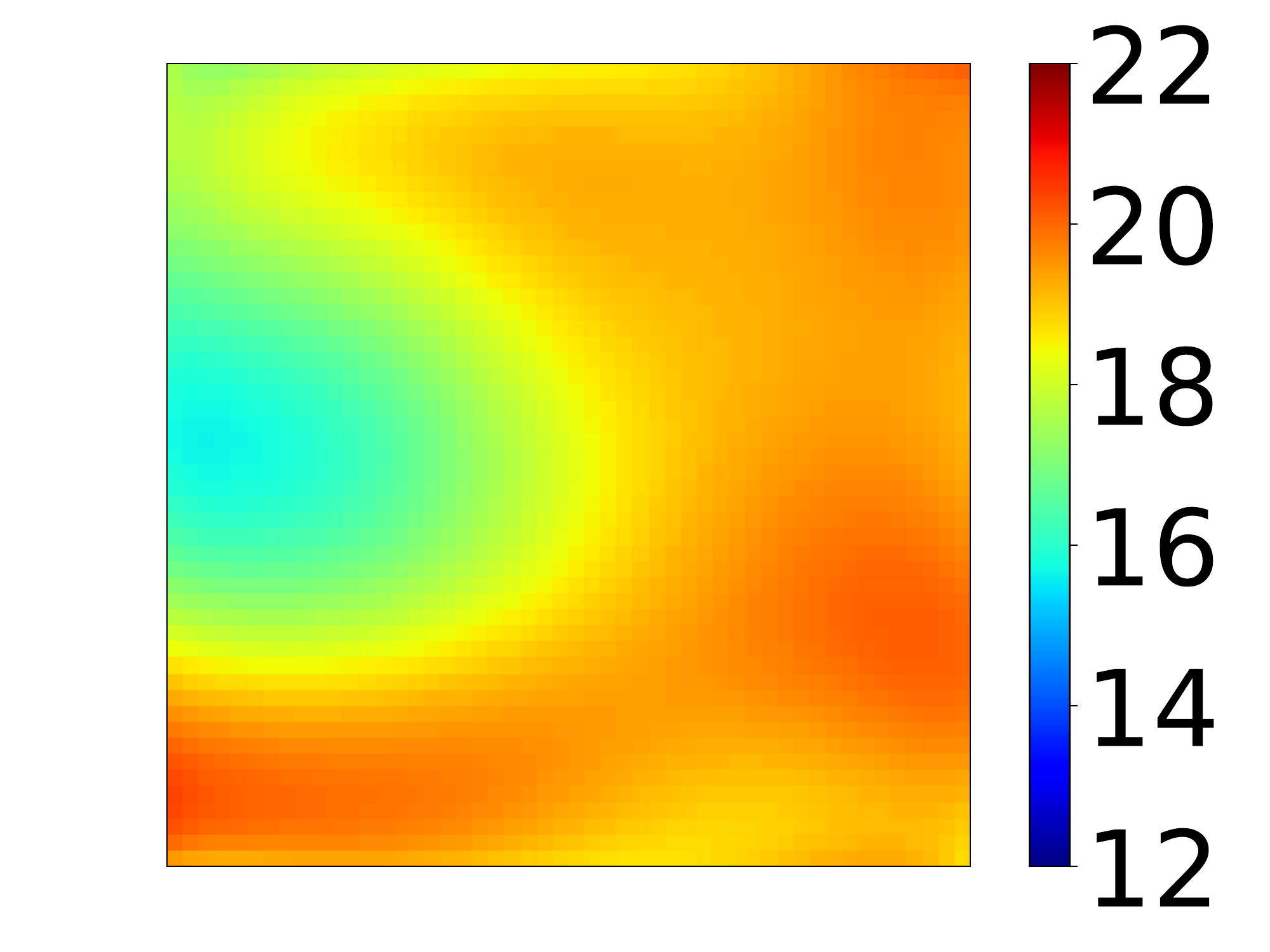}
        \caption{ }
        \label{fig:test_case_2_nwell_2_nt_2_kx0_2_ky0_0}
    \end{subfigure}
    \begin{subfigure}[b]{0.19\textwidth}
        \includegraphics[width=0.99\linewidth]{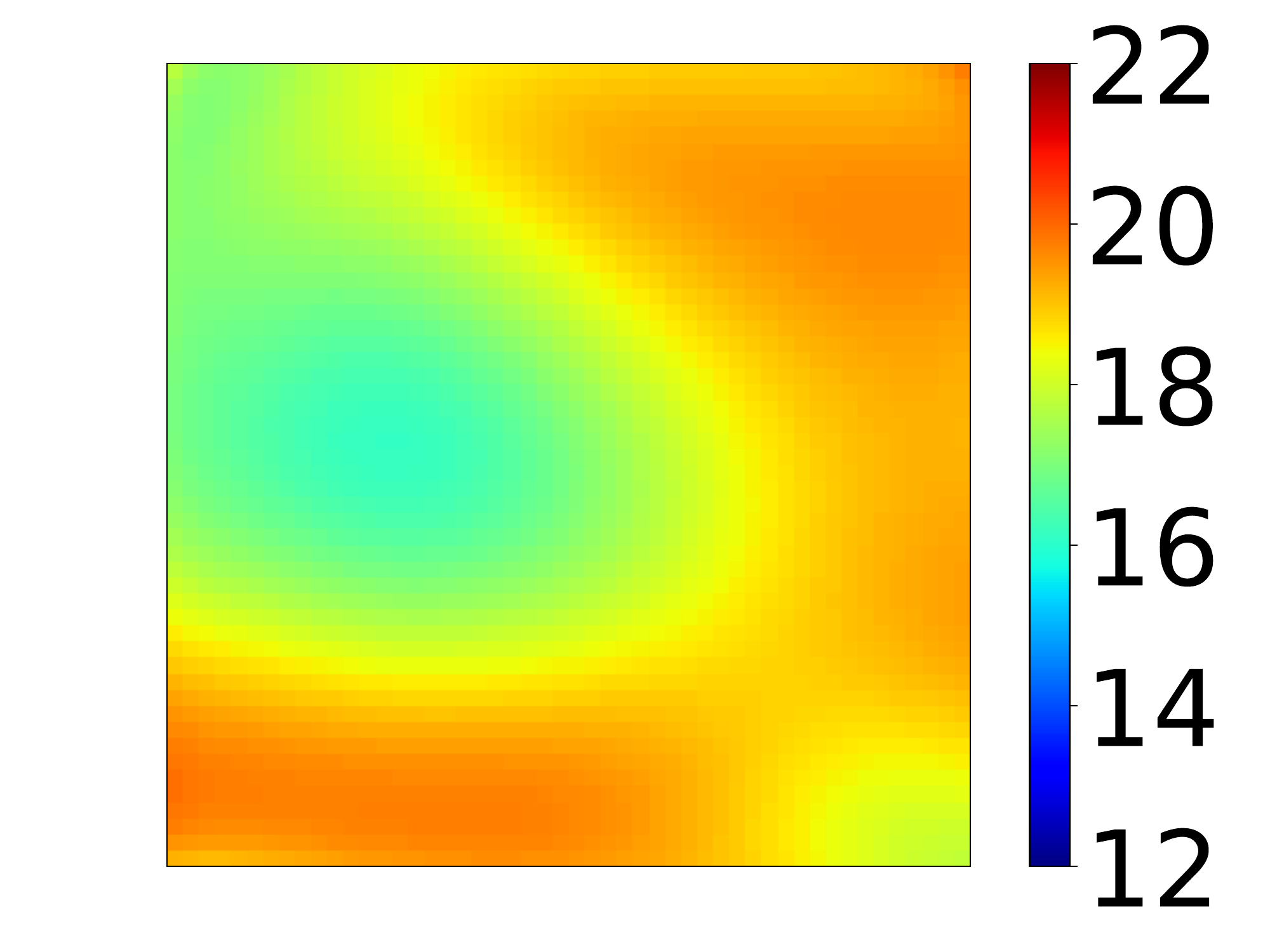}
        \caption{ }
        \label{fig:test_case_2_nwell_2_nt_2_kx0_2_ky0_1}
    \end{subfigure}
    \begin{subfigure}[b]{0.19\textwidth}
        \includegraphics[width=0.99\linewidth]{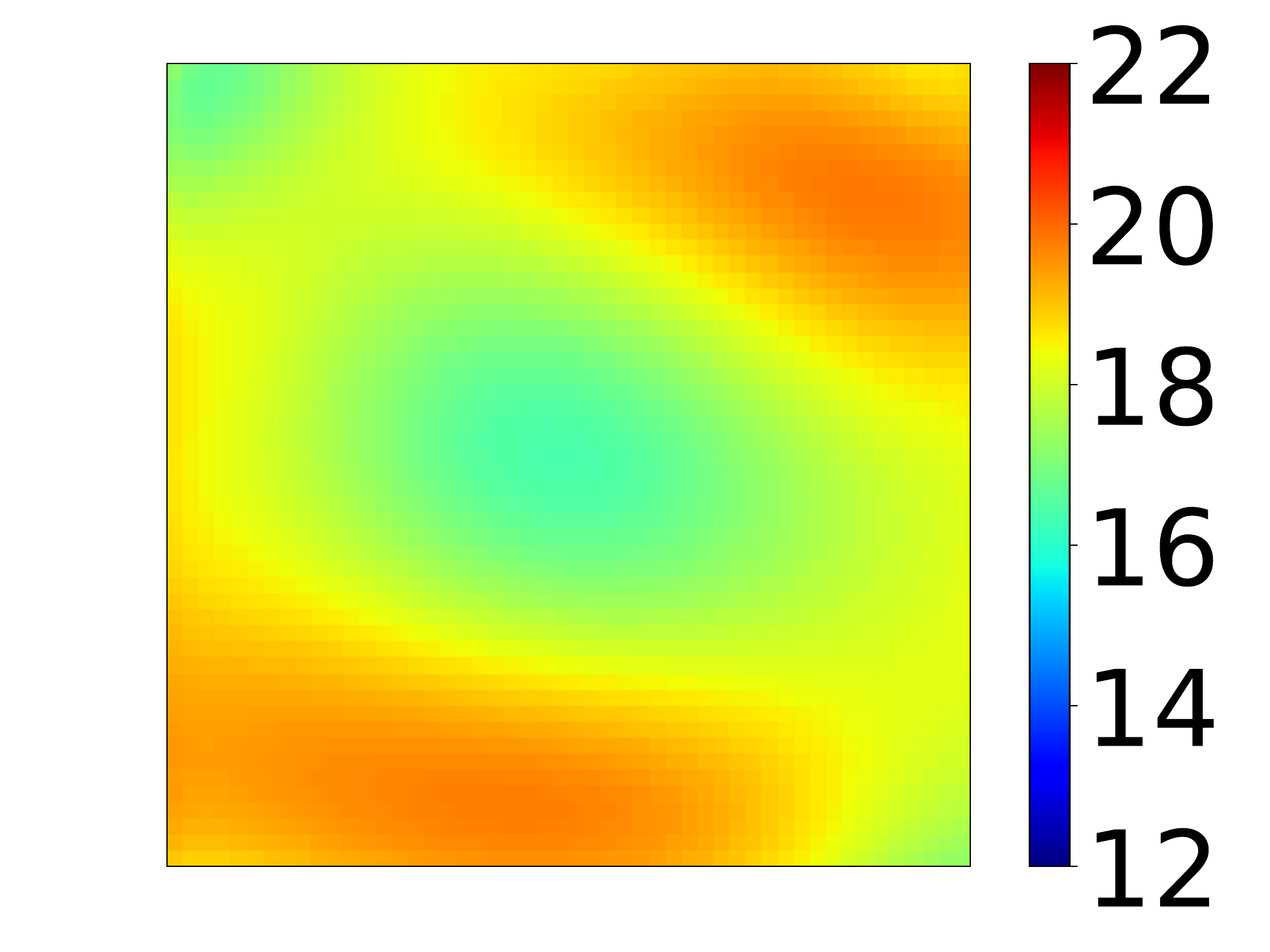}
        \caption{ }
        \label{fig:test_case_2_nwell_2_nt_2_kx0_2_ky0_2}
    \end{subfigure}
    \begin{subfigure}[b]{0.19\textwidth}
        \includegraphics[width=0.99\linewidth]{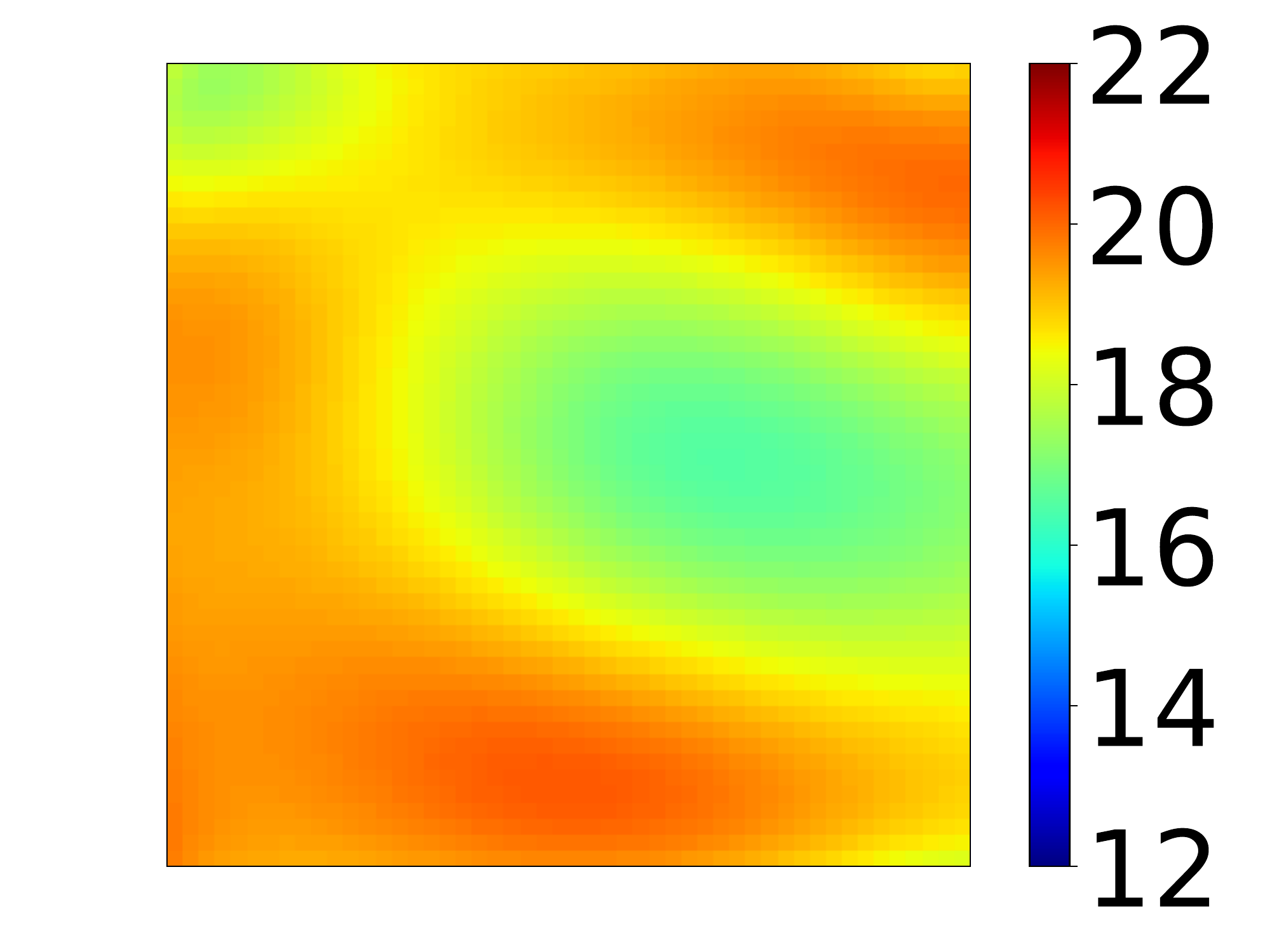}
        \caption{ }
        \label{fig:test_case_2_nwell_2_nt_2_kx0_2_ky0_3}
    \end{subfigure}
    \begin{subfigure}[b]{0.19\textwidth}
        \includegraphics[width=0.99\linewidth]{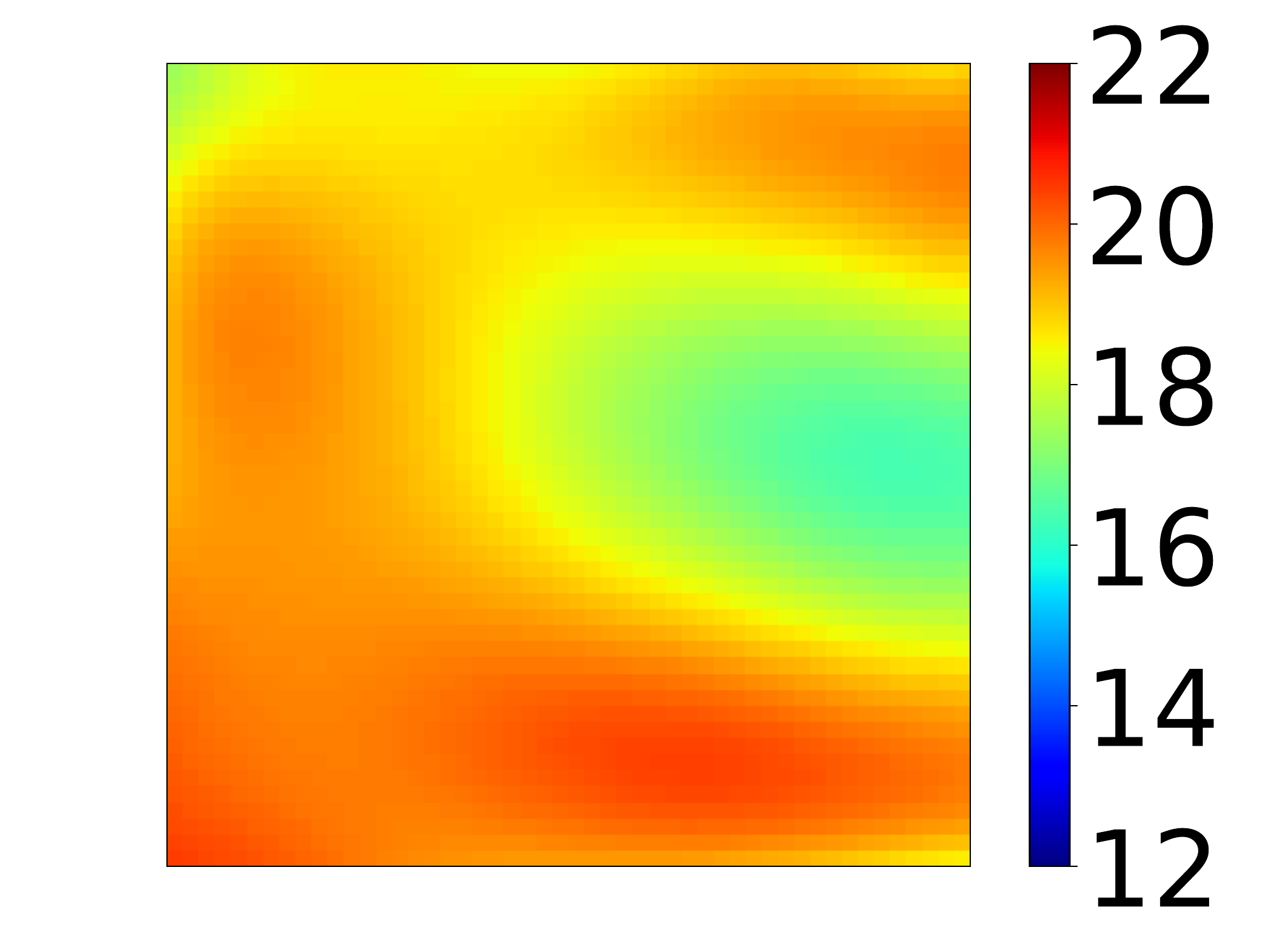}
        \caption{ }
        \label{fig:test_case_2_nwell_2_nt_2_kx0_2_ky0_4}
    \end{subfigure}

    \begin{subfigure}[b]{0.19\textwidth}
        \includegraphics[width=0.99\linewidth]{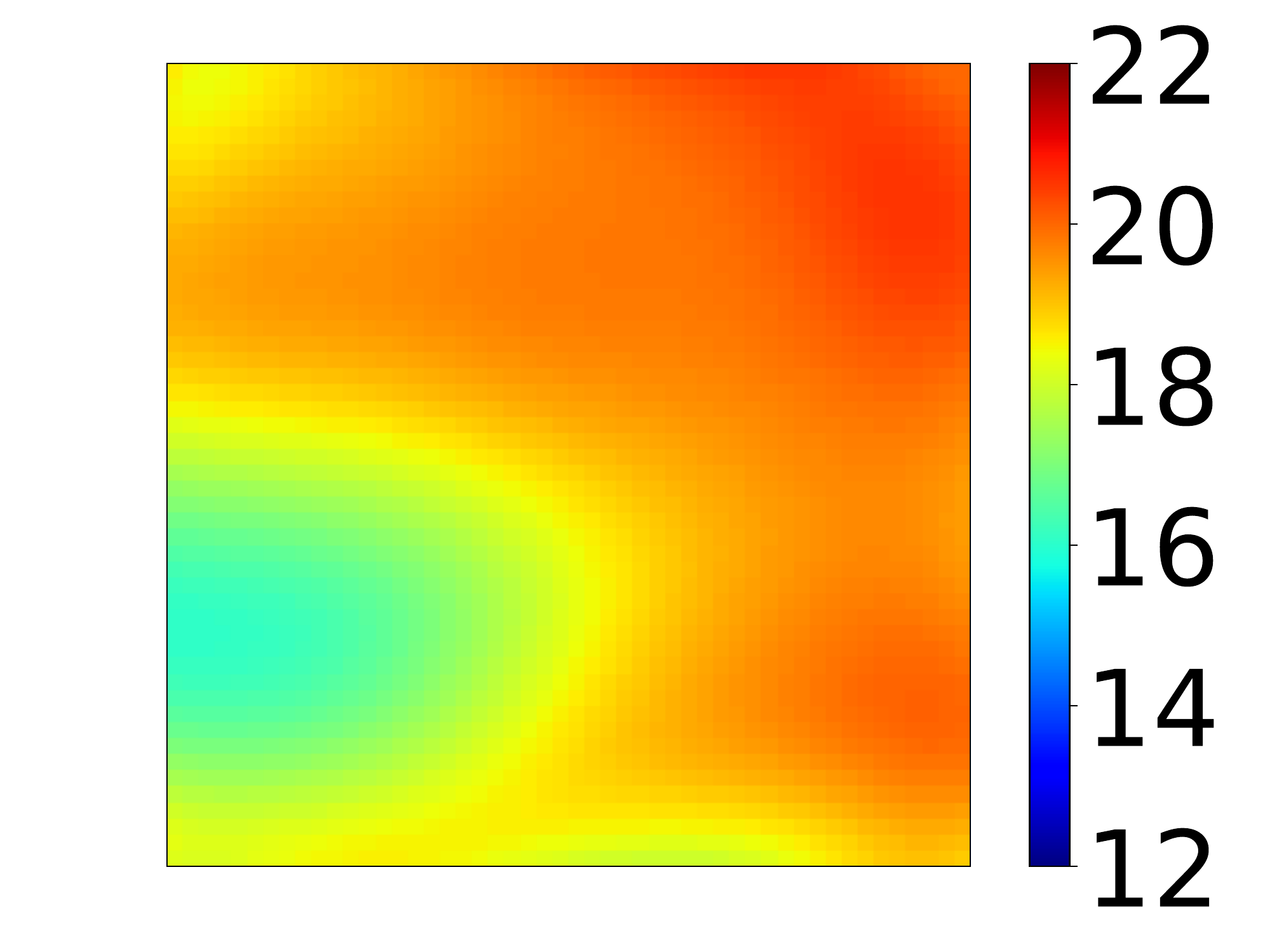}
        \caption{ }
        \label{fig:test_case_2_nwell_2_nt_2_kx0_3_ky0_0}
    \end{subfigure}
    \begin{subfigure}[b]{0.19\textwidth}
        \includegraphics[width=0.99\linewidth]{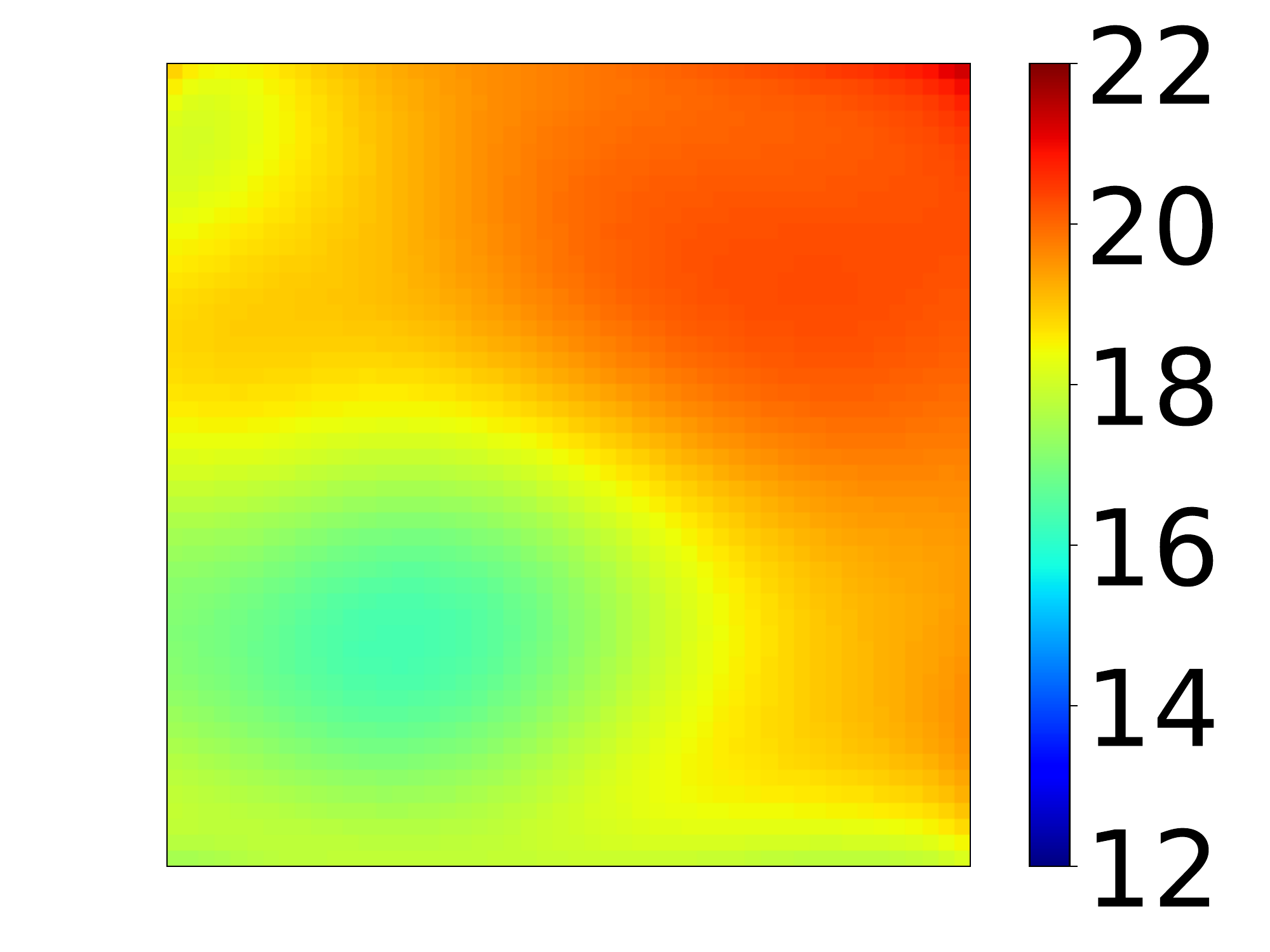}
        \caption{ }
        \label{fig:test_case_2_nwell_2_nt_2_kx0_3_ky0_1}
    \end{subfigure}
    \begin{subfigure}[b]{0.19\textwidth}
        \includegraphics[width=0.99\linewidth]{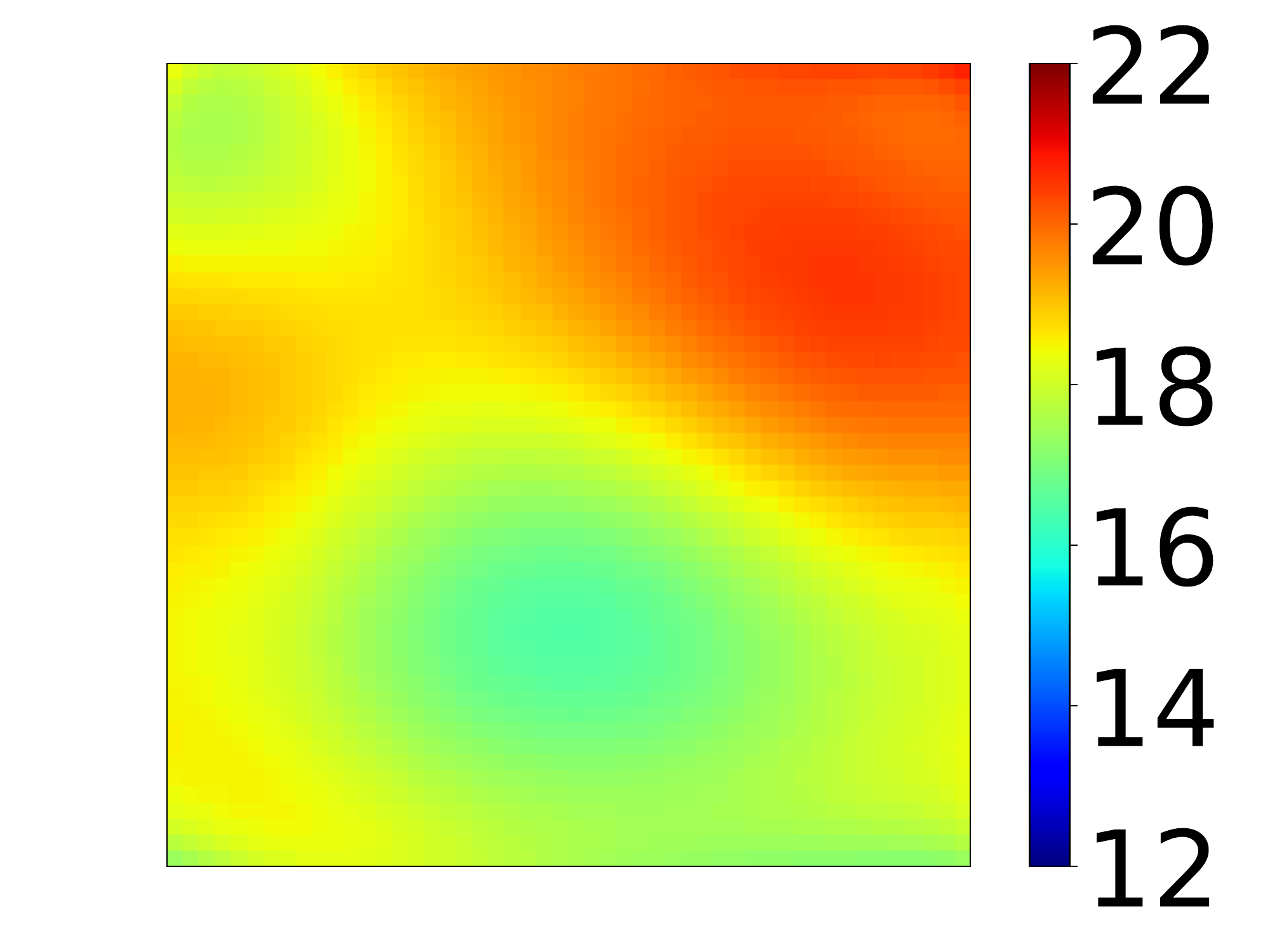}
        \caption{ }
        \label{fig:test_case_2_nwell_2_nt_2_kx0_3_ky0_2}
    \end{subfigure}
    \begin{subfigure}[b]{0.19\textwidth}
        \includegraphics[width=0.99\linewidth]{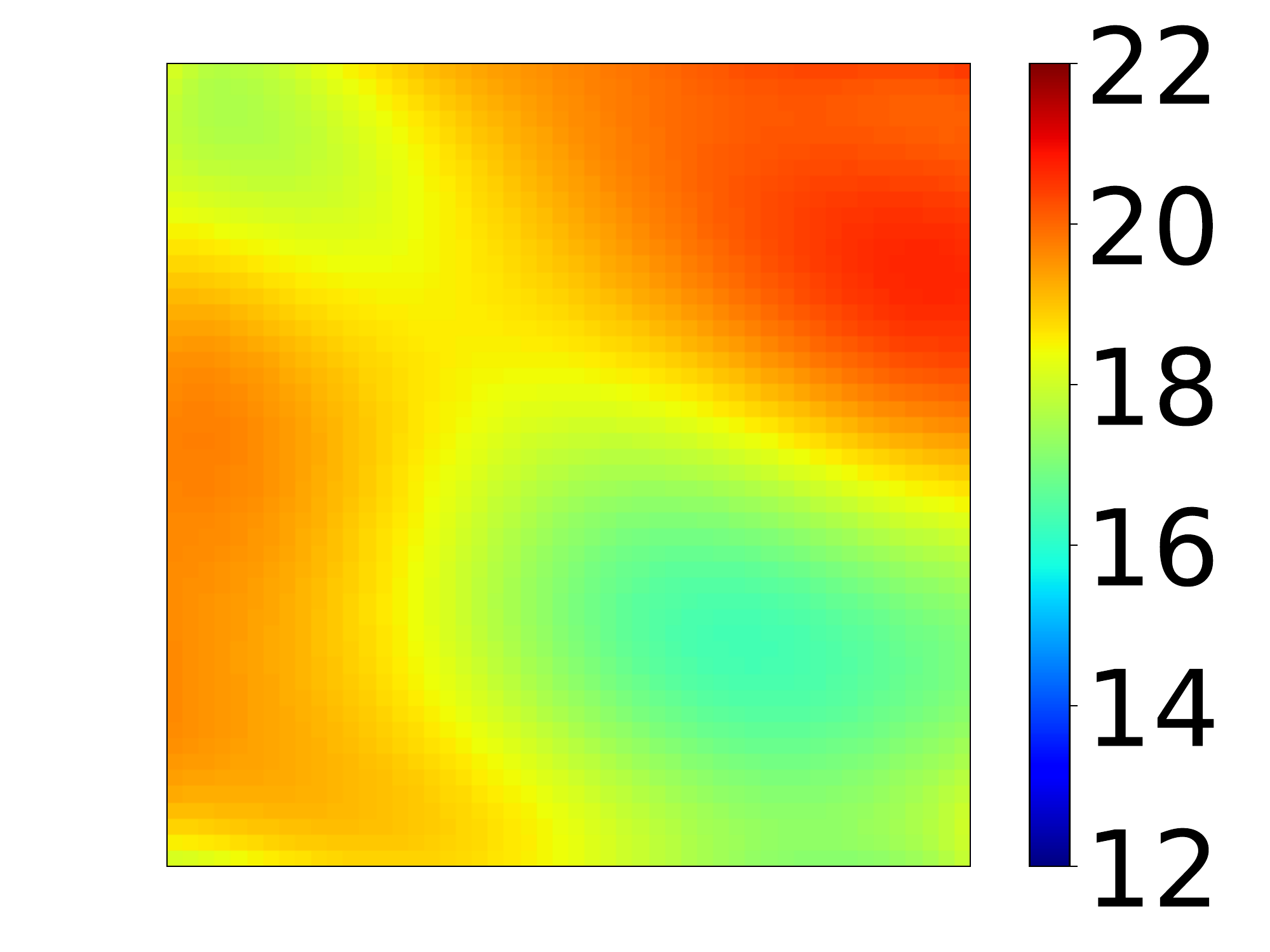}
        \caption{ }
        \label{fig:test_case_2_nwell_2_nt_2_kx0_3_ky0_3}
    \end{subfigure}
    \begin{subfigure}[b]{0.19\textwidth}
        \includegraphics[width=0.99\linewidth]{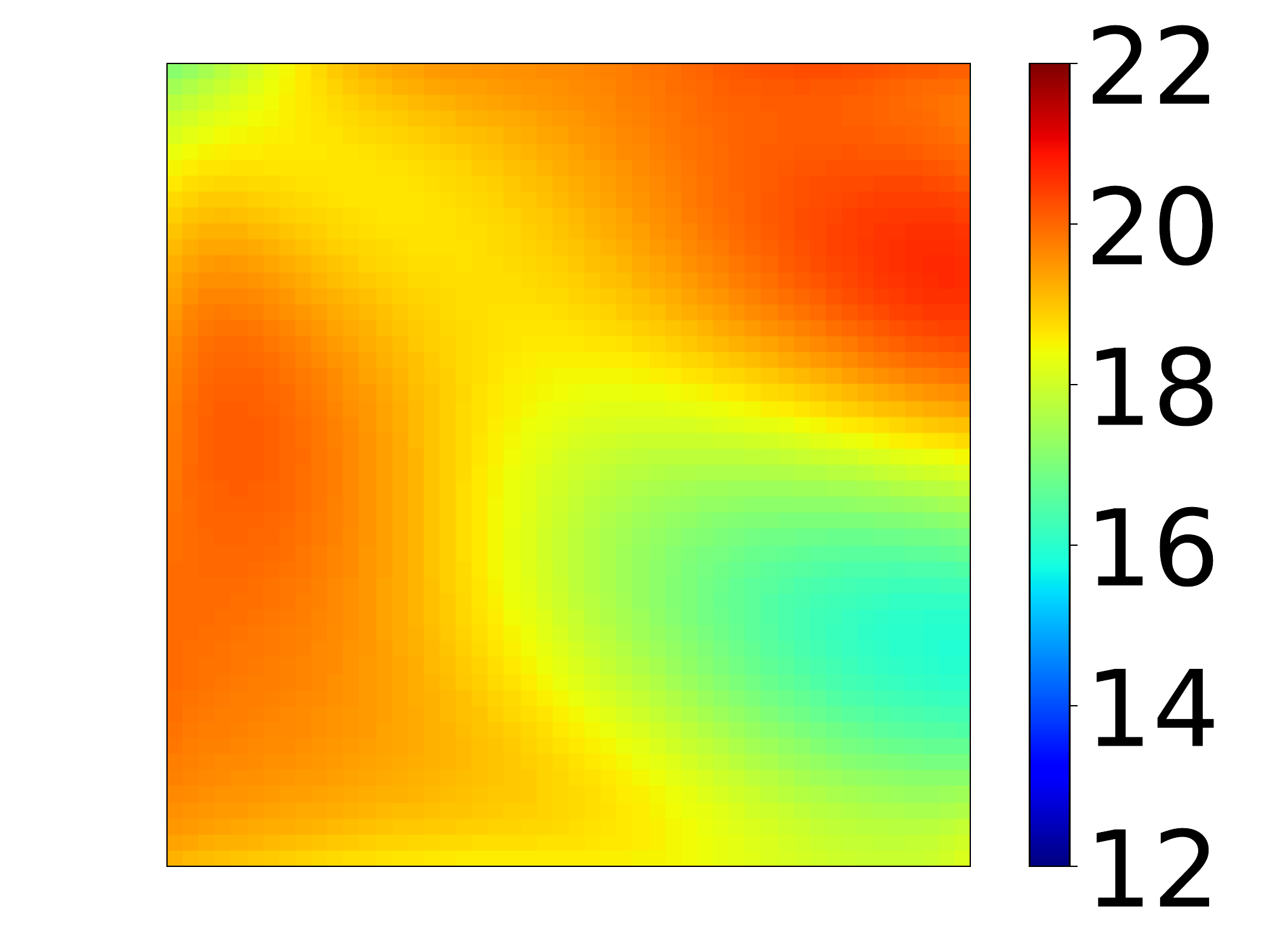}
        \caption{ }
        \label{fig:test_case_2_nwell_2_nt_2_kx0_3_ky0_4}
    \end{subfigure}

    \begin{subfigure}[b]{0.19\textwidth}
        \includegraphics[width=0.99\linewidth]{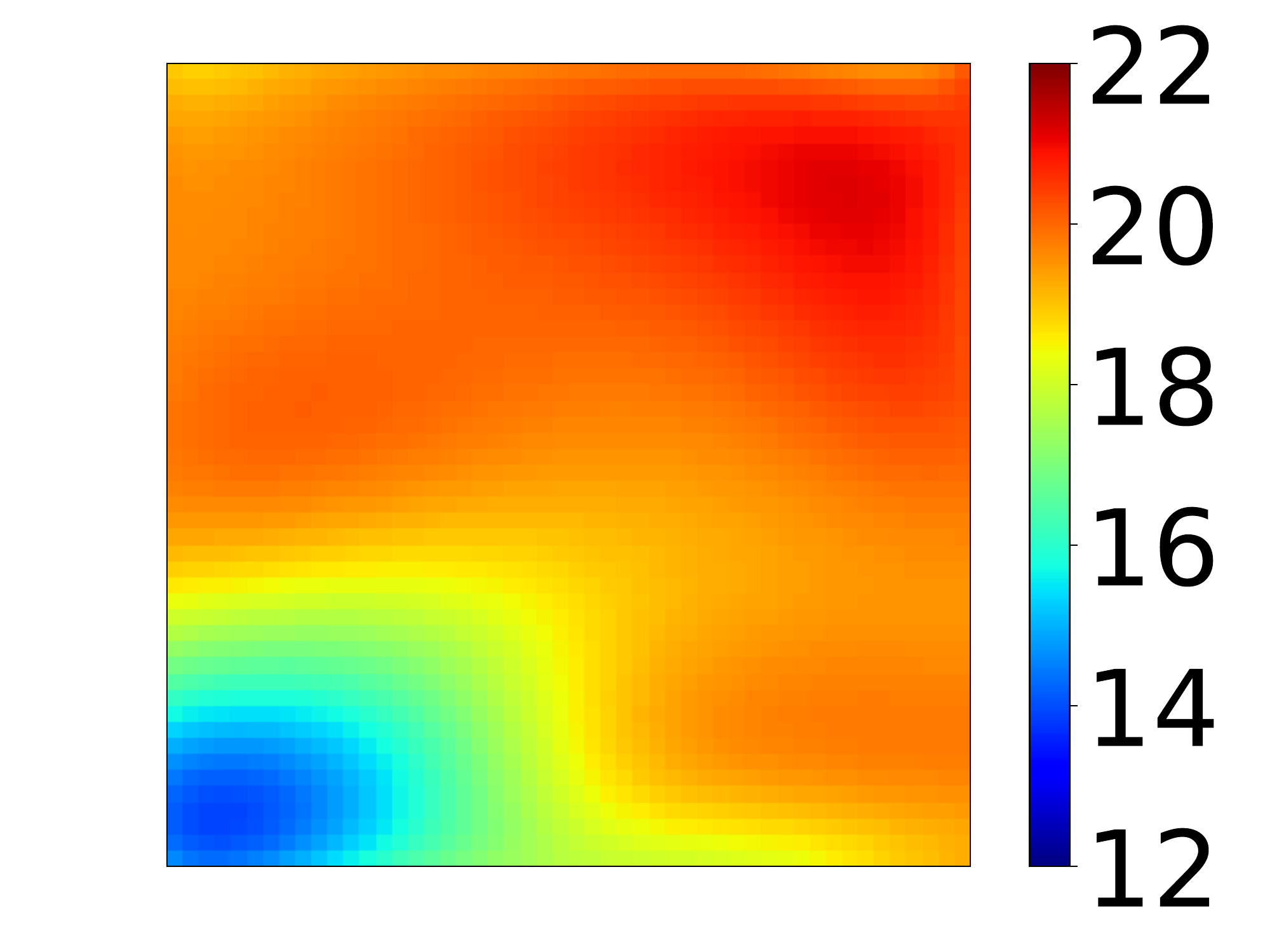}
        \caption{ }
        \label{fig:test_case_2_nwell_2_nt_2_kx0_4_ky0_0}
    \end{subfigure}
    \begin{subfigure}[b]{0.19\textwidth}
        \includegraphics[width=0.99\linewidth]{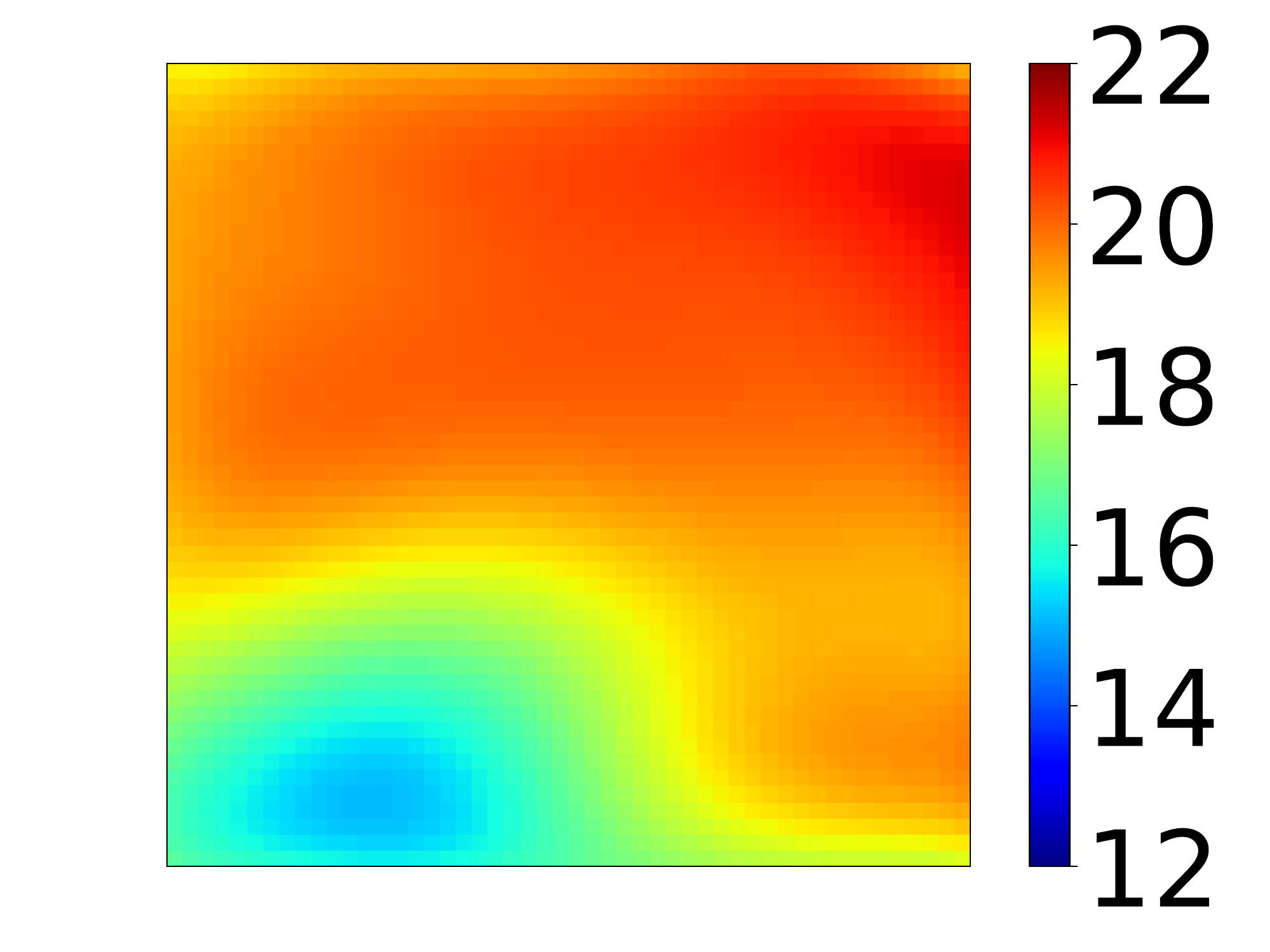}
        \caption{ }
        \label{fig:test_case_2_nwell_2_nt_2_kx0_4_ky0_1}
    \end{subfigure}
    \begin{subfigure}[b]{0.19\textwidth}
        \includegraphics[width=0.99\linewidth]{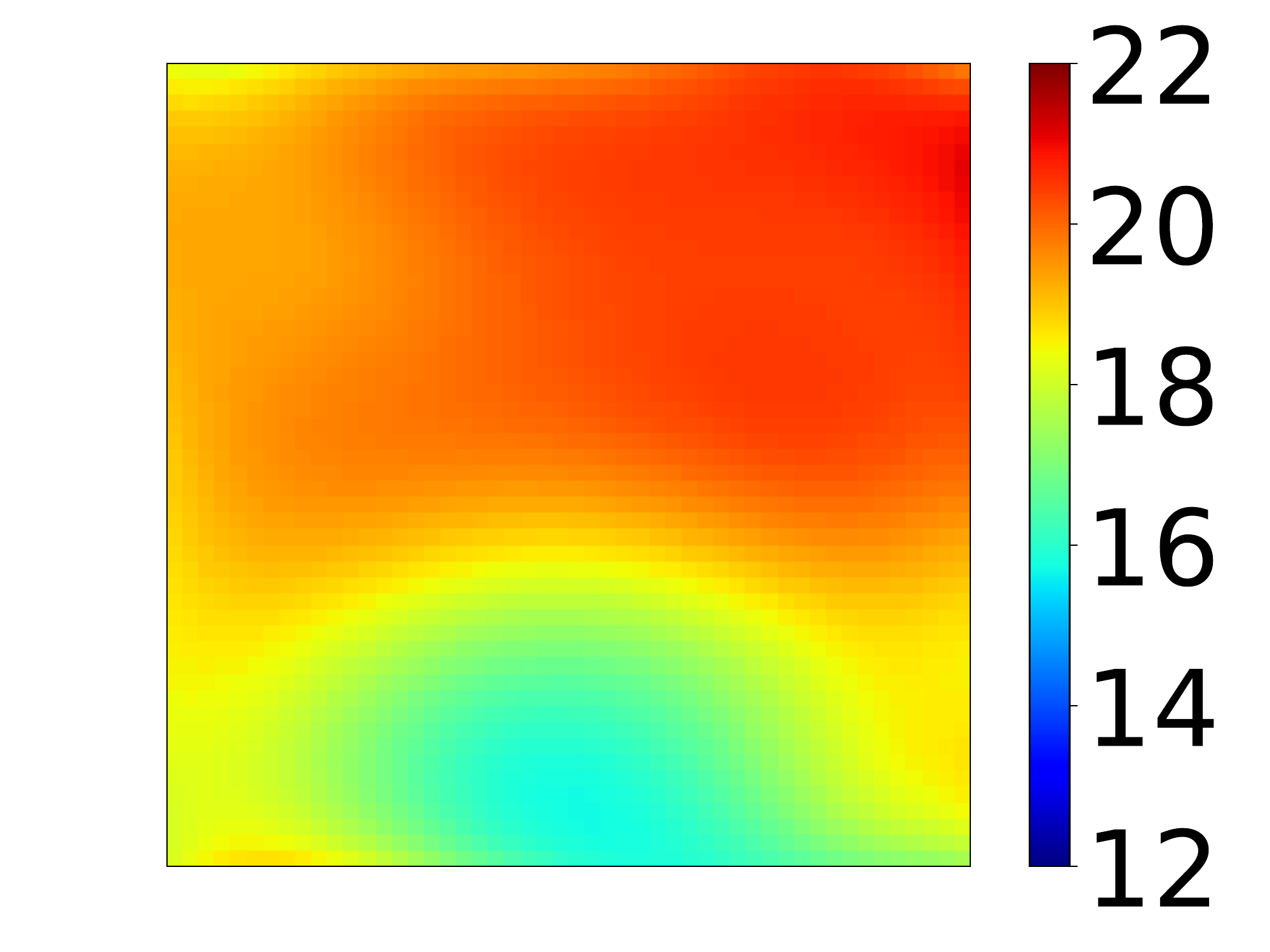}
        \caption{ }
        \label{fig:test_case_2_nwell_2_nt_2_kx0_4_ky0_2}
    \end{subfigure}
    \begin{subfigure}[b]{0.19\textwidth}
        \includegraphics[width=0.99\linewidth]{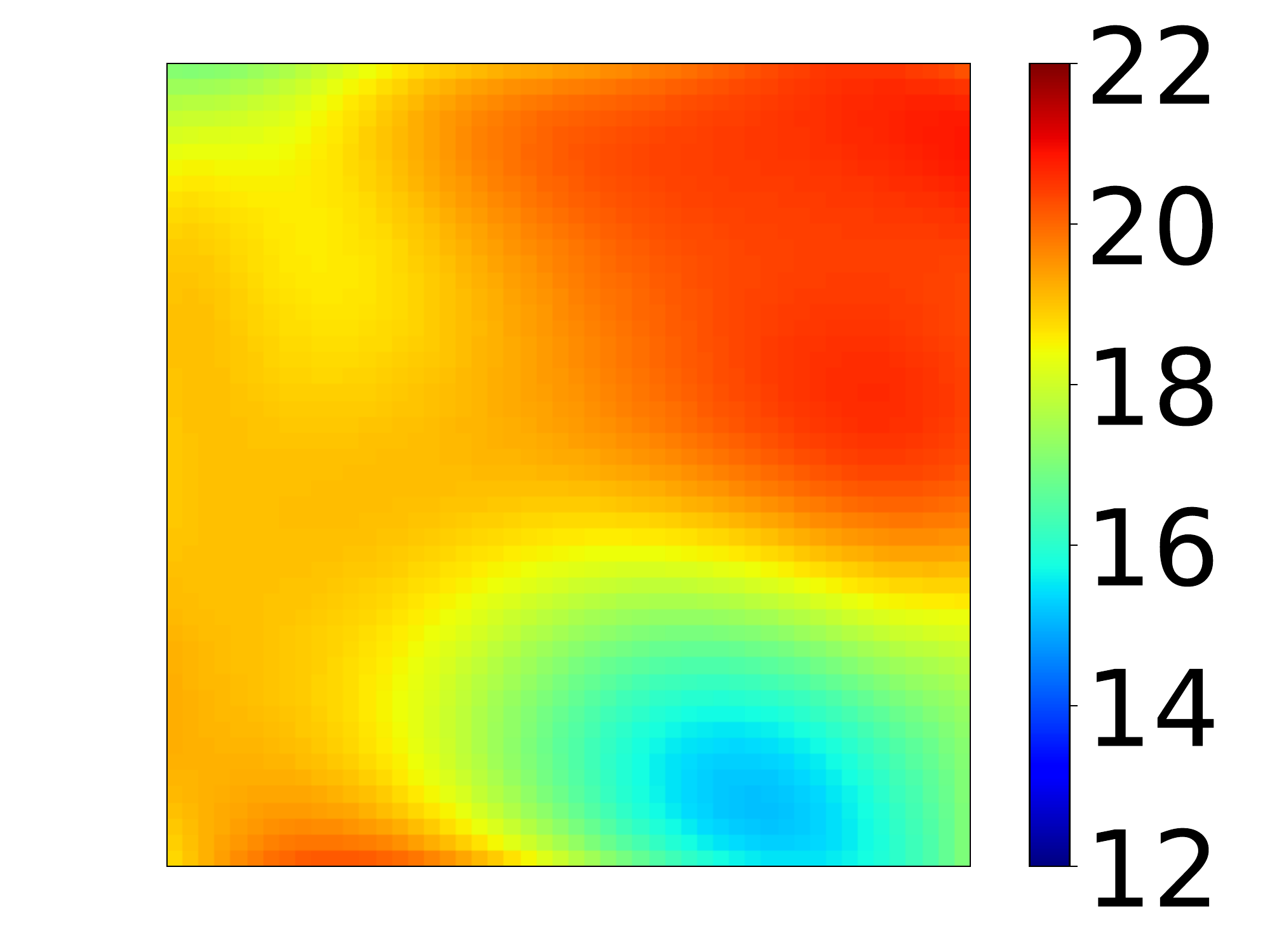}
        \caption{ }
        \label{fig:test_case_2_nwell_2_nt_2_kx0_4_ky0_3}
    \end{subfigure}
    \begin{subfigure}[b]{0.19\textwidth}
        \includegraphics[width=0.99\linewidth]{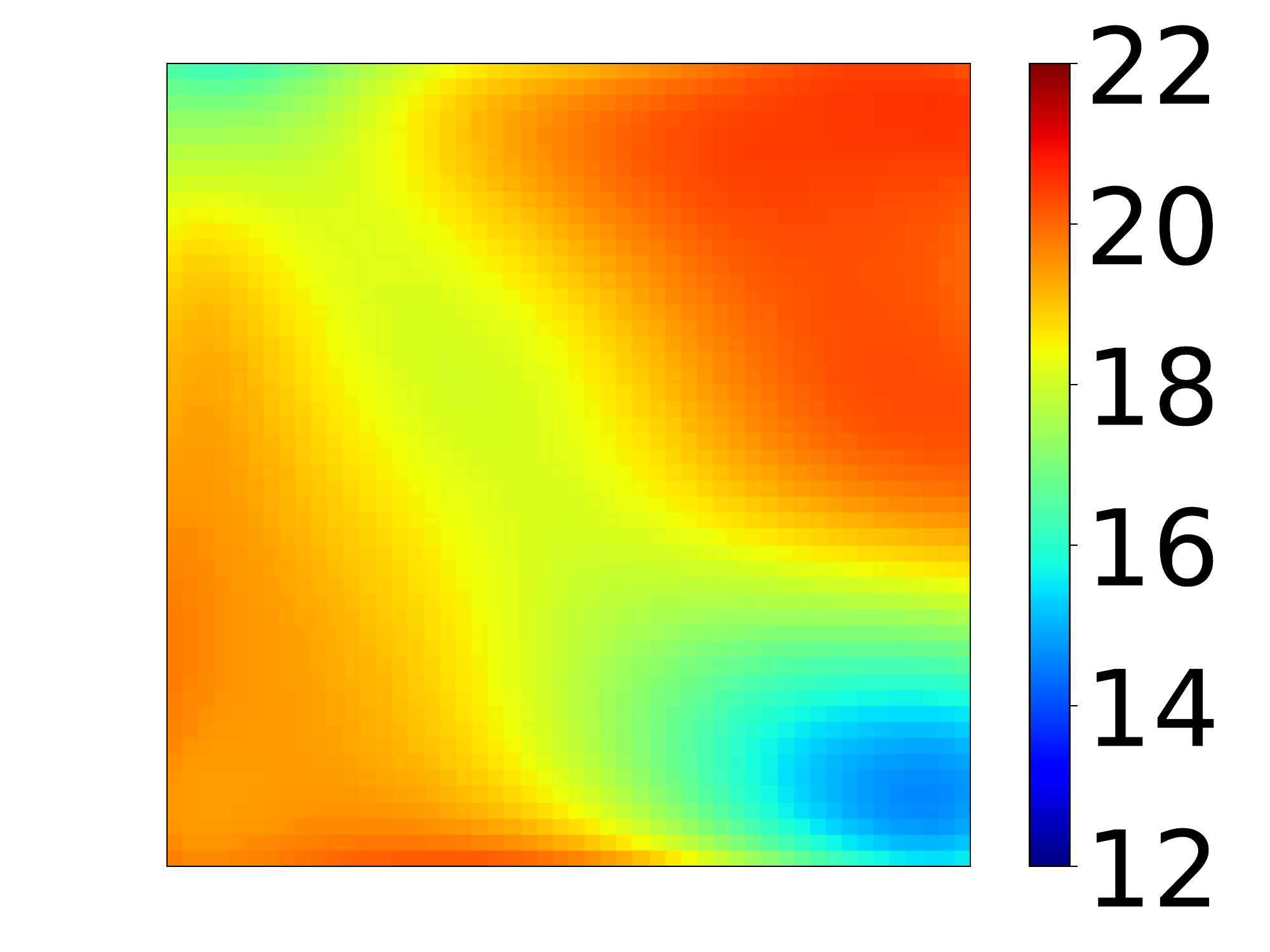}
        \caption{ }
        \label{fig:test_case_2_nwell_2_nt_2_kx0_4_ky0_4}
    \end{subfigure}
    \caption{Values of the expected information gain as a function the first well position if the location of the second well is fixed.
             Each of the figures (a) - (y) corresponds to different coordinates of the second well that corresponds to the minimum of the utility function (dark blue). Case of single pressure and permeability measurement is presented.}
    \label{fig:test_case_2_nwell_2_nt_2}
\end{figure}

\begin{figure}[H]
    \centering
    \begin{subfigure}[b]{0.19\textwidth}
        \includegraphics[width=0.99\linewidth]{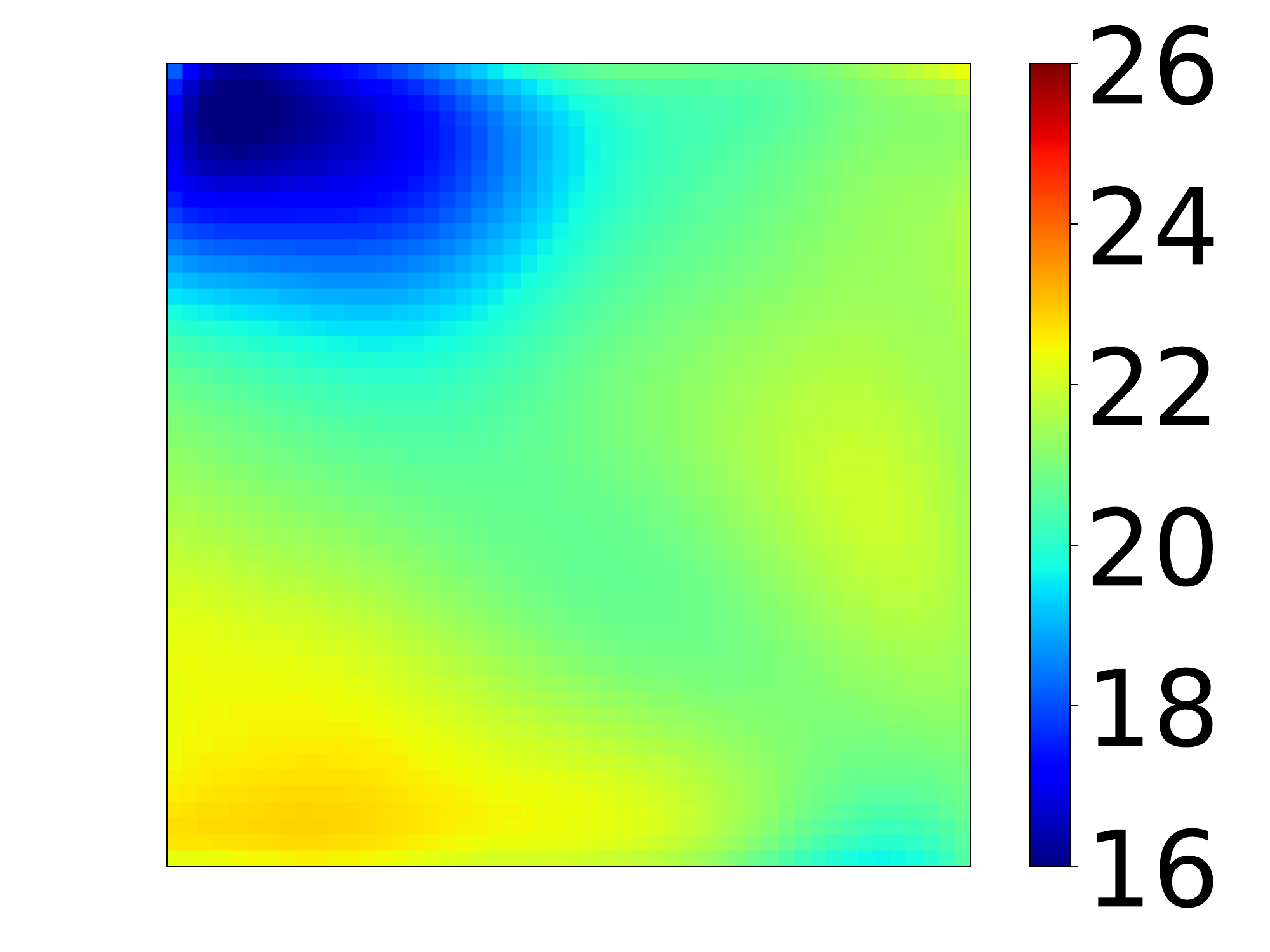}
        \caption{ }
        \label{fig:test_case_2_nwell_2_nt_5_kx0_0_ky0_0}
    \end{subfigure}
    \begin{subfigure}[b]{0.19\textwidth}
        \includegraphics[width=0.99\linewidth]{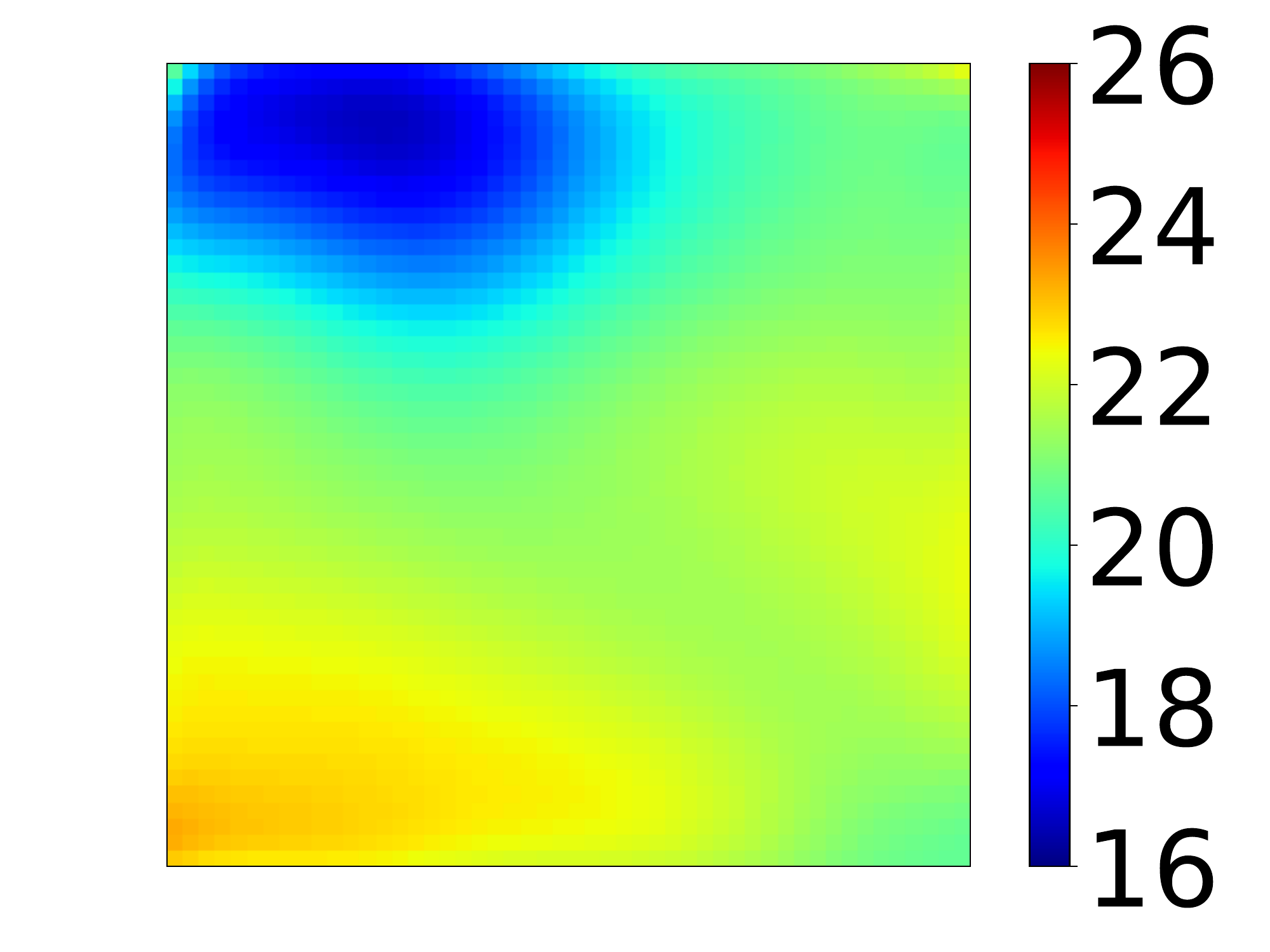}
        \caption{ }
        \label{fig:test_case_2_nwell_2_nt_5_kx0_0_ky0_1}
    \end{subfigure}
    \begin{subfigure}[b]{0.19\textwidth}
        \includegraphics[width=0.99\linewidth]{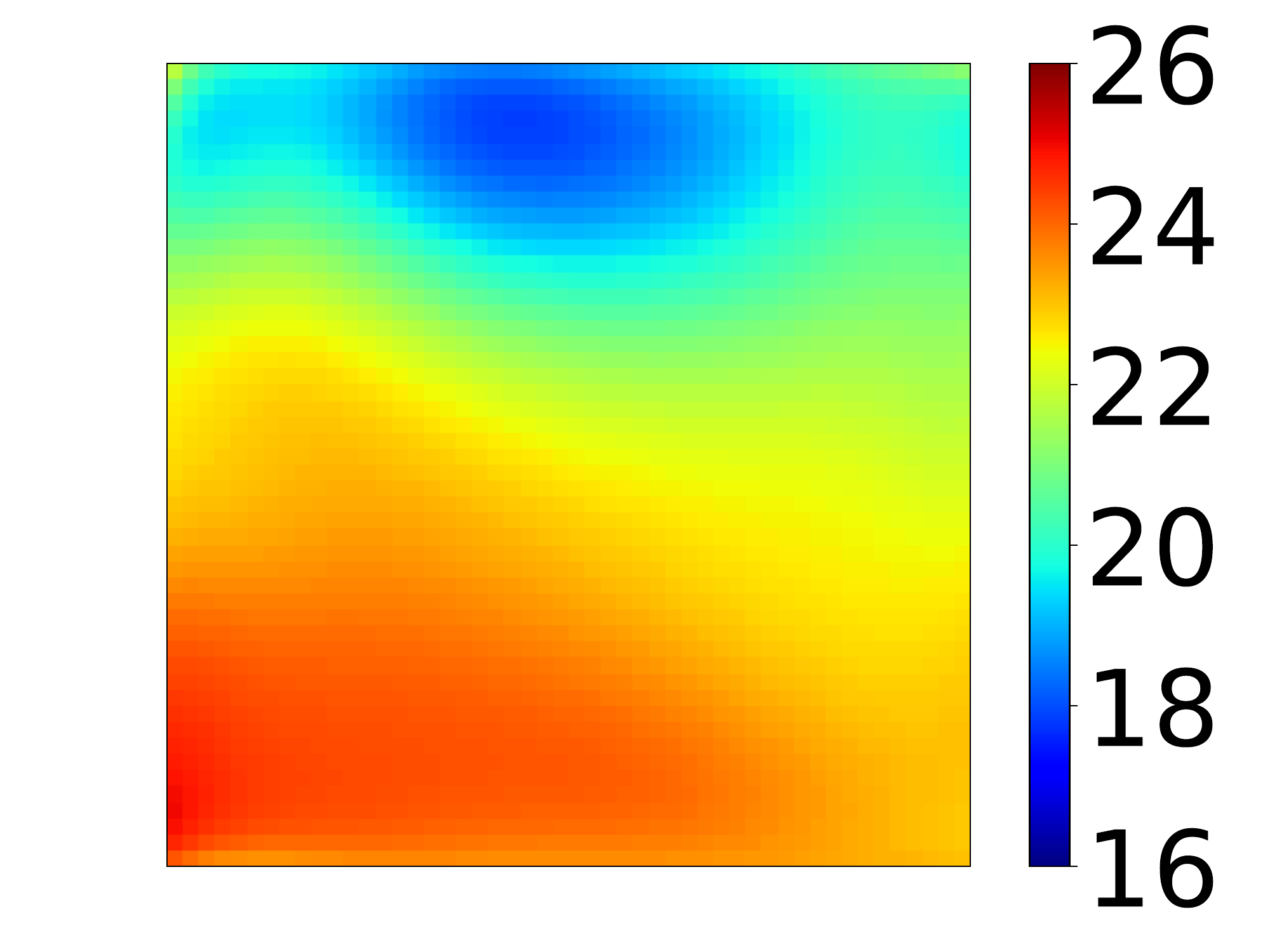}
        \caption{ }
        \label{fig:test_case_2_nwell_2_nt_5_kx0_0_ky0_2}
    \end{subfigure}
    \begin{subfigure}[b]{0.19\textwidth}
        \includegraphics[width=0.99\linewidth]{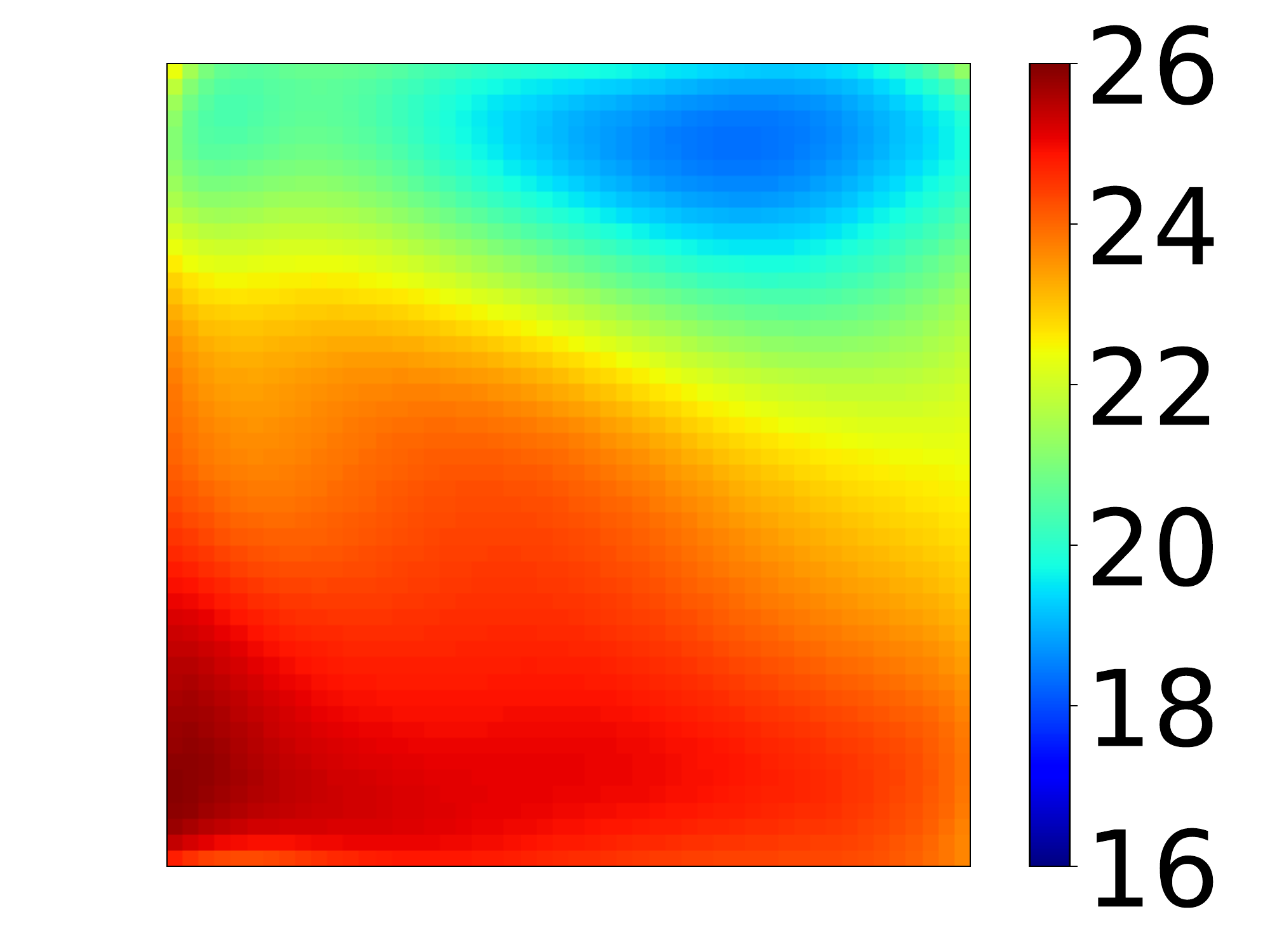}
        \caption{ }
        \label{fig:test_case_2_nwell_2_nt_5_kx0_0_ky0_3}
    \end{subfigure}
    \begin{subfigure}[b]{0.19\textwidth}
        \includegraphics[width=0.99\linewidth]{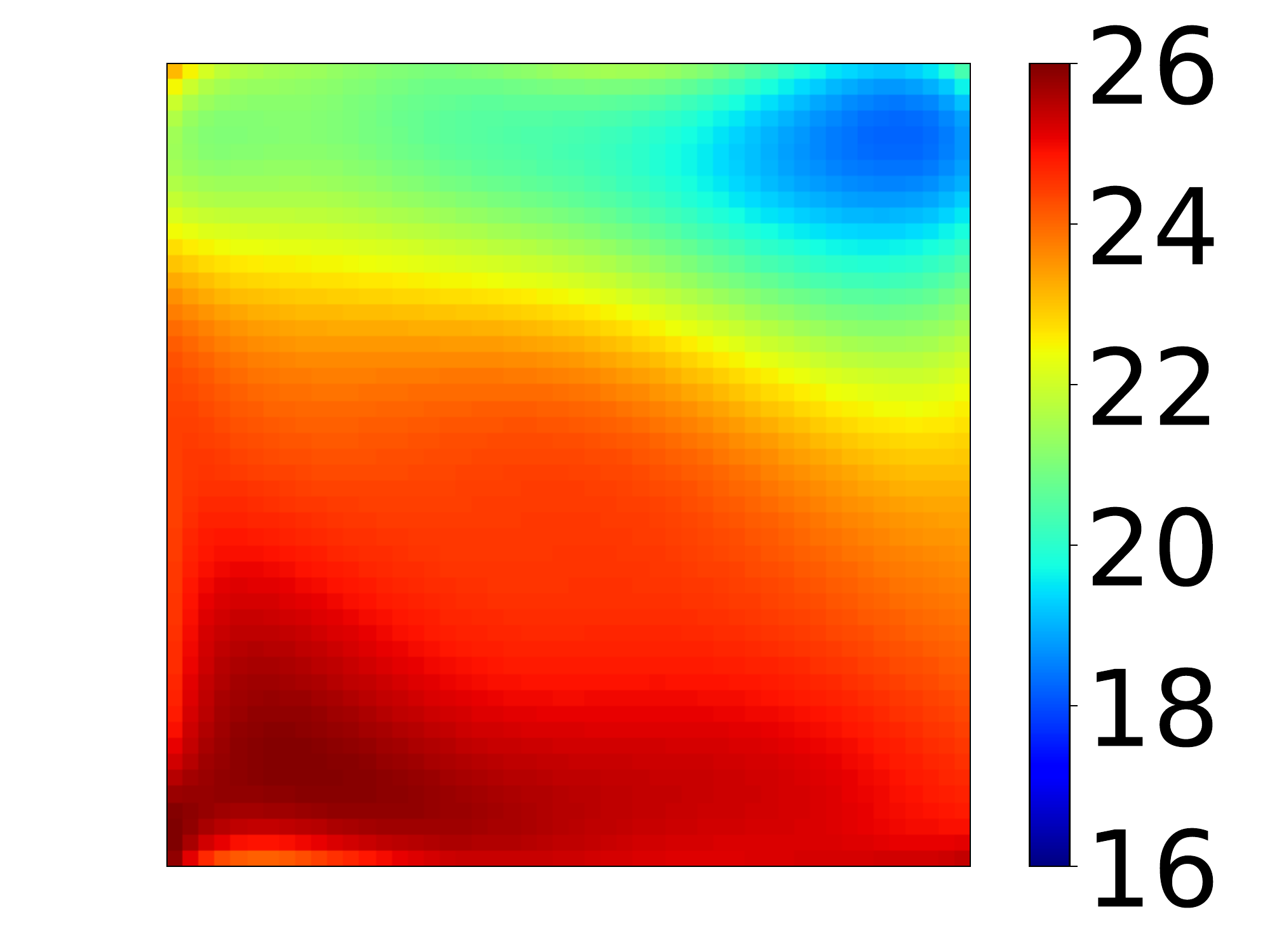}
        \caption{ }
        \label{fig:test_case_2_nwell_2_nt_5_kx0_0_ky0_4}
    \end{subfigure}

    \begin{subfigure}[b]{0.19\textwidth}
        \includegraphics[width=0.99\linewidth]{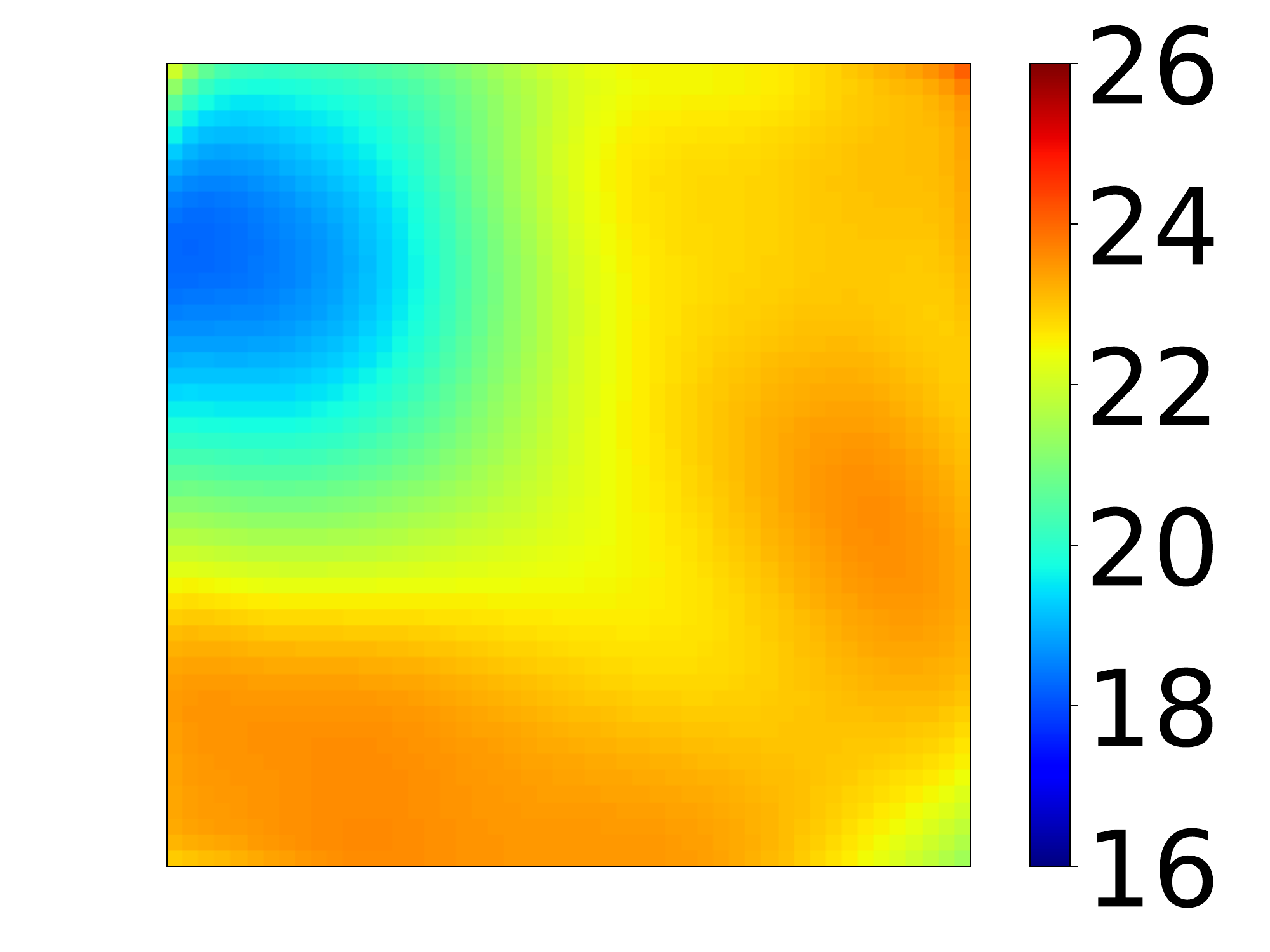}
        \caption{ }
        \label{fig:test_case_2_nwell_2_nt_5_kx0_1_ky0_0}
    \end{subfigure}
    \begin{subfigure}[b]{0.19\textwidth}
        \includegraphics[width=0.99\linewidth]{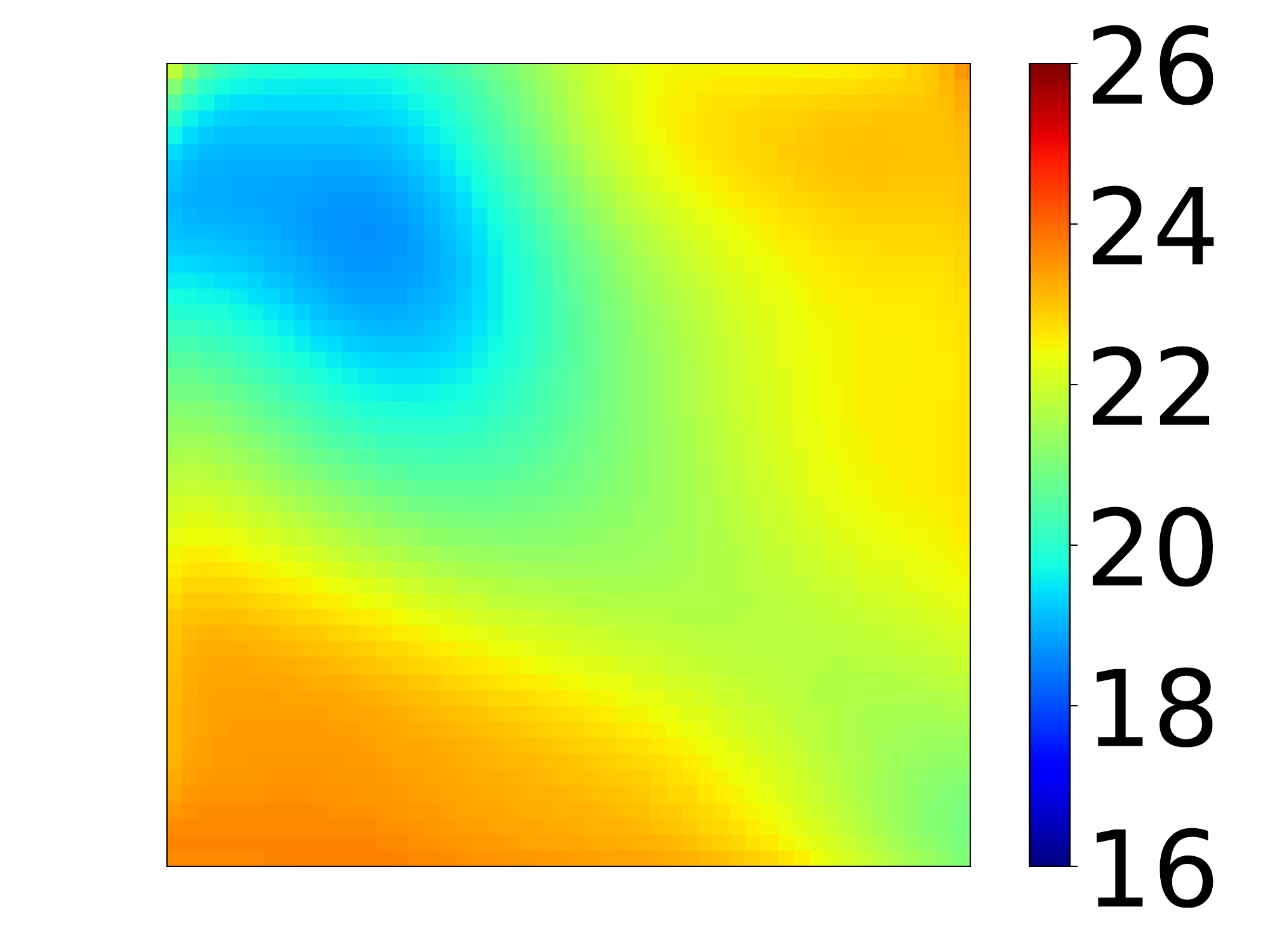}
        \caption{ }
        \label{fig:test_case_2_nwell_2_nt_5_kx0_1_ky0_1}
    \end{subfigure}
    \begin{subfigure}[b]{0.19\textwidth}
        \includegraphics[width=0.99\linewidth]{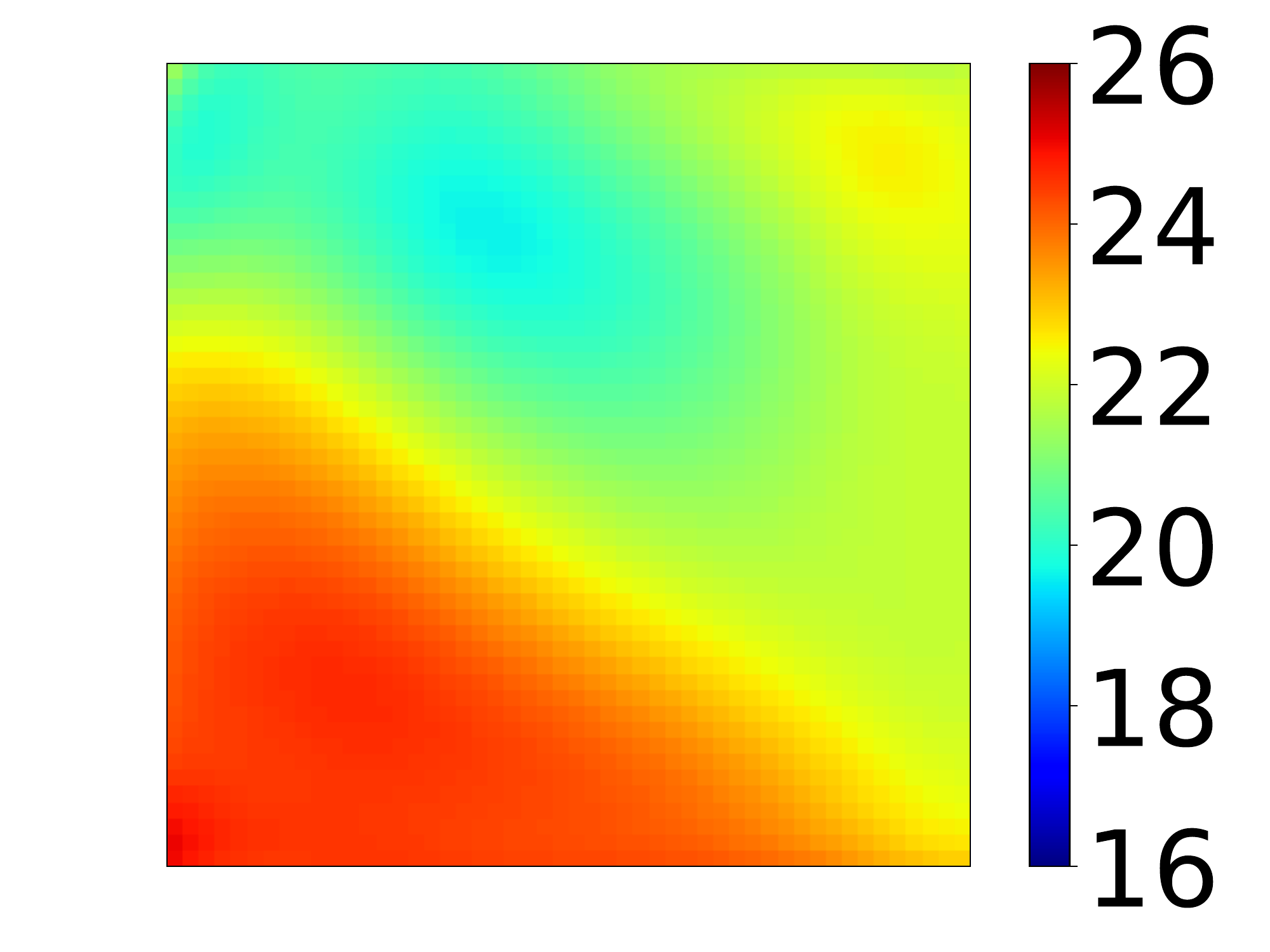}
        \caption{ }
        \label{fig:test_case_2_nwell_2_nt_5_kx0_1_ky0_2}
    \end{subfigure}
    \begin{subfigure}[b]{0.19\textwidth}
        \includegraphics[width=0.99\linewidth]{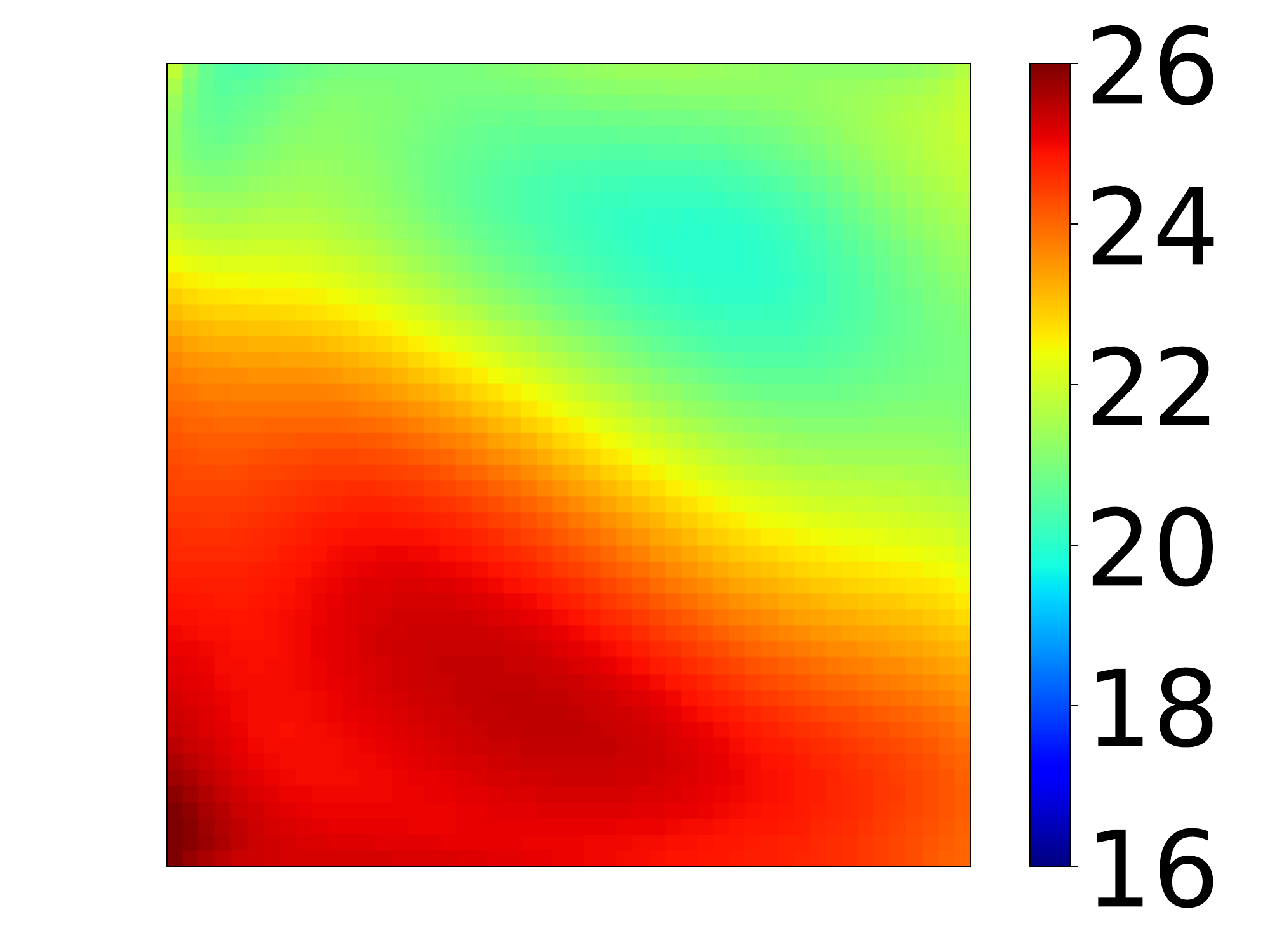}
        \caption{ }
        \label{fig:test_case_2_nwell_2_nt_5_kx0_1_ky0_3}
    \end{subfigure}
    \begin{subfigure}[b]{0.19\textwidth}
        \includegraphics[width=0.99\linewidth]{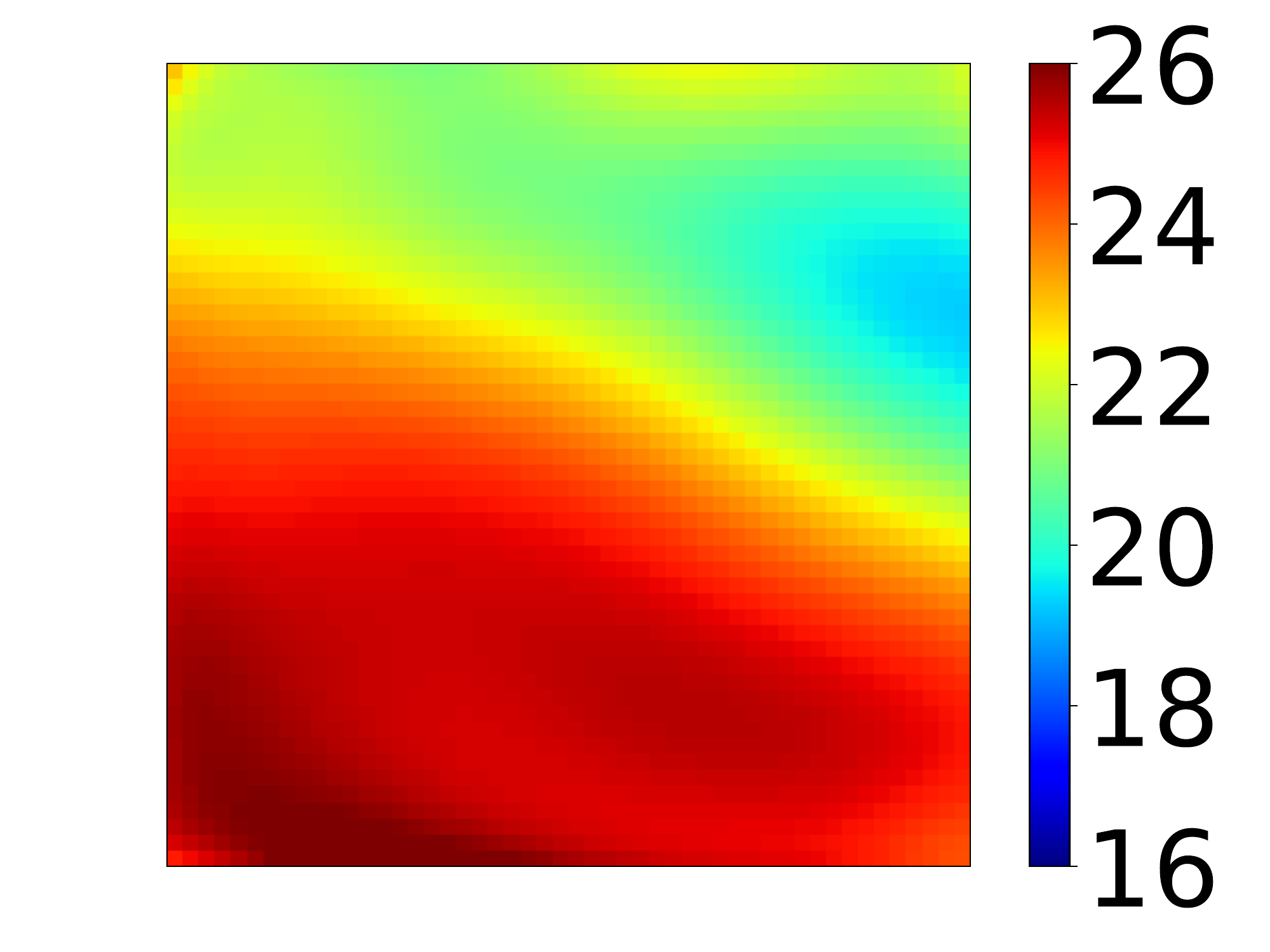}
        \caption{ }
        \label{fig:test_case_2_nwell_2_nt_5_kx0_1_ky0_4}
    \end{subfigure}

    \begin{subfigure}[b]{0.19\textwidth}
        \includegraphics[width=0.99\linewidth]{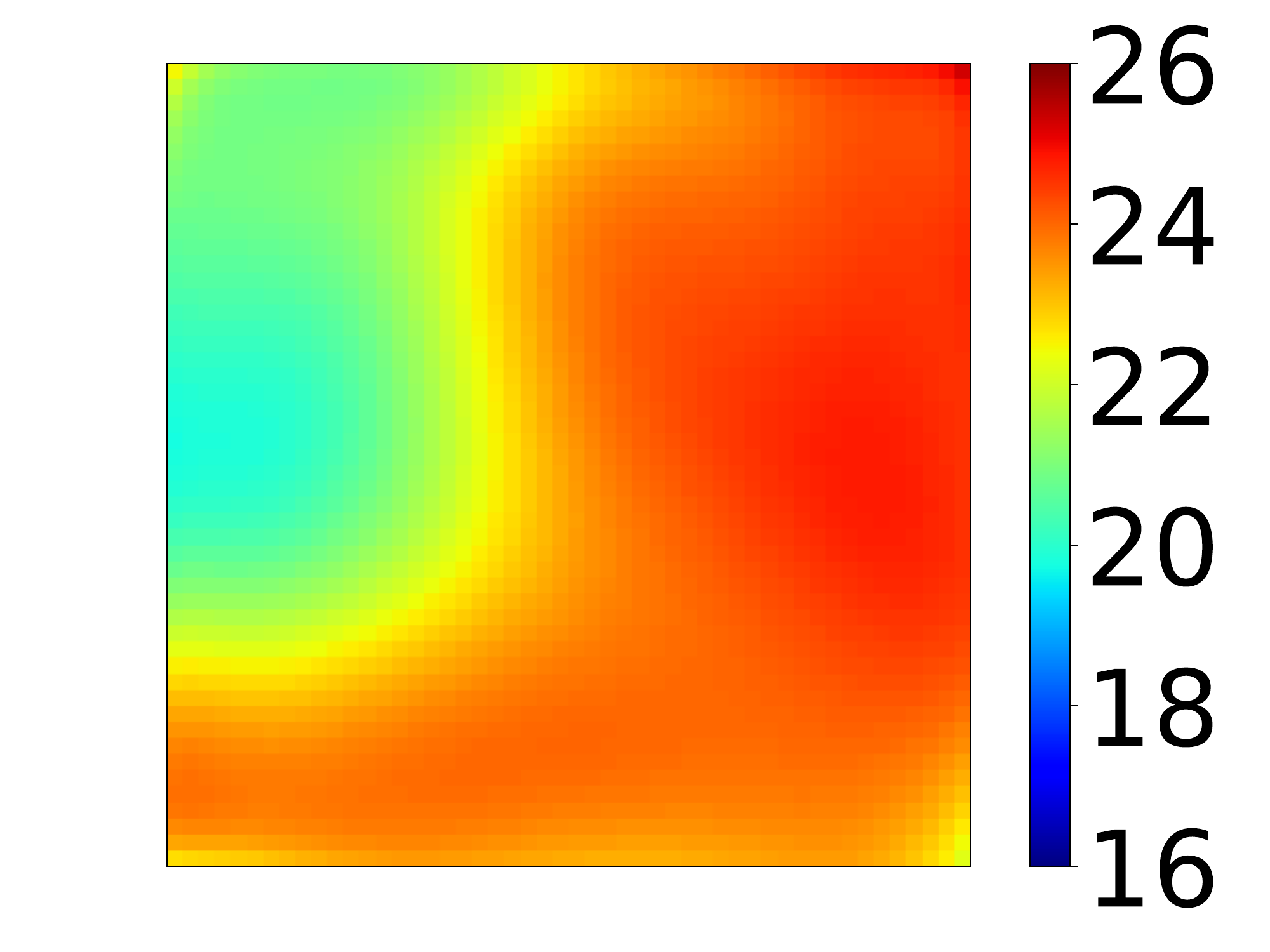}
        \caption{ }
        \label{fig:test_case_2_nwell_2_nt_5_kx0_2_ky0_0}
    \end{subfigure}
    \begin{subfigure}[b]{0.19\textwidth}
        \includegraphics[width=0.99\linewidth]{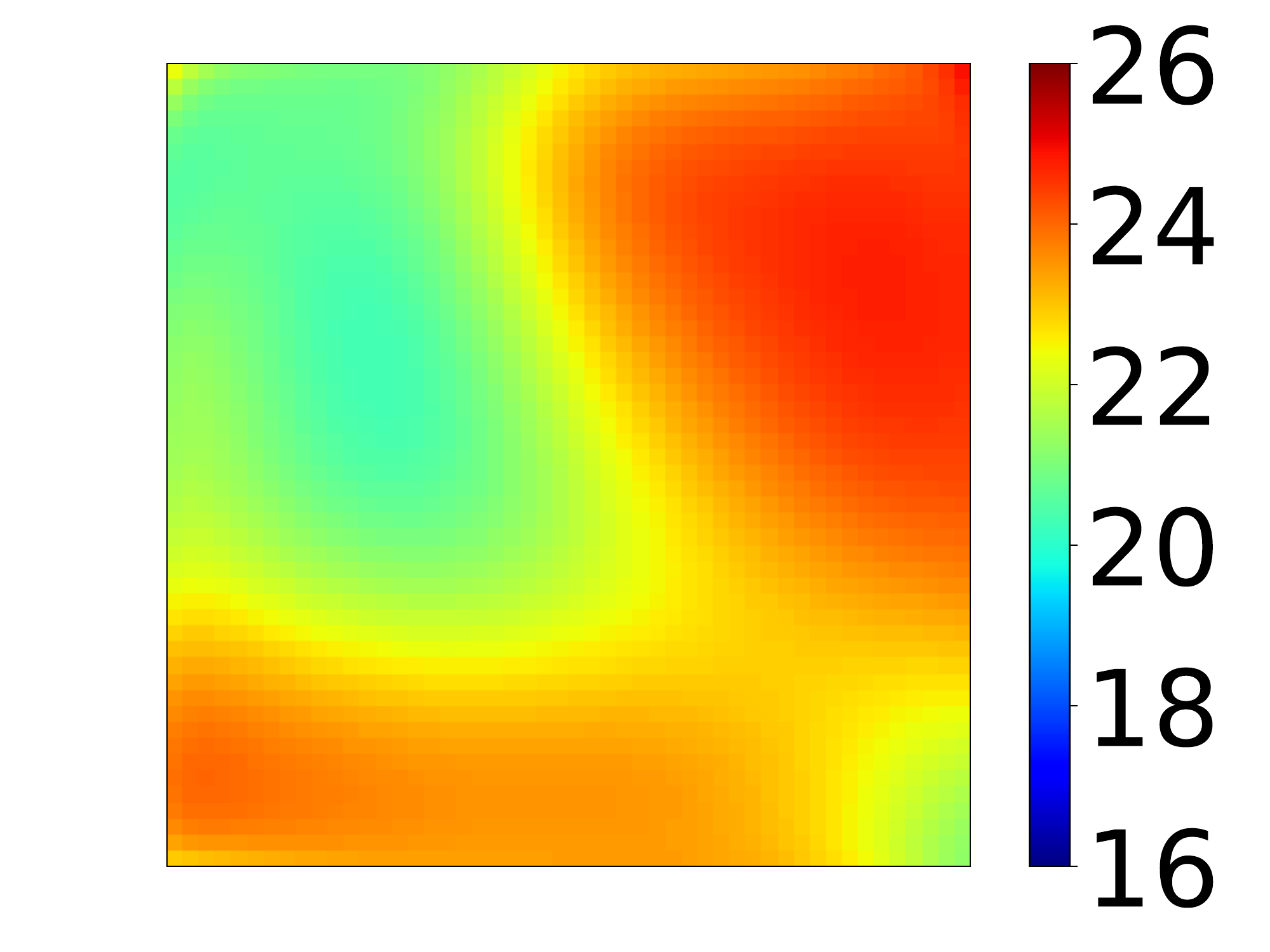}
        \caption{ }
        \label{fig:test_case_2_nwell_2_nt_5_kx0_2_ky0_1}
    \end{subfigure}
    \begin{subfigure}[b]{0.19\textwidth}
        \includegraphics[width=0.99\linewidth]{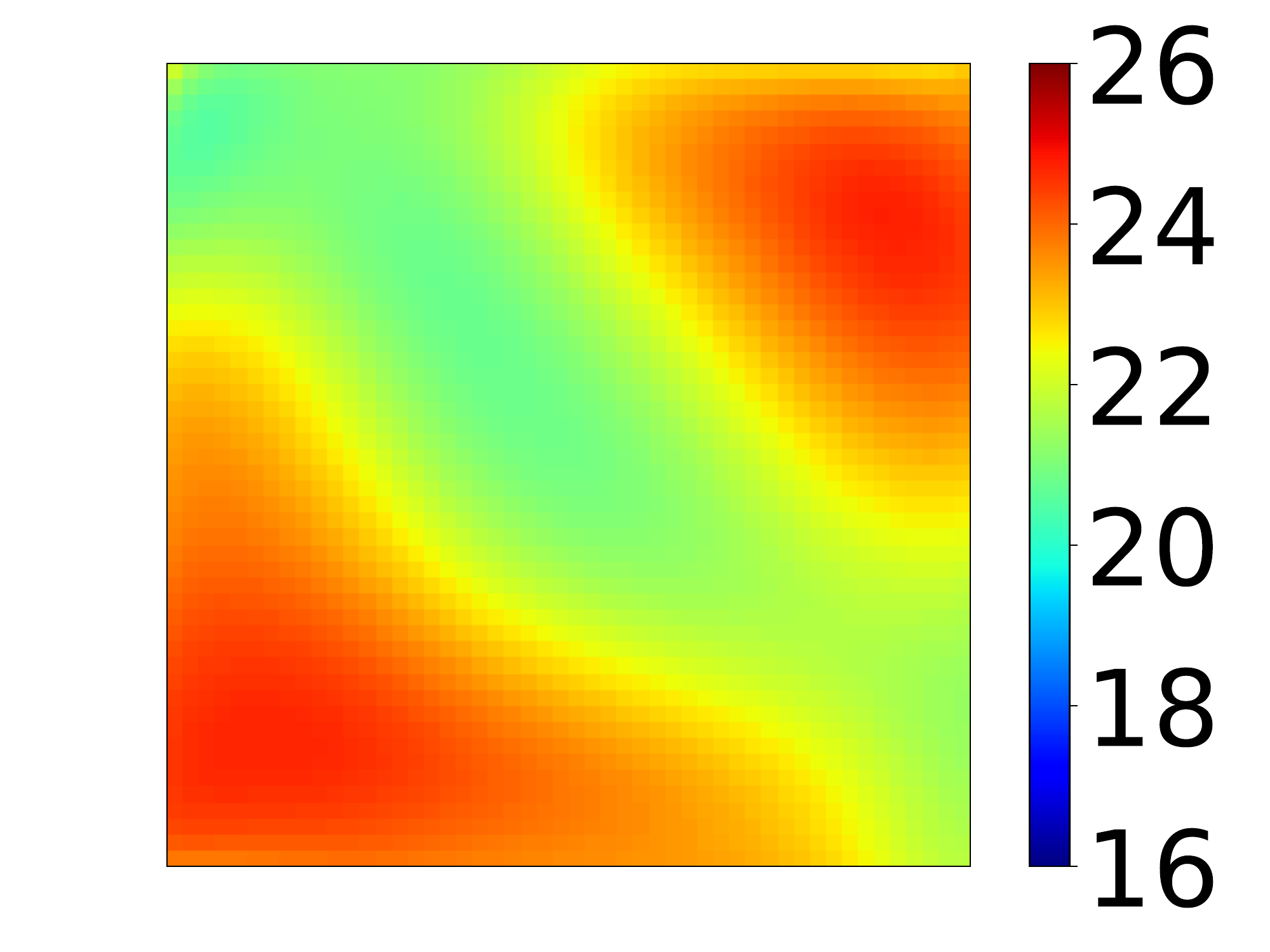}
        \caption{ }
        \label{fig:test_case_2_nwell_2_nt_5_kx0_2_ky0_2}
    \end{subfigure}
    \begin{subfigure}[b]{0.19\textwidth}
        \includegraphics[width=0.99\linewidth]{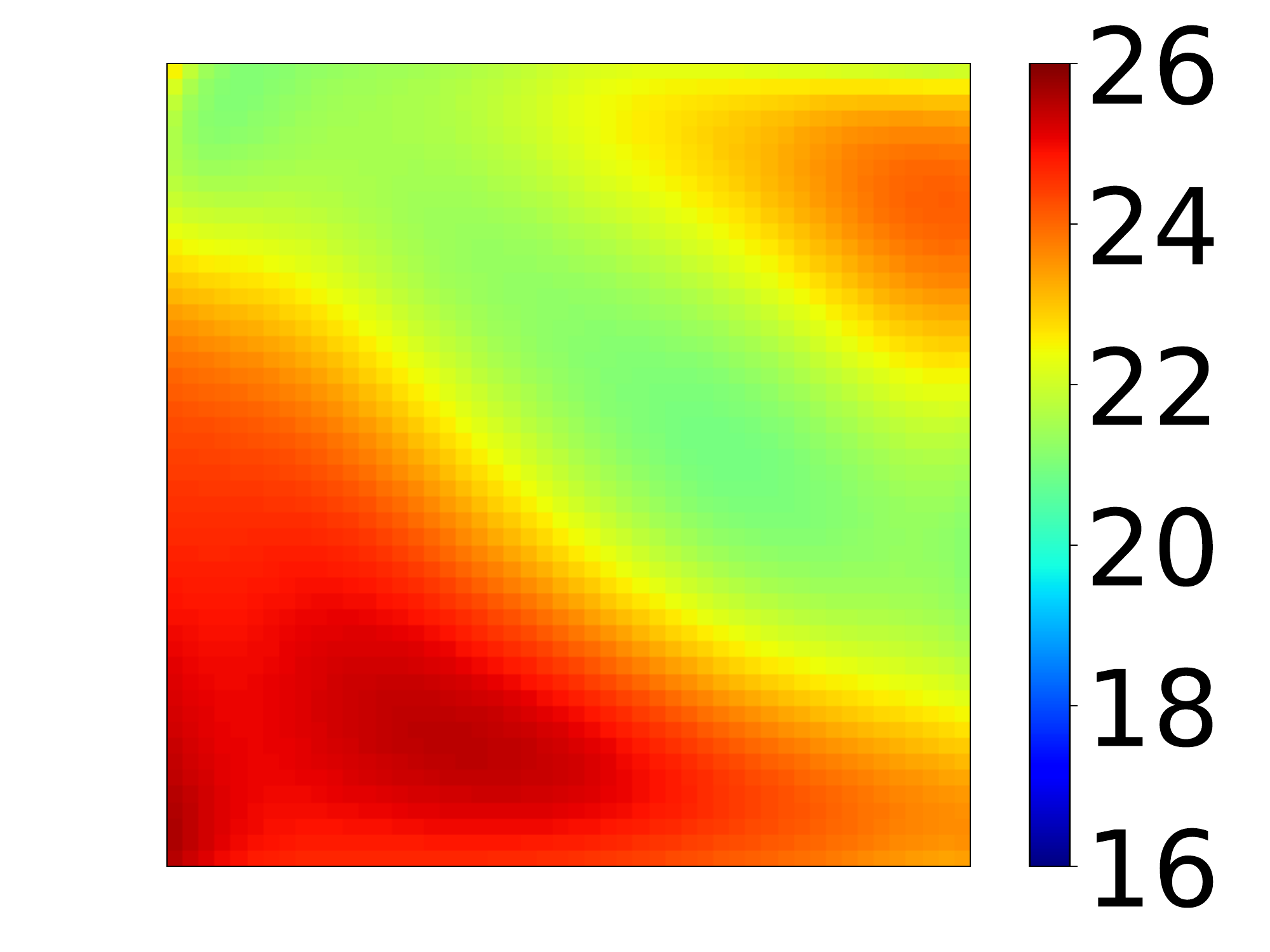}
        \caption{ }
        \label{fig:test_case_2_nwell_2_nt_5_kx0_2_ky0_3}
    \end{subfigure}
    \begin{subfigure}[b]{0.19\textwidth}
        \includegraphics[width=0.99\linewidth]{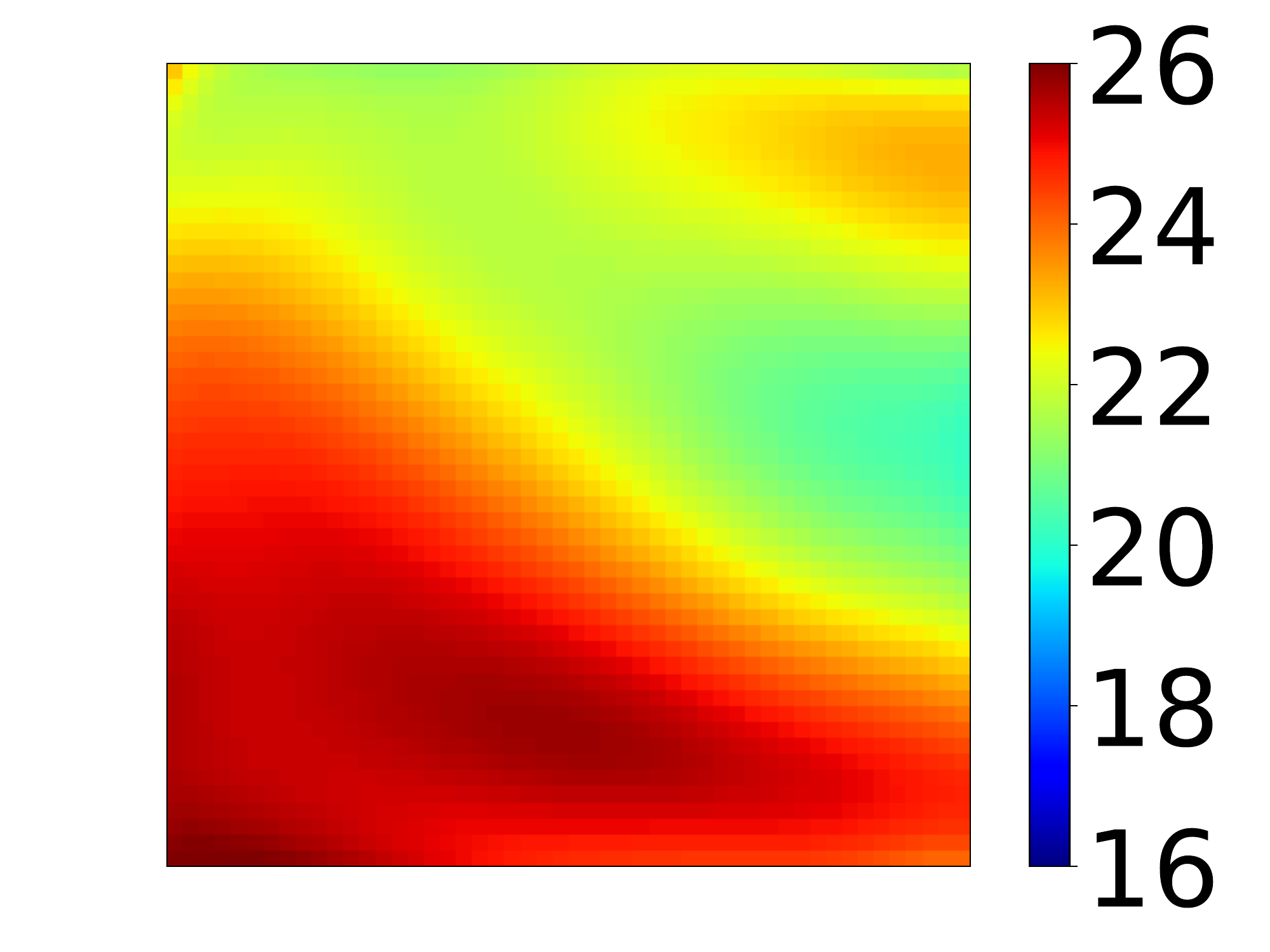}
        \caption{ }
        \label{fig:test_case_2_nwell_2_nt_5_kx0_2_ky0_4}
    \end{subfigure}

    \begin{subfigure}[b]{0.19\textwidth}
        \includegraphics[width=0.99\linewidth]{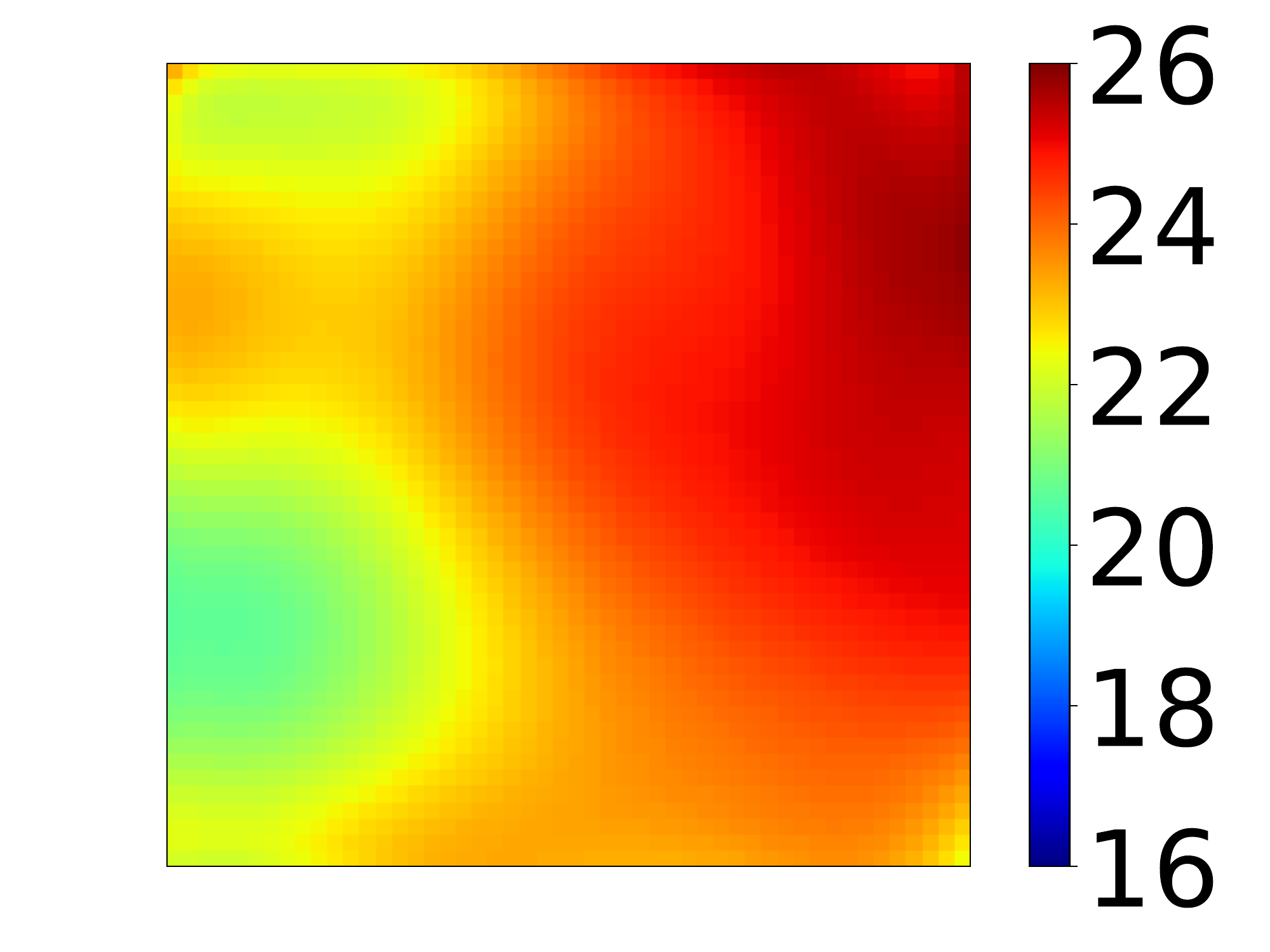}
        \caption{ }
        \label{fig:test_case_2_nwell_2_nt_5_kx0_3_ky0_0}
    \end{subfigure}
    \begin{subfigure}[b]{0.19\textwidth}
        \includegraphics[width=0.99\linewidth]{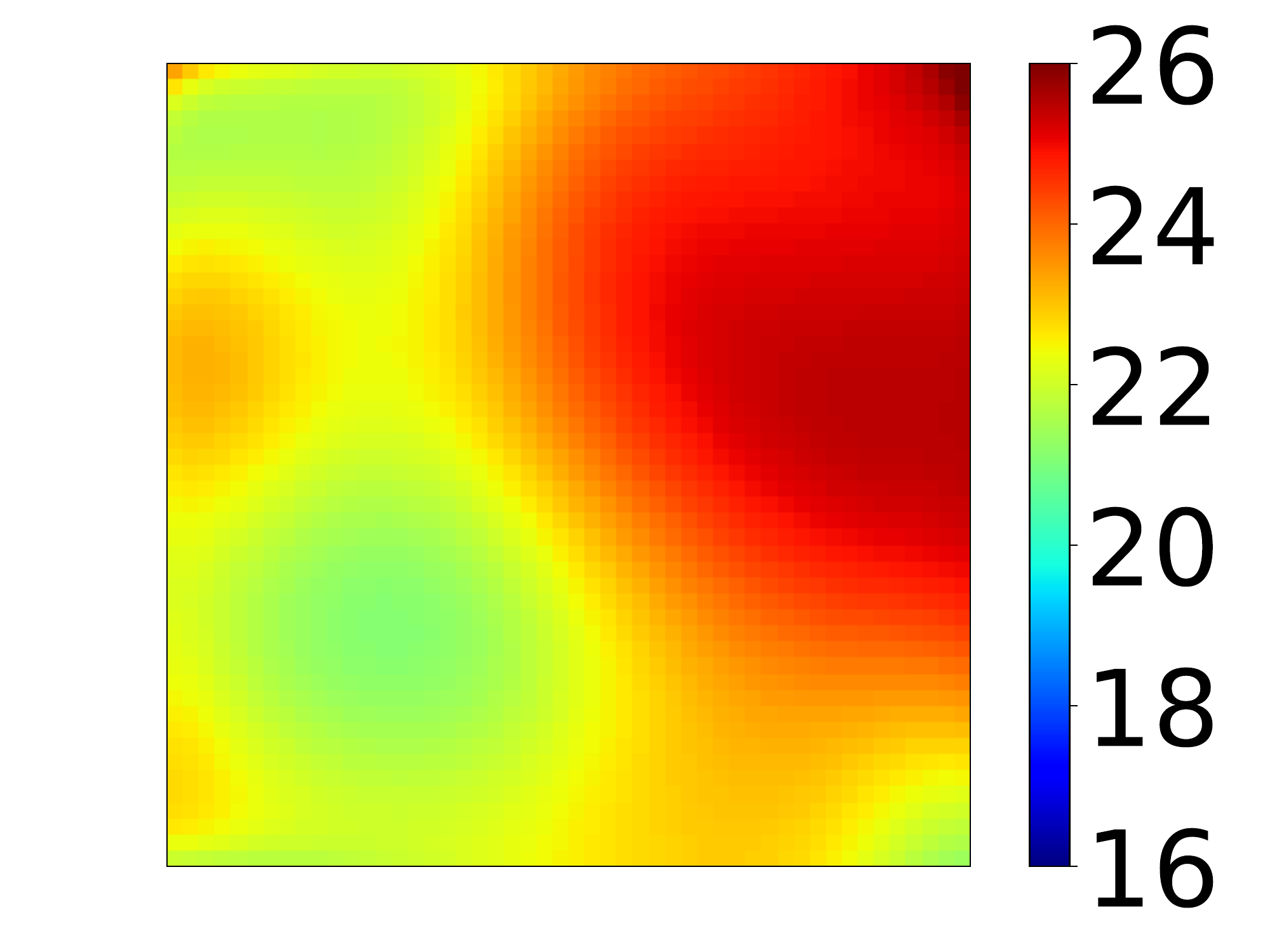}
        \caption{ }
        \label{fig:test_case_2_nwell_2_nt_5_kx0_3_ky0_1}
    \end{subfigure}
    \begin{subfigure}[b]{0.19\textwidth}
        \includegraphics[width=0.99\linewidth]{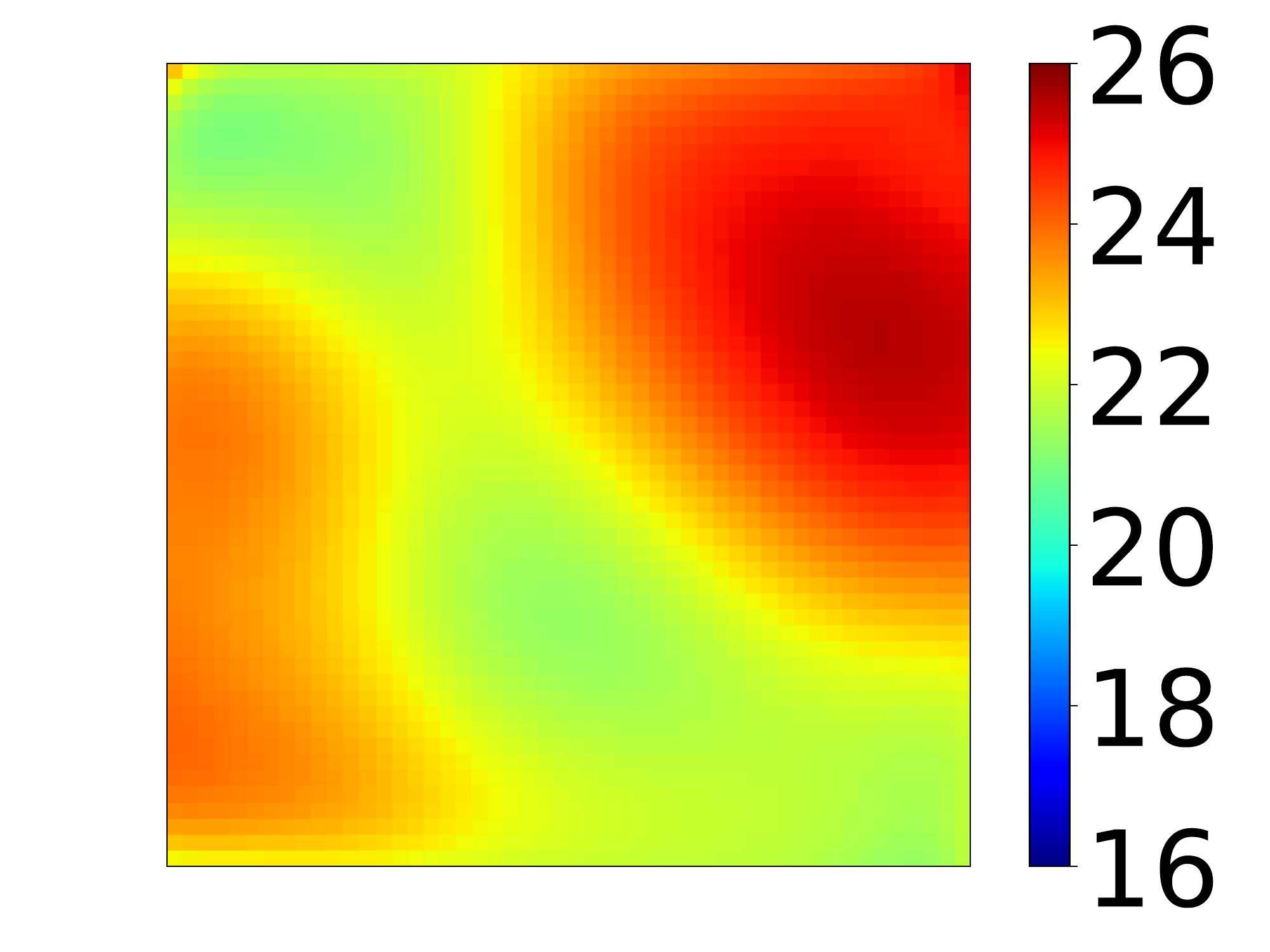}
        \caption{ }
        \label{fig:test_case_2_nwell_2_nt_5_kx0_3_ky0_2}
    \end{subfigure}
    \begin{subfigure}[b]{0.19\textwidth}
        \includegraphics[width=0.99\linewidth]{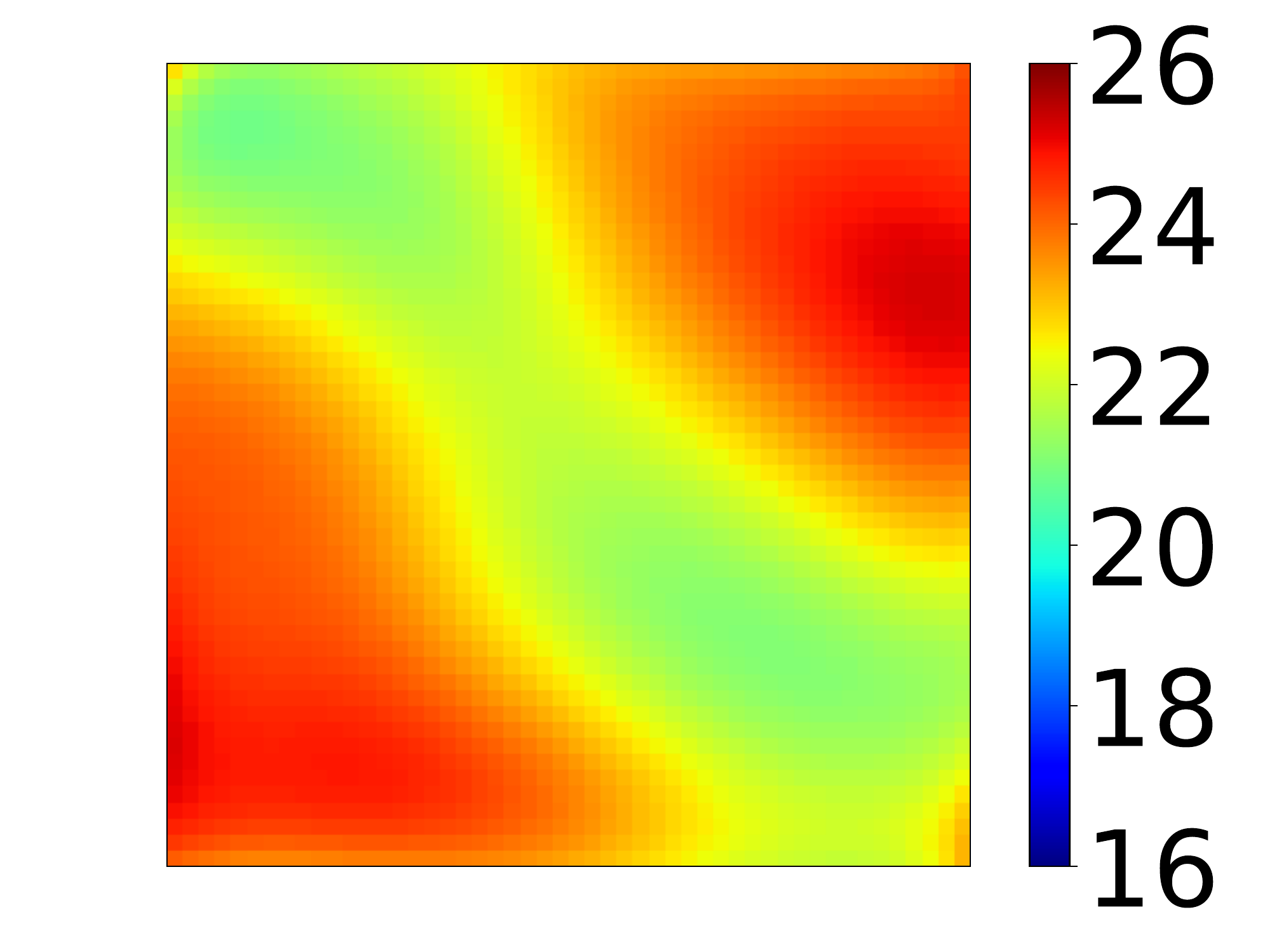}
        \caption{ }
        \label{fig:test_case_2_nwell_2_nt_5_kx0_3_ky0_3}
    \end{subfigure}
    \begin{subfigure}[b]{0.19\textwidth}
        \includegraphics[width=0.99\linewidth]{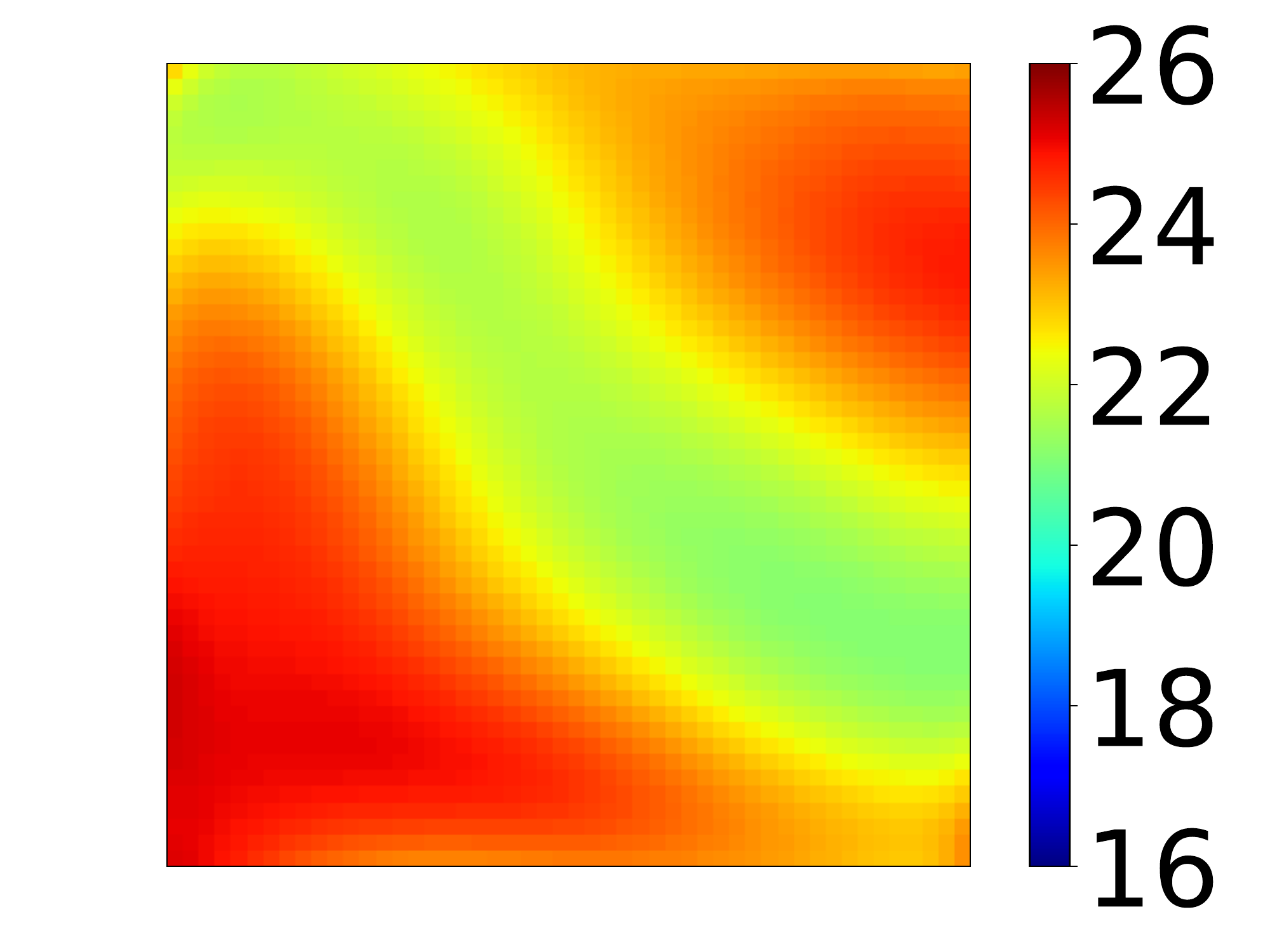}
        \caption{ }
        \label{fig:test_case_2_nwell_2_nt_5_kx0_3_ky0_4}
    \end{subfigure}

    \begin{subfigure}[b]{0.19\textwidth}
        \label{fig:test_case_2_nwell_2_nt_5_kx0_4_ky0_0}
        \includegraphics[width=0.99\linewidth]{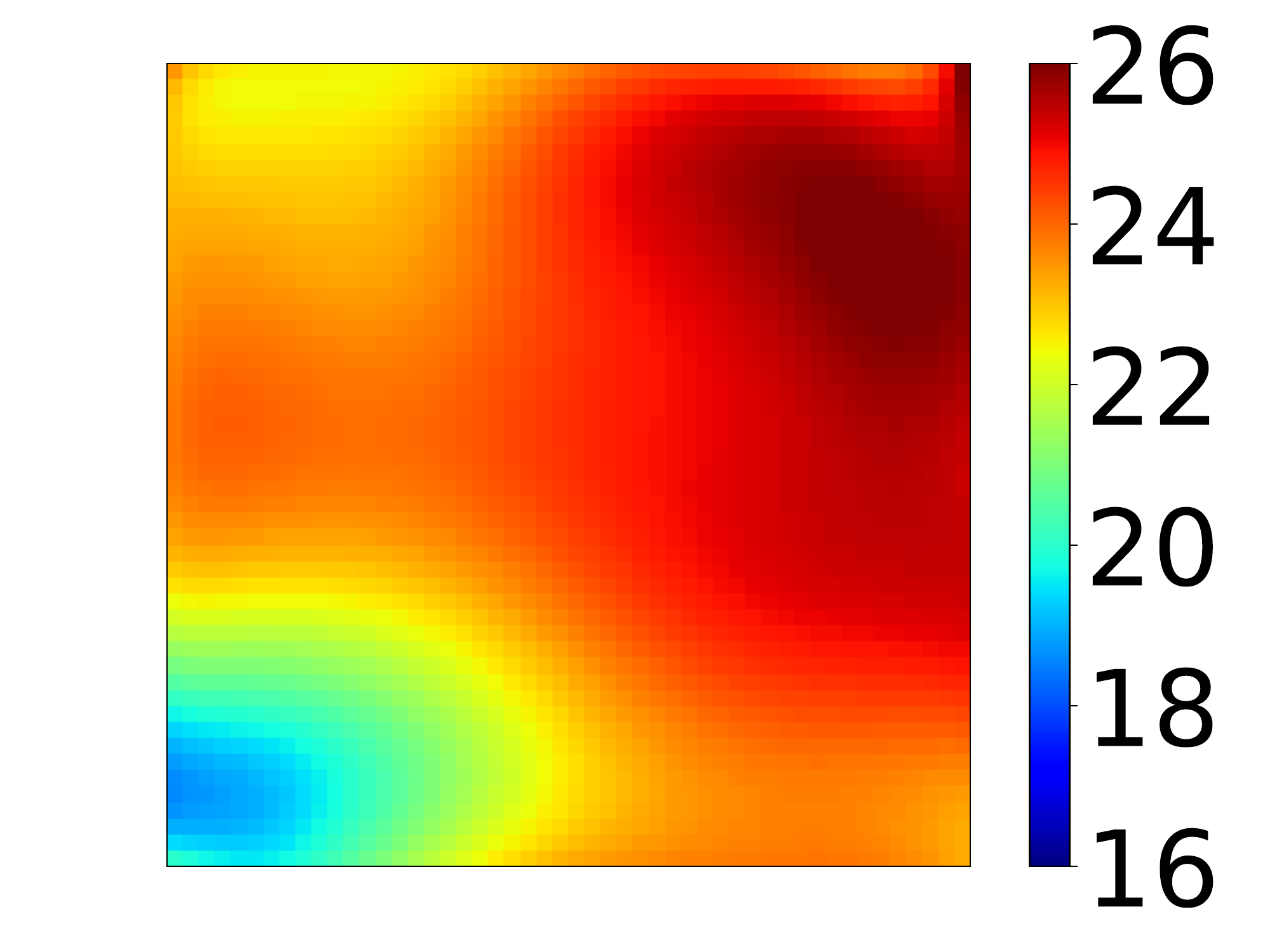}
        \caption{ }
    \end{subfigure}
    \begin{subfigure}[b]{0.19\textwidth}
        \includegraphics[width=0.99\linewidth]{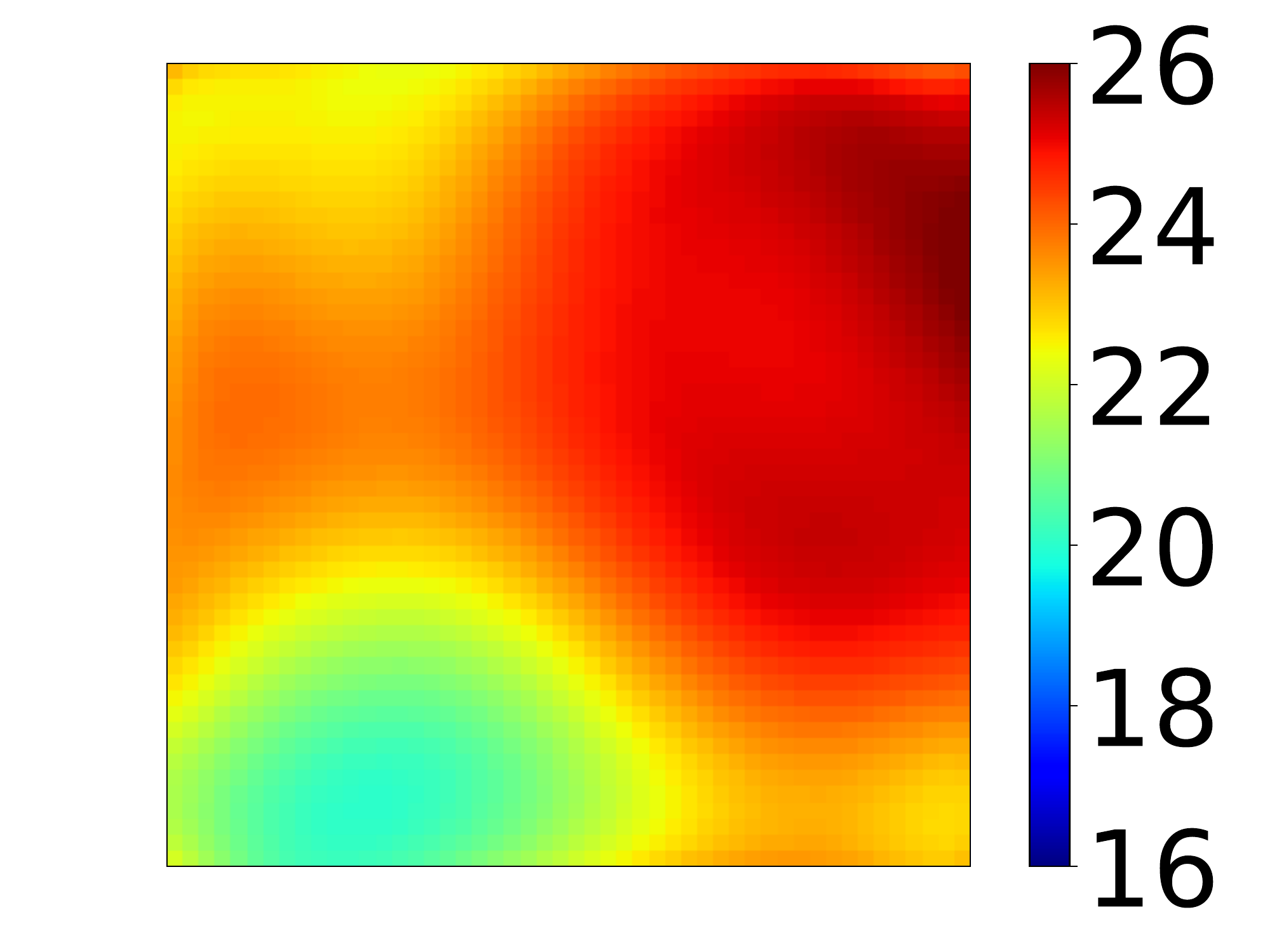}
        \caption{ }
        \label{fig:test_case_2_nwell_2_nt_5_kx0_4_ky0_1}
    \end{subfigure}
    \begin{subfigure}[b]{0.19\textwidth}
        \includegraphics[width=0.99\linewidth]{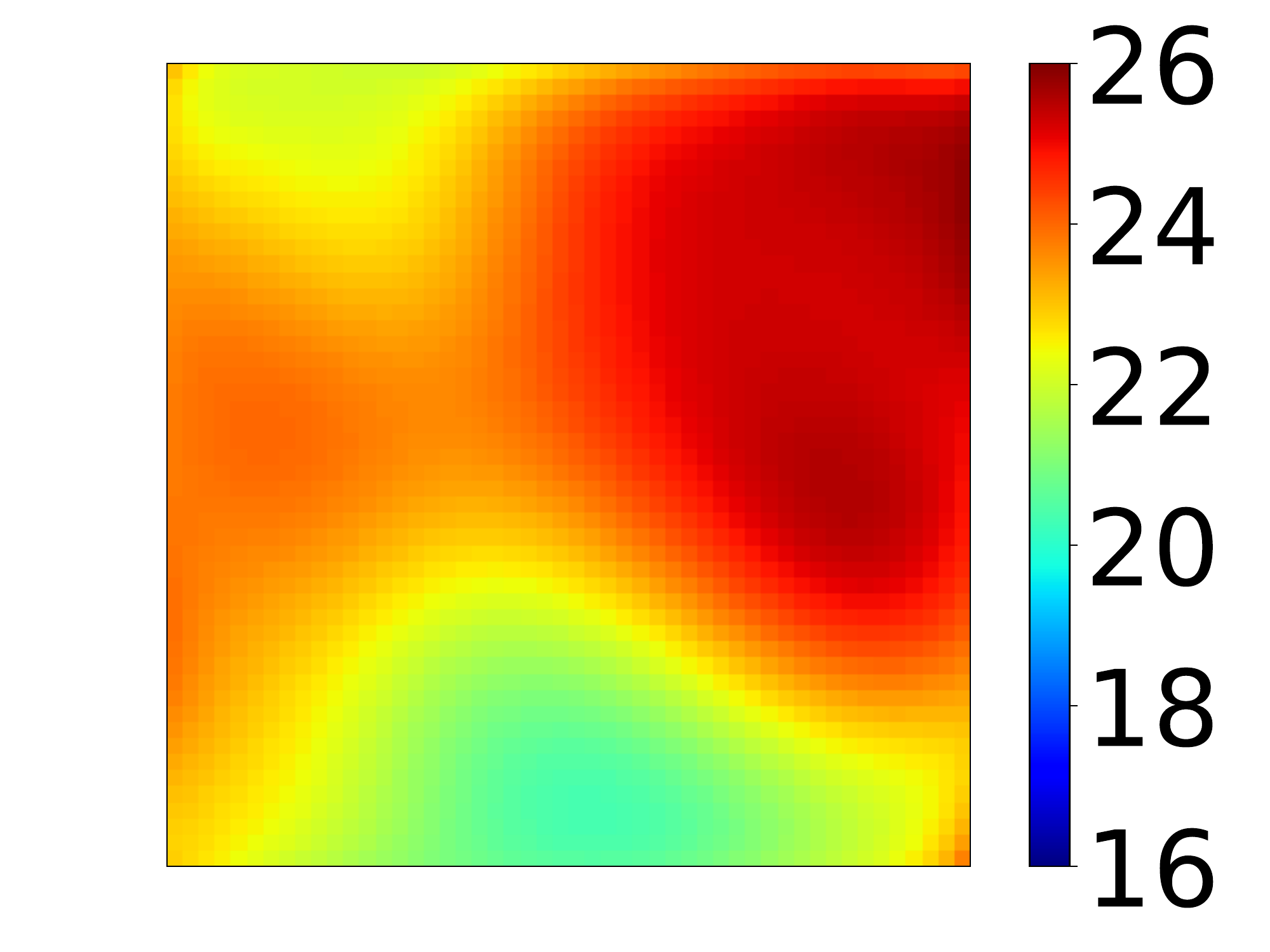}
        \caption{ }
        \label{fig:test_case_2_nwell_2_nt_5_kx0_4_ky0_2}
    \end{subfigure}
    \begin{subfigure}[b]{0.19\textwidth}
        \includegraphics[width=0.99\linewidth]{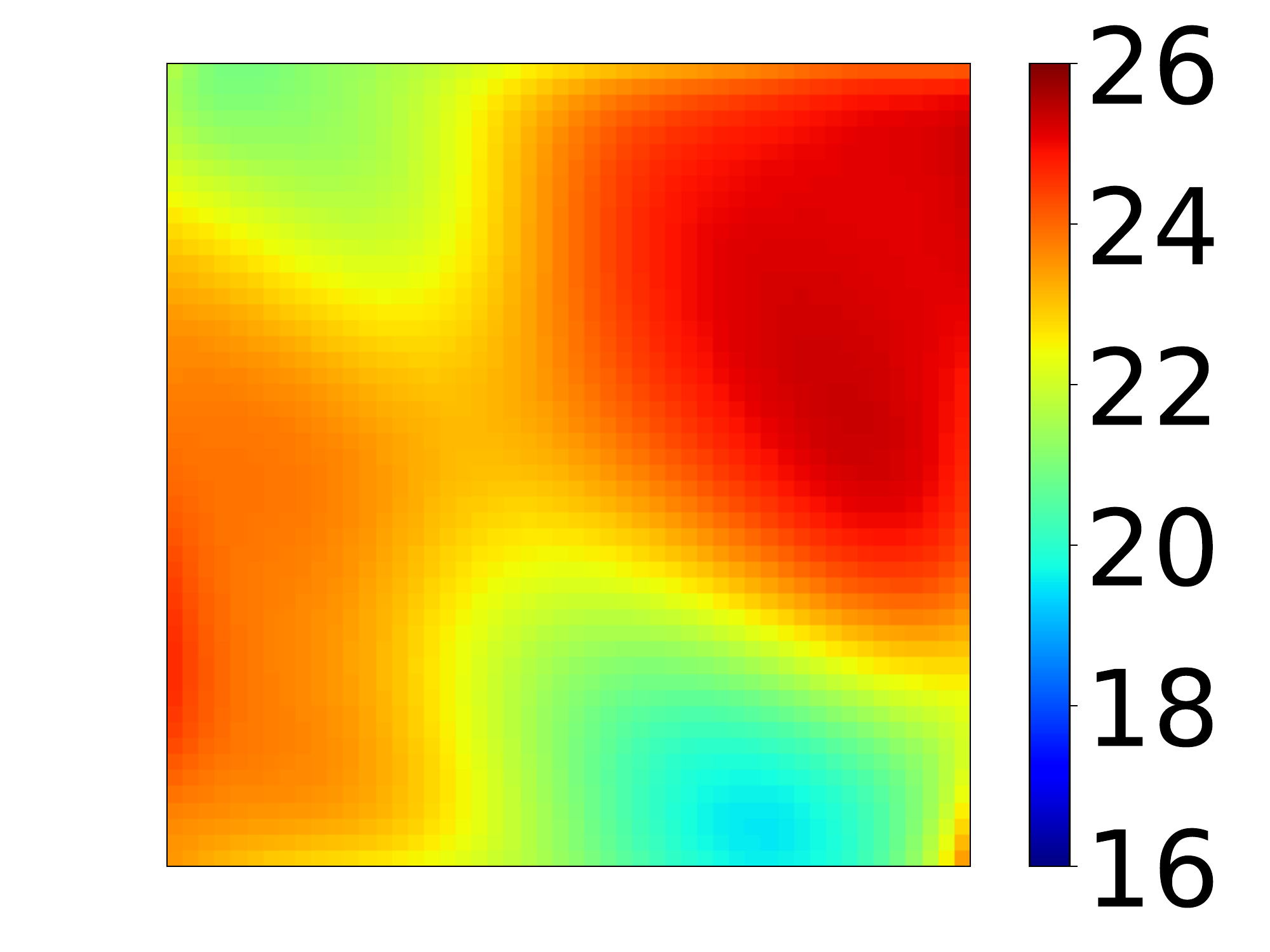}
        \caption{ }
        \label{fig:test_case_2_nwell_2_nt_5_kx0_4_ky0_3}
    \end{subfigure}
    \begin{subfigure}[b]{0.19\textwidth}
        \includegraphics[width=0.99\linewidth]{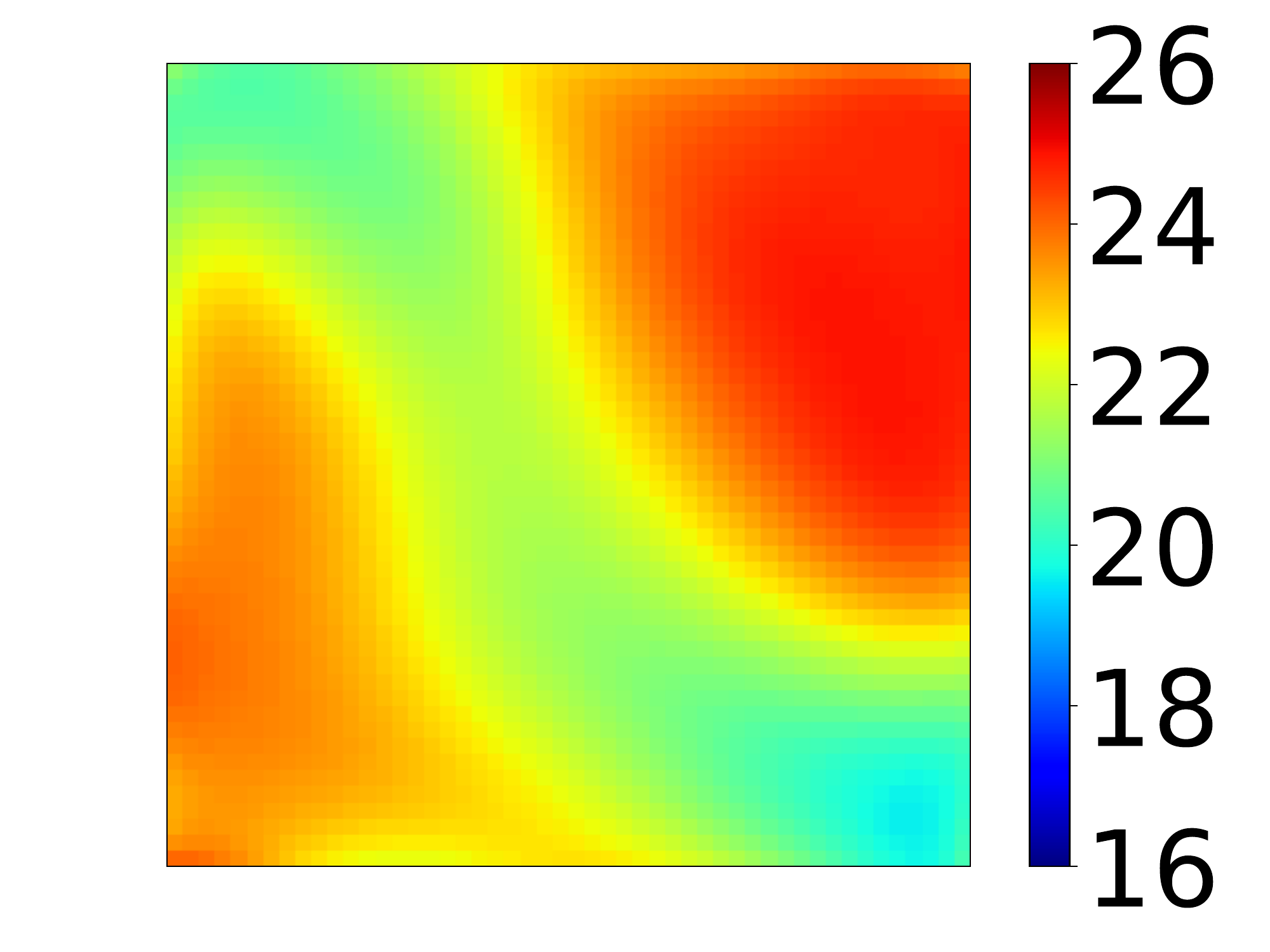}
        \caption{ }
        \label{fig:test_case_2_nwell_2_nt_5_kx0_4_ky0_4}
    \end{subfigure}
    \caption{Values of the expected information gain as a function the first well position if the location of the second well is fixed.
             Each of the figures (a) - (y) corresponds to different coordinates of the second well that corresponds to the minimum of the utility function (dark blue). Case of single permeability and four pressure measurements is presented.}
    \label{fig:test_case_2_nwell_2_nt_5}
\end{figure}

The calculation of $U(\mathbf{d})$ for the case of two new measurement wells is performed in a similar fashion. In the present scenario of measurements, we focus on the examination of PCE-based response surface for $U(\mathbf{d})$ rather then on optimization of utility function. Therefore, optimal design parameters are not provided for the current test case. Instead, the quality of response surface is assessed visually, given the low dimension of the design parameter space and clear geometric meaning of those parameters (aka well location). For that purpose, a $5$ by $5$ uniform lattice of possible locations of the second well has been generated and the expected information gain as a function of the location of the first well is plotted in Fig.~\ref{fig:test_case_2_nwell_2_nt_2} where a single pressure measurement is performed at each new well. For four pressure measurements, the results are shown in Fig.~\ref{fig:test_case_2_nwell_2_nt_5}.

It can be observed from Fig.~\ref{fig:test_case_2_nwell_2_nt_2} and Fig.~\ref{fig:test_case_2_nwell_2_nt_5} that the minimum of expected information gain is achieved when the exploration wells are drilled close to each other or close to either the production or injection wells. Moreover, the expected information gain is high when all those wells are far from each other. In addition to that, Fig.~\ref{fig:test_case_2_nwell_2_nt_2} and Fig.~\ref{fig:test_case_2_nwell_2_nt_5} demonstrate that optimal experimental design corresponds to the case when measurements are collected at vicinity of domian corners that are far from the location of injection and production wells. The latter is in agreement with the variance distribution of logarithm of permeability and pressure shown in Fig.~\ref{fig:test_case_2_logperm_sigma} and Fig.~\ref{fig:test_case_2_pressure_sigma}, respectively. Therefore, the proposed technique provides reasonable estimates for $U(\mathbf{d})$ in the scenario concerned.

Finally, the proposed PCE-based expected information gain provides reasonable approximation of the utility function in both cases (single and multiple pressure measurements) and for single and two additional wells utilized for measurements. In all of the cases, the estimates concerned are in agreement with variation of permeability and pressure measurements collected at locations with higher variance of permeability and pressure provide more information about permeability distribution. Additionally, PCE-based expected information gain $U(\mathbf{d})$ reflects generic dependencies between location of exploration wells and magnitude of $U(\mathbf{d})$. Taking measurements at spatially close points of the reservoir or at the neighborhood of injection and production wells is definitely not the optimal strategy of experiment. Therefore, PCE-based response surface provides reasonable representation of $U(\mathbf{d})$ and allows one to determine parameters of the optimal experimental design.

\section{Concluding Remarks}
\label{sec:conclusion}
In the current manuscript, we introduced a novel PCE approach to optimal Bayesian experimental design.
The central idea of the method is to derive a surrogate model for the expected information gain $U(\mathbf{d})$ without direct evaluations of $U(\mathbf{d})$ itself. For that purpose, PCE is utilized to derive response surface for the utility function via values of KL-divergence computed for various values of model parameters, noise and design parameters that have been sampled in agreement with prior distribution, normal distribution for the noise and a uniform distribution for design parameters. Given the PCE properties and the utilized sampling strategy, we were able to build a PCE approximation for the utility function $U(\mathbf{d})$ from the computed KL-divergence values as a projection of the PCE for the KL-divergence on the space of functions that depends only on design parameters $\mathbf{d}$. Moreover, the projection concerned can be computed numerically with the standard PCE tools such as minimization of regularized mean-square error as defined in Eq.~\eqref{eq:minimization_problem}. Therefore, the proposed approach could be easily implemented using standard machine-learning libraries~\citep{Scikit_Learn}.

The computational advantages of the proposed approach is evident as the cost of constructing a response surface for $U(\mathbf{d})$ is comparable to several dozens evaluations of the utility function with standard MCMC techniques. Therefore, the overall computational cost for finding the optimal experimental design can be significantly reduced. We notice that PCE-based optimal Bayesian experimental design should be preferably applied to problems with high degree of smoothness. We demonstrated this numerically for models with discontinuous derivatives of the utility function $U(\mathbf{d})$. However, the design parameters $\mathbf{d}$ are well-approximated even for problems with discontinuous derivatives. % Therefore, applicability of proposed PCE-based technique to that class of problems requires further research.

Finally, we believe that further development of numerical integration techniques based on PCE in the context of Bayesian experimental design is a promising area of research and applications because of several reasons. First of all, in typical setting of the experimental design optimization problem the process of measurements is controlled by moderate number of parameters. Therefore, high quality PCE response surface for expected information gain could be easily developed. Secondly, the assumption about normal distribution of noise is quite common in practice. Therefore, the problem setup for Eq.~\eqref{eq:central_equation} has strong connection to engineering practice and allows one to utilize the developed technique to optimize experimental design in a realistic setting.

\newpage
\appendix
\section{Approximation for KL-divergence}
\label{sec:approximation}
For a normally distributed noise $\mathbf{\eta}$ with a standard deviation $\sigma$, it is possible to derive an approximate expression for the KL-divergence between the prior and the posterior distributions which is valid for relatively small $\sigma$ and for one-dimensional problems. For that purpose, Laplace approximation technique~\citep{Azevedo-filho94laplace'smethod} is utilized. It is supposed that the observations vector $\mathbf{m}_0$ can be represented as combination of the noise $\mathbf{\eta}$ and the signal $f(\mathbf{\theta}_0, \mathbf{d})$:
\begin{equation}
    \label{eq:observation_vector}
    \mathbf{m}_0 = f(\mathbf{\theta}_0, \mathbf{d}) + \mathbf{\eta}.
\end{equation}
If $\sigma$ is small enough, then the model function $f(\mathbf{\theta}, \mathbf{d})$ can be linearized as following:
\begin{equation}
    \label{eq:linearization}
    f(\mathbf{\theta}, \mathbf{d}) \approx f(\mathbf{\theta}_0, \mathbf{d}) + \frac{\partial f(\mathbf{\theta}_0, \mathbf{d})}{\partial \mathbf{\theta}} \delta \mathbf{\theta}
\end{equation}
where $\delta\mathbf{\theta} = \mathbf{\theta} - \mathbf{\theta}_0$. The linearization in Eq.\eqref{eq:linearization} can be utilized to transform the expression for the likelihood function to the following form:
\begin{equation}
    \label{eq:linearization_and_likelihood}
    p(\mathbf{m} | \mathbf{\theta}, \mathbf{d}) = p\big(f(\mathbf{\theta}, \mathbf{d}) + \mathbf{\eta} | \mathbf{\theta}, \mathbf{d}\big) = \frac{1}{(2\pi\sigma^2)^{\text{dim}(\mathbf{m})/2}} \exp\bigg(  -\frac{|\mathbf{\eta} - \frac{\partial f}{\partial \mathbf{\theta}}(\mathbf{\theta}_0, \mathbf{d})(\mathbf{\theta} - \mathbf{\theta_0})|^2 }{2\sigma^2}  \bigg).
\end{equation}
In the present test case, one dimensional system ($\dim(\mathbf{\theta}) = 1$) with a uniform prior distribution $p(\mathbf{\theta} | \mathbf{d}) = 1$ is considered. It can be shown that for the given prior distribution and the likelihood as in Eq.~\eqref{eq:linearization_and_likelihood} the posterior distribution is a normal distribution with the following mean and standard deviation:
\begin{equation}
    \label{eq:sigma_theta}
    \sigma_1 = \frac{1}{|\frac{\partial f(\mathbf{\theta}_0, \mathbf{d})}{\partial \mathbf{\theta}}|} \sigma.
\end{equation}
In other words, the posterior distribution $p(\mathbf{\theta}|\mathbf{m}_0, \mathbf{d})$ can be approximated as a normal distribution:
\begin{equation}
    \label{eq:normal_posterior}
    p(\mathbf{\theta}|\mathbf{m}_0, \mathbf{d}) = \mathcal{N}(\mathbf{\theta}, \mathbf{\theta}_{\text{mean}}, \sigma_1 )
\end{equation}
where $\mathcal{N}(\mathbf{\theta}, \mathbf{\theta}_{\text{mean}}, \sigma_1 )$ is the density of a normal probability distribution with a mean value $\mathbf{\theta}_{\text{mean}}$ and a standard deviation $\sigma_1$ evaluated at the point $\mathbf{\theta}$. Combined with the uniform prior distribution over the model parameter space $p(\mathbf{\theta}|\mathbf{d}) = 1$, the KL-divergence in Eq.~\eqref{eq:dkl_analytical} can be computed as following:
\begin{equation}
    \label{eq:dkl_derivation}
    \begin{gathered}
    D_{\text{KL}}\big(f(\mathbf{\theta}_0, \mathbf{d}), \mathbf{d}\big) \approx D_{\text{KL}}(\mathbf{m}_0, \mathbf{d}) = \int \mathcal{N}(\mathbf{\theta}, \mathbf{\theta}_{\text{mean}}, \mathbf{\sigma_1}) \log\bigg(\frac{\mathcal{N}(\mathbf{\theta}, \mathbf{\theta}_{\text{mean}}, \mathbf{\sigma_1})} {p(\mathbf{\theta} | \mathbf{d})}\bigg) d\, \mathbf{\theta} = \\
    = - \int \mathcal{N}(\mathbf{\theta}, \mathbf{\theta}_{\text{mean}}, \mathbf{\sigma_1}) \bigg(  \frac{\log(2\pi)}{2} + \log(\sigma_1) + \frac{|\mathbf{\theta} - \mathbf{\theta}_{\text{mean}}|^2}{2\sigma_1^2} \bigg)  d\, \mathbf{\theta} \\
    = \log\bigg( \frac{\partial f(\mathbf{\theta}_0, \mathbf{d})}{\partial \mathbf{\theta}} \bigg) - \frac{1}{2}\log\big(2\pi\big) - \frac{1}{2} - \log\big(\sigma\big)
    \end{gathered}
\end{equation}

\newpage
\section*{Acknowledgments}

This work is part of the ENOS project (Enabling onshore CO2 storage in Europe) funded by the European Union’s Horizon 2020 research and innovation programme under grant agreement No 653718. The contribution of Dr. Shing Chan from the Oxford Big Data Institute to the reservoir simulation code used in the second test case is greatly acknowledged.
\newpage

\bibliographystyle{unsrtnat}
\bibliography{sample}

\end{document}